\RequirePackage{fix-cm}
\RequirePackage{amsmath}
\documentclass[natbib,smallextended]{svjour3}
\smartqed  % flush right qed marks, e.g. at end of proof
\pdfoutput=1
\usepackage{appendix}
\usepackage{amsmath}
\usepackage{graphicx}
\usepackage[caption=false]{subfig}
\usepackage{lineno}
\usepackage{array}
\usepackage{longtable}
\usepackage{natbib}
\usepackage{hyperref}
\usepackage{multirow}
\usepackage{amsmath}
\usepackage{listings}
\usepackage{color}
\usepackage{graphics}
\usepackage{graphicx}
\usepackage{tabularx}
\usepackage{longtable}
\usepackage{array} % for extrarowheight
\usepackage{booktabs}
\usepackage[table]{xcolor}
\usepackage{epstopdf} 
\usepackage{multirow}
\usepackage{pdflscape}
\usepackage{afterpage}
\usepackage{capt-of}
\usepackage{rotating}
\usepackage{pgfplots}
\pgfplotsset{compat=1.14}
\usepackage{algorithmic}
\usepackage{algorithm}
\usepackage{multicol}
\usepackage[T1]{fontenc}
\usepackage[utf8]{inputenc}
\usepackage{placeins}
\usepackage{authblk}
\usepackage{placeins}
\usepackage{ragged2e}
\usepackage[export]{adjustbox}
\usepackage{adjustbox}
\usepackage[misc]{ifsym}

\vfill
\newcommand*\patchAmsMathEnvironmentForLineno[1]{%
\expandafter\let\csname old#1\expandafter\endcsname\csname #1\endcsname
\expandafter\let\csname oldend#1\expandafter\endcsname\csname end#1\endcsname
\renewenvironment{#1}%
{\linenomath\csname old#1\endcsname}%
{\csname oldend#1\endcsname\endlinenomath}}% 
\newcommand*\patchBothAmsMathEnvironmentsForLineno[1]{%
\patchAmsMathEnvironmentForLineno{#1}%
\patchAmsMathEnvironmentForLineno{#1*}}%
\AtBeginDocument{%
\patchBothAmsMathEnvironmentsForLineno{equation}%
\patchBothAmsMathEnvironmentsForLineno{align}%
\patchBothAmsMathEnvironmentsForLineno{flalign}%
\patchBothAmsMathEnvironmentsForLineno{alignat}%
\patchBothAmsMathEnvironmentsForLineno{gather}%
\patchBothAmsMathEnvironmentsForLineno{multline}%
}

\begin{document}

\title{A Bibliometric Analysis of Publications in Computer Networking Research
\thanks{\textbf{{\color{blue}****This work has been submitted to the Springer Scientometrics Journal for possible publication. Copyright may be transferred without notice, after which this version may no longer be accessible****}}}
}

\titlerunning{A Bibliometric Analysis of Publications in Computer Networking Research}        % if too long for running head

\authorrunning{Waleed Iqbal et al.}

%\author{Waleed Iqbal\inst{1}
% \and Junaid Qadir\inst{1} \and
% Gareth Tyson\inst{2} \and Adnan Noor Mian\inst{1} \inst{3} \and \\
% Saeed ul Hassan\inst{1} \and
% Jon Crowcroft\inst{3}
%\authorrunning{Waleed Iqbal et al.}
%
%\institute{Information Technology University, Lahore, Pakistan.\\
% \email{\{waleed.iqbal, junaid.qadir, adnan.noor, saeed-ul-hassan\}@itu.edu.pk}\\
% \and
% Queen Mary University of London, United Kingdom.\\
% \email{g.tyson@qmul.ac.uk}
% \and
% Computer Laboratory, University of Cambridge, United Kingdom\\
%\email{\{anm42, jon.crowcroft\}@cam.ac.uk}}

\author{Waleed Iqbal$^1$  \and
  Junaid Qadir$^1$   \and \newline Gareth Tyson$^2$    \and 
  Adnan Noor Mian$^1$ $^3$    \and
  Saeed-ul-Hassan$^1$ \and 
  Jon Crowcroft$^3$
}


\institute{ \Letter \space Waleed Iqbal (\email{waleed.iqbal@itu.edu.pk})\\\\
  $^1$ Information Technology University, Lahore, Pakistan\\
  $^2$ Queen Mary University of London, United Kingdom\\
  $^3$ Computer Laboratory, University of Cambridge, United Kingdom
}

%\date{Received: DD Month YEAR / Accepted: DD Month YEAR}

\maketitle
\newcommand\gareth[1]{\textbf{\textcolor{blue}{GT: #1}} }

\begin{abstract}
Computer networking is a major research discipline in computer science, electrical engineering, and computer engineering. The field has been actively growing, in terms of both research and development, for the past hundred years. This study uses the article content and metadata of four important computer networking periodicals---IEEE Communications Surveys and Tutorials (COMST), IEEE/ACM Transactions on Networking (TON), ACM Special Interest Group on Data Communications (SIGCOMM), and IEEE International Conference on Computer Communications (INFOCOM)---obtained using ACM, IEEE Xplore, Scopus and CrossRef, for an 18-year period (2000--2017) to address important bibliometrics questions. All of the venues are prestigious, yet they publish quite different research. The first two of these periodicals (COMST and TON) are highly reputed journals of the fields while SIGCOMM and INFOCOM are considered top conferences of the field. SIGCOMM and INFOCOM publish new original research. TON has a similar genre and publishes new original research as well as the extended versions of different research published in the conferences such as SIGCOMM and INFOCOM, while COMST publishes surveys and reviews (which not only summarize previous works but highlight future research opportunities). In this study, we aim to track the co-evolution of trends in the COMST and TON journals and compare them to the publication trends in INFOCOM and SIGCOMM. Our analyses of the computer networking literature include: (a) metadata analysis; (b) content-based analysis; and (c) citation analysis. In addition, we identify the significant trends and the most influential authors, institutes and countries, based on the publication count as well as article citations. Through this study, we are proposing a methodology and framework for performing a comprehensive bibliometric analysis on computer networking research. To the best of our knowledge, no such study has been undertaken in computer networking until now.


\end{abstract}

\section{Introduction}
\label{intro}

Bibliometric analysis of a literature is a crucially important source of objective knowledge and information about the quantity and quality of scientific work \citep{narin1994bibliometrics}. In this work we perform a bibliometric analysis of the the literature of the field of computer networking, which is a major research domain in electrical and computer engineering and science. This breadth-wise knowledge saves ample amount of time for researchers to get started with the research of a domain and helps inform about the major trends observed in computer networking publications.

There are several article genres in computer networking, such as conference articles, letters, editorials, surveys, and empirical studies. To keep the scope of this study to manageable proportions, we have focused on journal publications principally (survey and empirical studies) but have also compared journal publications to conference publications in this area. We have selected four exemplar venues that represent the highest standard of research in the field of computer networking---namely, IEEE Communications Surveys and Tutorials (COMST), IEEE/ACM Transactions on Networking (TON), ACM Special Interest Group on Data Communications (SIGCOMM), and IEEE International Conference on Computer Communications (INFOCOM). COMST and TON are among the top ranked journals in the field of computer networking while SIGCOMM and INFOCOM represent the top ranked conferences of the field. 

Towards this end, we statistically analyze 18 years of accepted articles published in the two journals (IEEE TON and COMST), explore various bibliometric questions, and examine the publication behaviors of several research entities and how these are affected by the elements of articles. We also analyze popular topics in periodicals on computer networking and the effects of several parameters on the citations of an article. We also compare and contrast the publication standards and practices of the two journals (TON and COMST) and the two conferences (INFOCOM and SIGCOMM) that we consider. We believe that a deep study of the articles published in these venues can not only provide insight into current publication practice, but can also inform about the temporal evolution of the publishing trends in these venues. 

We structure our work around three major comparisons. First, we directly compare publication trends in TON vs.\ COMST, to understand how these two distinct publication types differ. Second, we compare trends over time to understand how they have evolved. Thirdly, we compare trends from TON and COMST with INFOCOM and SIGCOMM to map out the differences between the trends of top conferences and journals. 

Our aim is to investigate changes in publication behavior and collaboration patterns of distinctive authors, institutes and countries in the various computer networking publications, and the distribution of various mathematical and graphical elements (figures, tables, and equations) within them. Our goal is therefore to provide generalized insights into the publication trends in the field of networking. We also aim to answers questions such as the following: Which topics are popular in which regions of the world? What are the topics discussed by the top authors in their articles in the various publications? Which parameters affect the citations of an article? 

The \textit{key contribution of this article} is to develop a methodology and framework for performing a comprehensive bibliometric analysis on computer networking research and the public release of a comprehensive dataset. To the best of our knowledge, no such comprehensive study has been undertaken to study the publication trends in the field of computer networking. To facilitate future research in this area, we have publicly released our dataset including metadata, content, and citation related data for the articles published in IEEE COMST, TON, ACM SIGCOMM, and IEEE INFOCOM from 2000 to 2017\footnote{https://github.com/waleediqbal411/Scientometrics-paper-data2019}.

The rest of this article is structured as follows. In section \ref{sec:relatedwork}, we discuss related previous research work. The bulk of our investigations focus on the publication trends in computer networking journal publications in COMST and TON (Sections \ref{sec:methodology}--\ref{sec:citation}), but to make our analysis complete we also compare these trends with those observed in top ranked conferences (INFOCOM and SIGCOMM) in the area (Section \ref{sec:confcomp}). In Section \ref{sec:methodology}, our dataset is described and our methodology is broadly outlined. A detailed bibliographic focused on comparison of TON and COMST is presented in Sections \ref{sec:metadata}, \ref{sec:content}, \ref{sec:citation} in which metadata analyses, content-based analyses, citation-based analyses are presented respectively. A detailed comparison of publication trends in top networking journals and conferences (TON/COMST vs. INFOCOM/SIGCOMM) is presented in Section \ref{sec:confcomp}. We discuss future directions of this study in section \ref{sec:future_directions}. The paper is finally concluded in Section \ref{sec:conclusion}.

%We present our results in the next four sections---in particular, our metadata analyses, content-based analyses, citation-based analyses, and comparison of different analysis results in top journals and top conferences are presented in sections \ref{sec:metadata}, \ref{sec:content}, \ref{sec:citation}, \ref{sec:confcomp} respectively.  We will finally conclude the paper in section  \ref{sec:conclusion}.

%Sections \ref{sec:metadata}, \ref{sec:content}, \ref{sec:citation} are more focused on comparison of journals. Our analysis on conference publications is mainly discussed in section \ref{sec:confcomp}.
\section{Related Work}
\label{sec:relatedwork}

In this section, we present related work and highlight the novelty of this article. Bibliometrics is an established field in which the major trends of research fields are studied rigorously. A number of bibliometrics studies have been conducted in various fields to gain useful insights through the analysis of authorship and publication trends of different research outlets and areas  \citep{nobre2017scientific,fernandes2017evolution,serenko2009scientometric,chiu2010publish, rajendran2011scientometric,nattar2009indian,yin2017dancing}. These bibliometric analyses are not confined to the authorship based meta-data analysis of venues. 

Some authors have also undertaken quantitative analysis on the top ACM conferences. The purpose of these studies is to determine the genre of the article and to understand the publication culture of these conferences \citep{flittner2018survey}. These related studies do not explain which factors of the article affect the productivity parameters and the information about the correlation between important parameters required to analyze the productivity of different entities. Many previous works have performed an analysis on the content of various research areas using topic modeling \citep{paul2009topic} and keyword-based analysis \citep{choi2011analysis}.

A number of studies have used social networking analysis for social sciences and medical science research to find the most significant collaborating entities \citep{savic2017analysis,wagner2017growth,didegah2018co,borgatti2009network,waheed2018bibliometric}, using social network analysis on generally social media data and altmetric data \citep{hassan2017measuring}. Social media analysis has not been used to determine the communities in computer networking research due to which we do not yet have complete insights into the collaborating patterns that exist in computer networking research. 

Limited work has focused on using bibliometric or scientometric techniques to analyze the publication mores of the field of computer networks. Chiu et al. \citep{chiu2010publish} have performned an analysis of author productivity in computer networking venues in 2010. Our work is different in that we perform a detailed bibliometric analysis on the computer networking literature including an analysis of the effects of various features of article (such as the graphical and mathematical elements and the numbers of references) on the article's productivity metrics as defined in the field of bibliometrics.

Bibliometric analyses can also be utilized to see the extent of the incorporation of related research. Reference count in a article is the simplest way to observe the inclusion of related research and literature review. Different researchers analyzed referencing patterns in research articles to identify incorporation of the latest studies relating to a research article \citep{heilig2014scientometric} and citation analysis of the productivity of various research entities \citep{hamadicharef2012scientometric,bartneck2009scientometric}. These studies do not explain how the references are affected by the type of article venue. 
\section{Data Collection and Methodology}
\label{sec:methodology}

%% GT: I've cut down the below introduction. There is quite a bit of repeat text that says roughly the same thing, but in a slightly different way. Hence, I've tried to filter this out. 
%Bibliometric analysis helps researchers to identify patterns in research, for instance, the most powerful entities and communities, the hot and emerging topics, the correlation between elements of the research article, and how each element affects the others' occurrences \citep{ferreira2016we, ferreira2017mapping}. 

We start by describing our data collection methodology. There are several article genres in the field of computer networking, including conference articles, letters, editorials, survey articles, and empirical studies. To capture a broad swathe of these, we sample from 4 different well known publication outlets.

\subsection{Dataset Collection}
%First, we discuss the collected dataset and their features from all of the four venues. We also discuss the pre-processing techniques for our collected data. We separated this subsection into further two parts: first part discusses the data collection and pre-processing for journals and second part discusses the data collection and pre-processing for conferences.

To perform the analysis of journals, we used a collection of 3,281 articles. This contains 842 articles from IEEE Communication Surveys and Tutorials (IEEE COMST)\footnote{  https://www.comsoc.org/cst} 2000--2017 and 2,439 articles from IEEE/ACM Transaction on Networking (IEEE/ACM TON)  2000--2017.\footnote{https://ton.lids.mit.edu/}
We chose COMST and TON because COMST leans towards publishing tutorials and survey-based literature, whereas TON leans towards original research containing analytical and experimental studies. Our dataset allows us to perform a comparative analysis of computer networking research based on surveys and experimental studies. Details of the features extracted from these articles are shown in Table \ref{tab:table1}. The data was obtained from various sources, including IEEE Xplore\footnote{https://ieeexplore.ieee.org/}, Scopus\footnote{https://www.scopus.com} and CrossRef\footnote{https://www.crossref.org}. Data from CrossRef repository were scraped using Harzing's 'Publish or Perish' utility\footnote{https://harzing.com/resources/publish-or-perish}.

We then repeat the above process for two popular conferences, SIGCOMM and INFOCOM. We chose SIGCOMM as it is a well known venue that publishes primarily experimental research. In contrast, INFOCOM (also well known) focuses on more theoretical aspects of computer networking. 
We  collect 8707 research articles from these top conferences during 2000--2017. This collection of articles contains 1962 articles from SIGCOMM and 6745 research articles from INFOCOM. 
In total, we have gathered a collection of 11988 articles from these top journals and conferences.

\begin{table*}
	\centering
	\scriptsize
	\caption{Features of dataset extracted from COMST \& TON articles}
	\label{tab:table1}
	\begin{adjustbox}{width=1\textwidth}
		\begin{tabular}{m{4cm}m{1.5cm}m{1cm}m{1cm}m{0.9cm}m{0.9cm}m{0.9cm}m{0.9cm}}
			\hline
			\textbf{Attribute Name} & \textbf{Type of Attribute} & \multicolumn{2}{c}{\textbf{Count}}& \multicolumn{2}{c}{\textbf{Avg over article}}& \multicolumn{2}{c}{\textbf{Std. Dev.}}  \\\hline
			& &COMST&TON&COMST&TON&COMST&TON\\\hline
			Number of Articles & Numerical & 842&2439 & N/A&N/A &N/A&N/A \\
			Number of Authors & Numerical & 2451&5302 & 3.63&3.36 &1.67&1.42 \\
			Names of Authors & String &  2451&5302 &N/A&N/A  &N/A&N/A\\
			Number of Institutes & Numerical & 823&899 & 2.01&1.98 & 1.16&1.04\\
			Names of Institutes & String & 823&899&N/A&N/A&N/A&N/A\\
			Institutes from Same Country of Lead Author & Numerical & 1145&3659 & 1.36&1.5 & 0.66&0.765\\
			Institutes from Different Country of Lead Author & Numerical & 563&1213 & 0.67&0.5 & 0.996&0.846\\
			Flesch Kincaid Ease Score & Numerical & 41579&144067 & 49.38&59.06 & 7.45&6.01 \\
			Flesch Kincaid Grade Score & Numerical & 8404&22969 & 9.98&9.4 & 1.29&1.05\\
			Coleman Liau Score & Numerical & 11205&32540 & 13.3&13.34 & 2.57&1.23\\
			SMOG Readability Score & Numerical&10537&31330&12.51&12.84&1.47&0.76\\
			Number of Figures & Numerical&9843&26430&11.69&10.84&7.97&5.268\\
			Number of Tables & Numerical&4245&4985&5.04&2.04&4.23&2.534\\
			Number of Equations & Numerical&5914&44882&7.02&16.95&18.41&19.1\\
			Section of Pitfalls & Dichotomous&396 Yes \& 446 No&259 Yes \& 2180 No&0.47&0.11&0.5&0.31\\
			Number of References & Numerical&107084&76277&127.1&31.27&75.5&10.83\\
			References from last 10 years  & Numerical&78746&44511&93.52&18.25&63.75&9.6\\
			Citations of Articles & Numerical &56104&90301&66.63&37.02&137.7&105.21\\
			Number of Participating countries&Numerical&1344&3473&1.6&1.42&0.862&0.684\\
			Names of Participating countries&String&1344&3473&N/A&N/A&N/A&N/A\\
			Number of international authors & Numerical&653&1373&0.78&0.56&1.23&0.986\\
			Number of local authors & Numerical &2400&6669&2.85&2.73&1.32&1.28\\
			Authors from top 100 universities & Numerical& 437&2403&0.52&0.99&1.07&1.382\\
			Lead author's institute in top 100 universities & Dichotomous & 110 Yes \& 732 No& 731 Yes \& 1708 No  &0.13&0.3&0.337&0.605\\
			\hline 
		\end{tabular}
	\end{adjustbox}
\end{table*}

\subsection{Feature Extraction}

We next describe how we perform feature extraction across the two datasets. We describe the pre-processing for journals and conferences separately, as they are naturally associated with different metadata.

\subsubsection{Journal Dataset Pre-processing}

The journal data was obtained in PDF (Portable Document Format) and CSV (Comma Separated Values) formats from the aforementioned scientific repositories. The CSV files contain bibliographic details such as authors' name, affiliation, citation count and publication year. These details of articles were supplemented by manually extracted metadata such as the number of foreign authors and local authors, the number of authors from the top 100 universities of the world and the number of foreign and local institutes. 

For the extraction of text from the PDF files, we used Poppler's pdf2text utility\footnote{www.poppler.freedesktop.org}. Two further pre-processing tasks were performed on the extracted text: (a) calculation of readability scores (Flesch Kincaid \citep{kincaid1975derivation}; Coleman Liau \citep{coleman1975computer}; SMOG \citep{mc1969smog}); (b) Finding the number of references in an article cited from the previous decade's published articles. For references, we used an in-house formula script in Microsoft Excel\footnote{https://products.office.com/en/excel}, which takes the list of all references for an article and outputs the total number of references, for the past decade.

To construct a collaboration network, we created an adjacency list from the entries of author names and their affiliations. Statistical details of the dataset are shown in Table \ref{tab:table1}

%\subsubsection{Collection of Conference Dataset}

\begin{table}
	\centering
	\caption{Features of dataset extracted from SIGCOMM and INFOCOM}
	\label{tab:table2}
	\begin{tabular}{llll} 
		\hline
		Attribute Name                    & Type of Attribute & \multicolumn{2}{c}{\textbf{Count} }                         \\ 
		\hline
		\textbf{}                         & \textbf{}         & \multicolumn{1}{c}{SIGCOMM} & \multicolumn{1}{c}{INFOCOMM}  \\ 
		\hline
		Number of Articles                 & Numerical         & 1962                        & 6745                          \\
		Number of Authors                 & Numerical         & 4196                        & 8415                          \\
		Names of Authors                  & String            & 4196                        & 8415                          \\
		Number of Institutes              & Numerical         & 576                         & 1678                          \\
		Names of Institutes               & String            & 576                         & 1678                          \\
		Number of References              & Numerical         & 46809                       & 142487                        \\
		References from last 10 years     & Numerical         & 33650                       & 105753                        \\
		Citations of Articles             & Numerical         & 76594                       & 210555                        \\
		Number of Participating Countries & Numerical         & 57                          & 70                            \\
		Names of Participating Countries  & String            & 57                          & 70                            \\
		\hline
	\end{tabular}
\end{table}

\subsubsection{Conference Dataset Pre-processing}
Again, data was obtained in CSV (Comma Separated Values) format from the aforementioned scientific repositories. The CSV files contain bibliographic details such as authors' name, affiliation, citation count, publication year and references used in an article. Incomplete and irrelevant entries were removed from the dataset. These entries include messages from editors, entries without references, and entries without relevant metadata such as author names, institute names and indexed keywords. Details of the features extracted from these articles are shown in Table \ref{tab:table2}.

Two further pre-processing tasks were performed on the extracted text: (a) calculation of number of metadata elements such as authors, institutes, countries; (b) Finding the number of references in an article cited and number of references in an article cited from the previous decade's published articles. For references, we used an in-house formula script in Microsoft Excel and a python scripts as final step, which takes the list of all references for an article and outputs the total number of references, for the past decade.

\subsection{Bibliometric Indicators}

%%GT: I've removed the below introductory text. I found that many of the sections were starting with verbose introductions that we largely repeating the same things/motivation. 
%The peer review process takes considerable time, energy, money, and expertise in a specific research area. To reduce the number of resources that have to be dedicated to peer review, the scientific community is focusing on using bibliometrics to enhance the review process \citep{belter2015bibliometric}. %Bibliometric indicators use far less time and resources and can give a more objective result \citep{durieux2010bibliometric}.

In this study, we used several bibliometric indicators in order to measure the impact of research published in COMST and TON. Details of these bibliometric indicators are shown in Table \ref{tab:table_indicators}. Here, we briefly list the methodologies we will use in the remainder of the paper. 
\begin{table*}[!h]
	\centering
	\footnotesize
	\caption{Bibliometric indicators used in this article}
	\label{tab:table_indicators}
	\begin{adjustbox}{width=1\textwidth}
		\begin{tabular}{m{3.3cm}m{3cm}m{6.5cm}} 
			\hline
			Dimension                                 & Indicator                           & Definition                                                                                                                         \\ 
			\hline
			\multirow{5}{*}{Metadata based Analysis} & Publication count (P) per author    & Number of articles published by an author                                                                                            \\ 
			
			& Publication count (P) per institute & Number of articles published by an institute \\ 
			
			& Publication count (P) per country   & Number of articles published by a country                                                                                            \\ 
			
			& h-index of an author                & h-index of a researcher (h) shows us that \textit{h} articles of a researcher have~got \textit{h} citations \\ 
			
			& Reference count per article           & Number of references used in an article                                                                                               \\ 
			\hline
			Content-based Analysis                    & Readability scores                  & Score indicates the difficulty level of language for intended audience              \\ 
			\hline
			\multirow{2}{*}{Citation based Analysis}  & Citation count per keyword          & Total number of citation against a keyword                                                                                         \\ 
			
			& Citation count per author           & Total number of citation obtained by an author                                                                                     \\
			\hline
		\end{tabular}
	\end{adjustbox}
\end{table*}

\begin{itemize}
	
	\item \textit{Statistical Analysis}: There are a number of analyses that come under the umbrella of statistical analysis, but our focus, for the most part, will be on occurrence-based analysis \citep{weatherburn1949first} in this study for finding significant entities either in terms of publications count or in terms of citation and h-index count. 
	
	\item \textit{Social Network Analysis}: Social network analysis is useful in finding connections and relations between various entities. These relations cannot be observed through statistical analysis. Social network analyses are useful in finding hidden communities within data, e.g., we used a modularity class-based clustering technique \citep{blondel2008fast} for finding various communities in our data. To find the significance of a single node, we used an average degree algorithm.
	
	\item \textit{Topic Modeling}: Another well-known method of extracting features from the raw text is Topic Modeling.  One of the best-known algorithms for topic modeling is Latent Dirichlet Allocation (LDA). LDA takes the raw text, the number of topics, and a dictionary of words as input, and then provides as an output the most significant topics \citep{blei2003latent}. We used LDA on our dataset to explore significant topics in COMST and TON. For LDA, we used a Python implementation of the Gensim\footnote{https://radimrehurek.com/gensim/} library. We kept the number of latent output topics to 10 and iterated our algorithms 400 times on our dataset in order to achieve converged results.    
\end{itemize}

The rest of this paper will explore our datasets through the lens of the above analytical techniques. 
We performed analysis over journals' data explicitly in section \ref{sec:metadata}, \ref{sec:content} and \ref{sec:citation} respectively. Readers will find analysis on conferences' data and their comparison with journals' data in section \ref{sec:confcomp}.
\section{Metadata Analysis and Findings}
\label{sec:metadata}

We start our analysis by exploring the key metadata attributes associated with the publications. Specifically, we focus on metadata associated with publications authors and their respective institutes, before inspecting the structural elements of the articles (e.g., presence of figures). In this section we focus on comparing these observations across the two journals under study. 

\subsection{Research Productivity of Authors and Countries}

\vspace{2mm}
\subsubsection{Author Based Productivity Analysis}

\begin{figure}[!h]
	\begin{center}
		\includegraphics[width=0.7\textwidth]{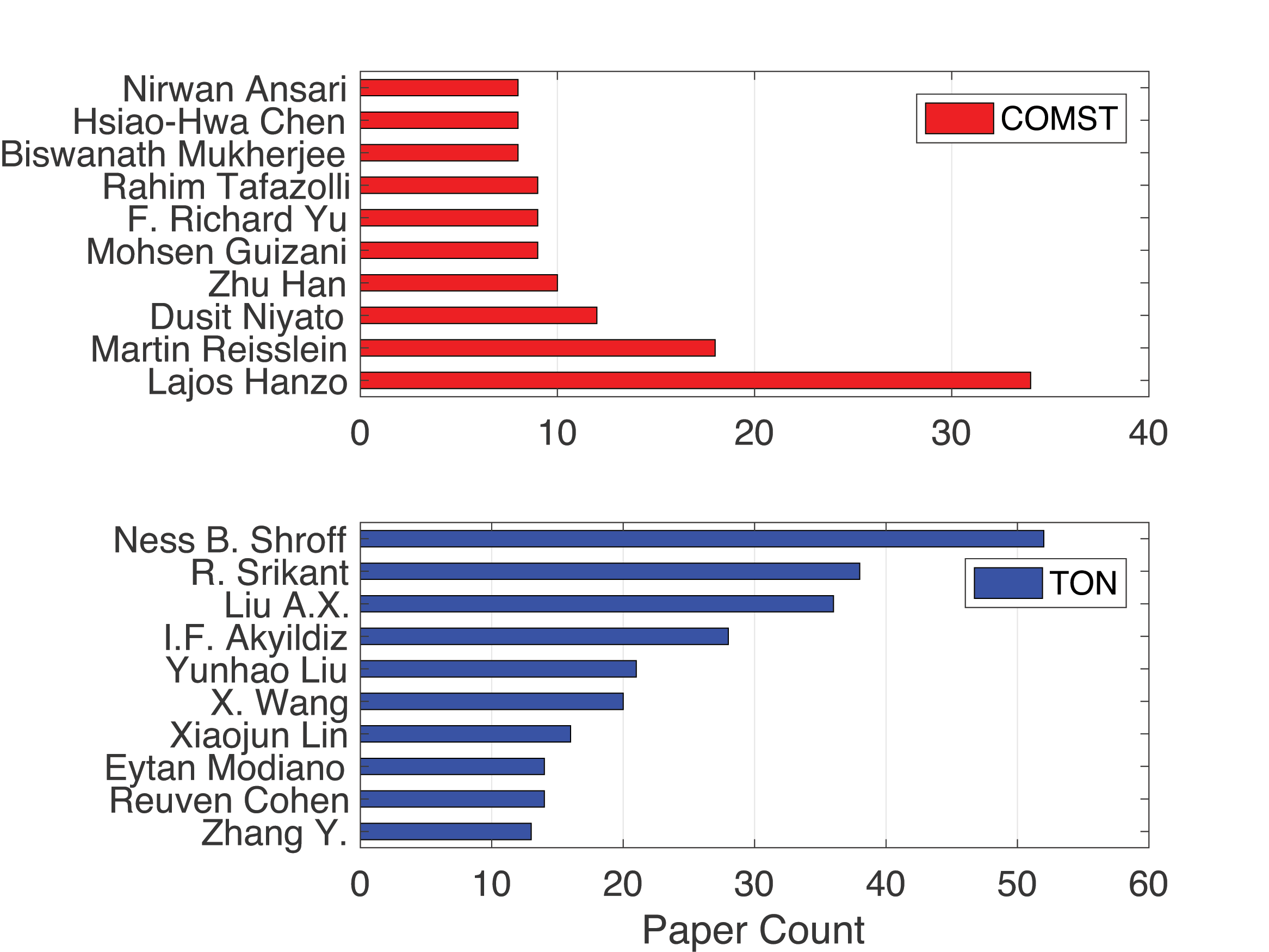}
	\end{center}
	\caption{Most-published authors during 2000--2017, according to article count. \textit{Interestingly, there is no overlap at all in the top 10 list, supporting a ``horses for courses'' hypothesis implying that it's rare to find an author who is extremely prolific in both these genres.}}
	\label{fig:authors_top}
\end{figure}

First, we investigate the most important authors of the two journals. There are many parameters to analyze the significance of a researcher's published work. A simple measure would be publication count is listed in Figure \ref{fig:authors_top}. 
The h-index is also another widely used metric where \textit{h} tells us that \textit{h} articles of a researcher have \textit{h} citations \citep{hirsch2005index}. Using the h-index of only COMST and TON, we can observe which authors are publishing highly cited research in COMST and TON. 

\begin{figure}[!h]
	\begin{center}
		\includegraphics[width=0.7\textwidth]{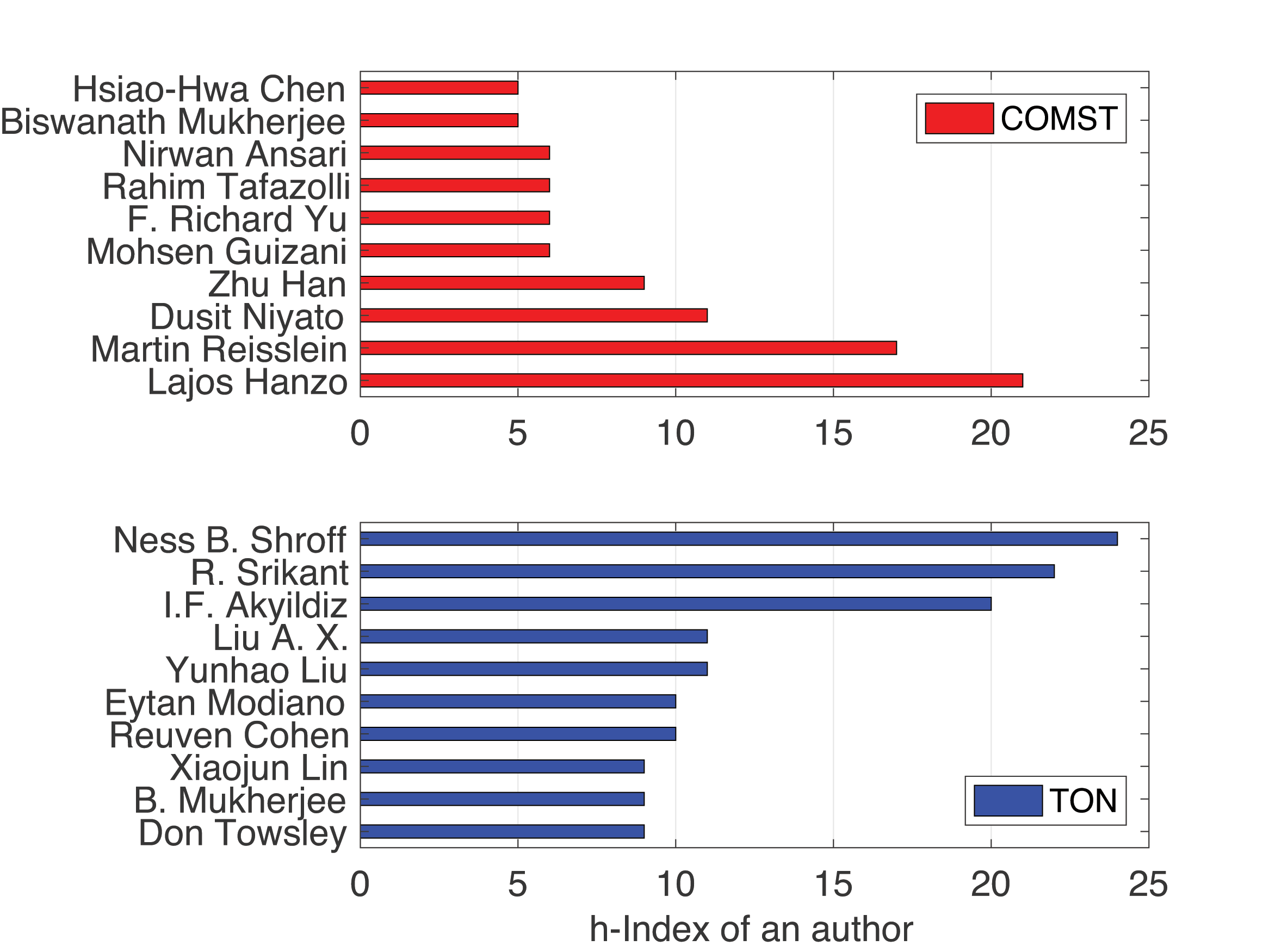}
	\end{center}
	\caption{Ten authors with the highest h-index during 2000--2017. \textit{The top 10 most-published list and the top 10 authors with the highest h-index are almost identical in both COMST and TON indicating a strong relationship between numbers of articles published and h-index.}}
	\label{fig:cited_top_hindex}
\end{figure}

Figure \ref{fig:cited_top_hindex} shows the authors in COMST and TON with the highest h-index, and how the top five highest publication counts are from the top ten authors with the highest h-index in COMST and TON. The data confirms that the top authors (measured by publication count) are the ones who have significant research contributions in terms of publication count as well as citation count.

\vspace{2mm}
\subsubsection{Country Based Productivity Analysis}

\begin{figure}[!h]
	\centering
	\subfloat[COMST]{\includegraphics[width=0.5\textwidth]{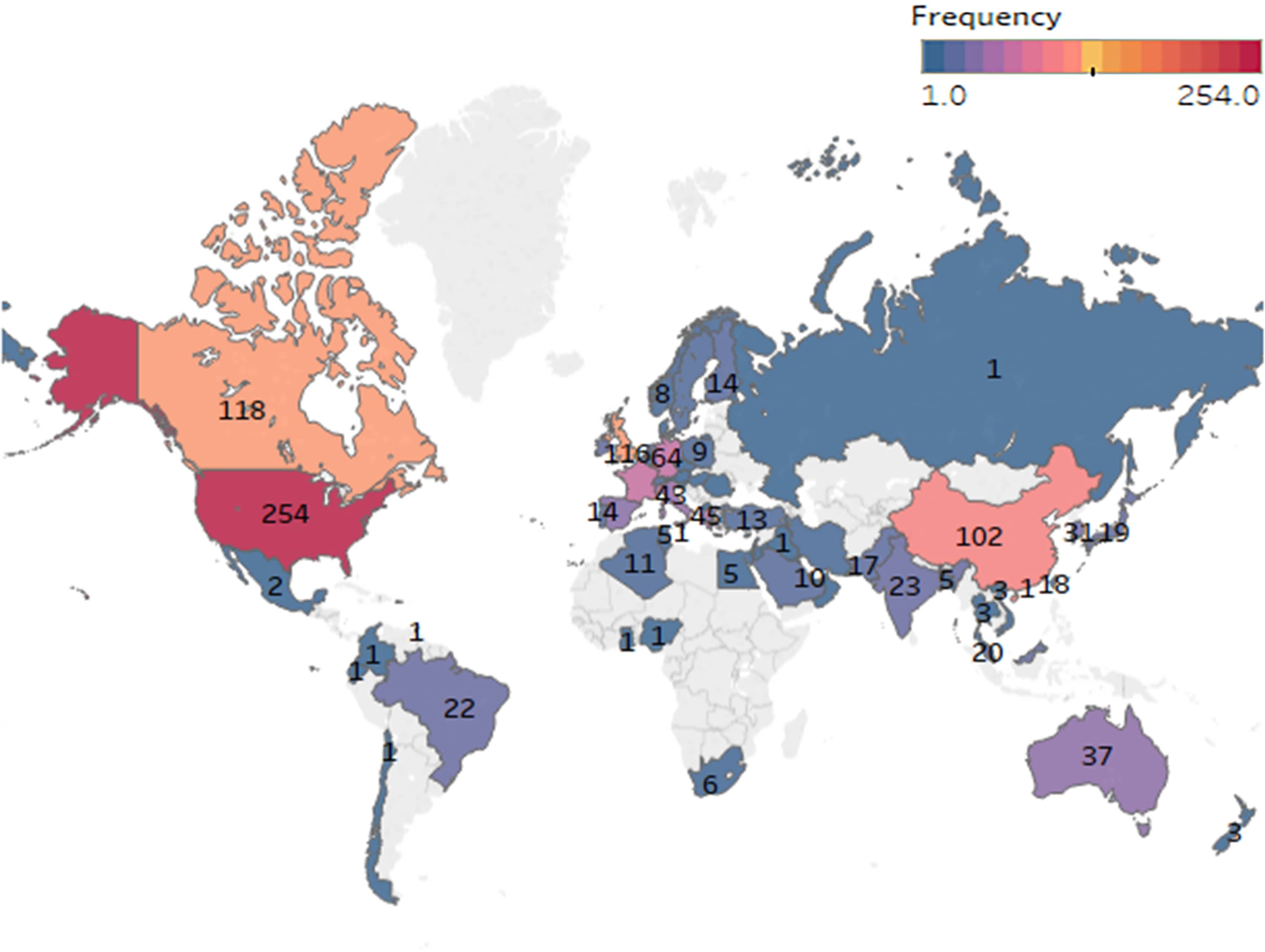}}
	\subfloat[TON]{\includegraphics[width=0.45\textwidth]{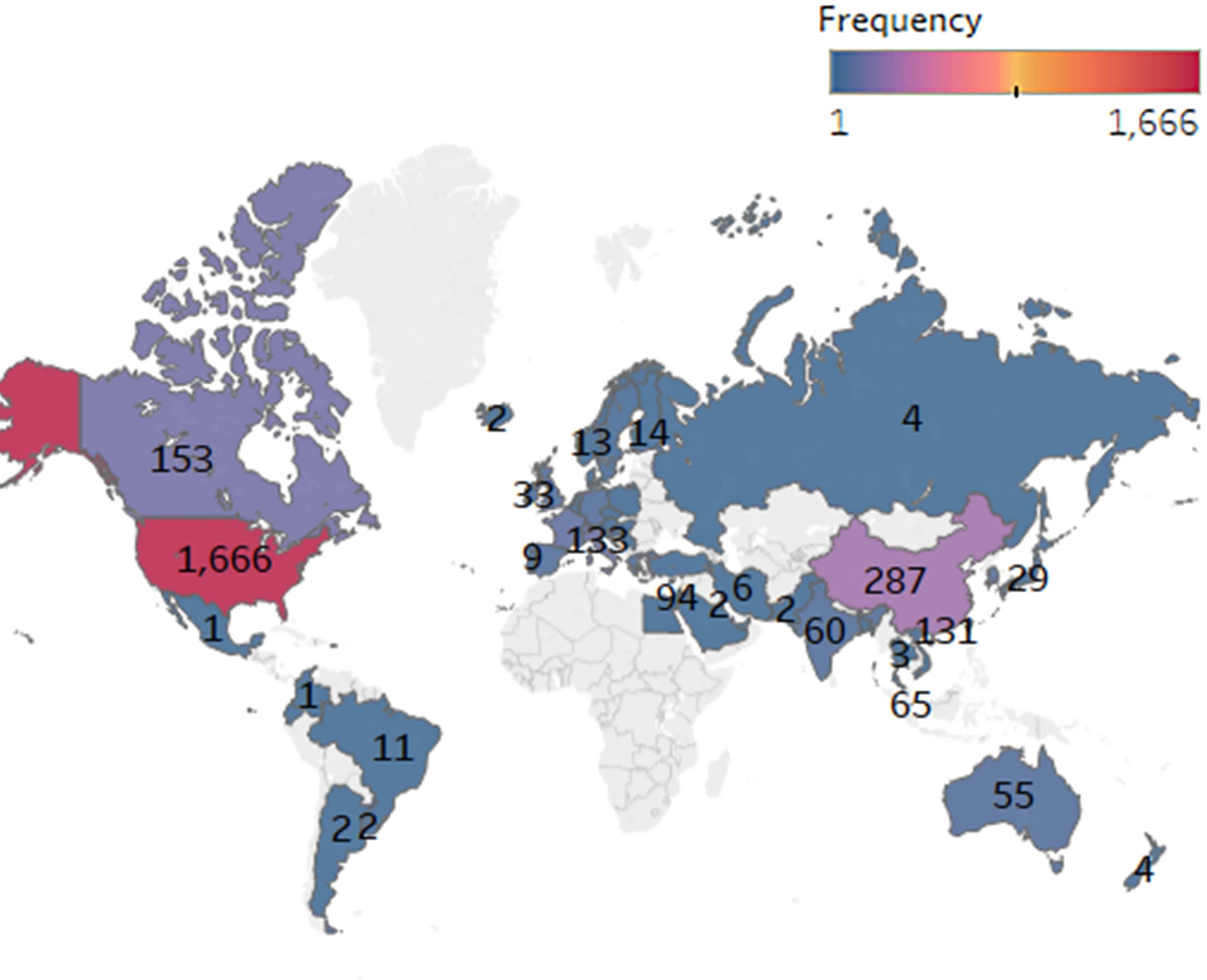}}
	\caption{Publication count of different countries in COMST and TON. \textit{Although most countries have similar productivity in these two journals, there are notable exceptions where the publication trends are quite dissimilar (also see the next figure).}}
	\label{fig:country_top}
\end{figure}

In a research domain, some countries play a pivotal role in driving the ongoing advancements in that field. Figure \ref{fig:country_top} shows the distribution of published articles in COMST and TON from different countries using a global heat map. As expected, the United States is in the highest position in COMST and TON in terms of publication count. Other top countries include Canada, China, France, and the United Kingdom in COMST. In TON, top countries remained the same but Italy replaced the United Kingdom in the list of top countries. 

\begin{figure}[!h]
	\centering
	\subfloat[COMST]{\includegraphics[width=0.5\textwidth]{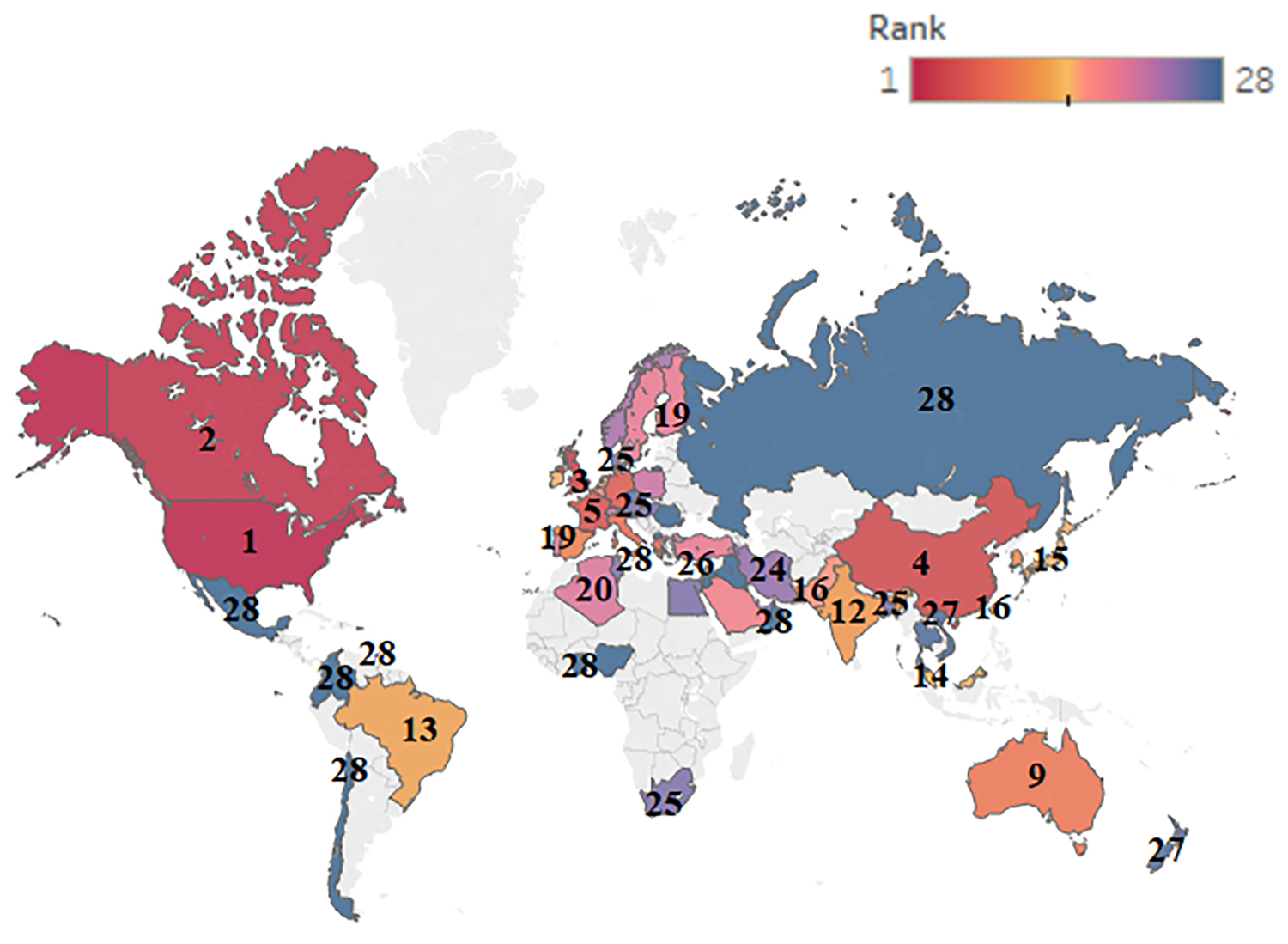}}
	\subfloat[TON]{\includegraphics[width=0.5\textwidth]{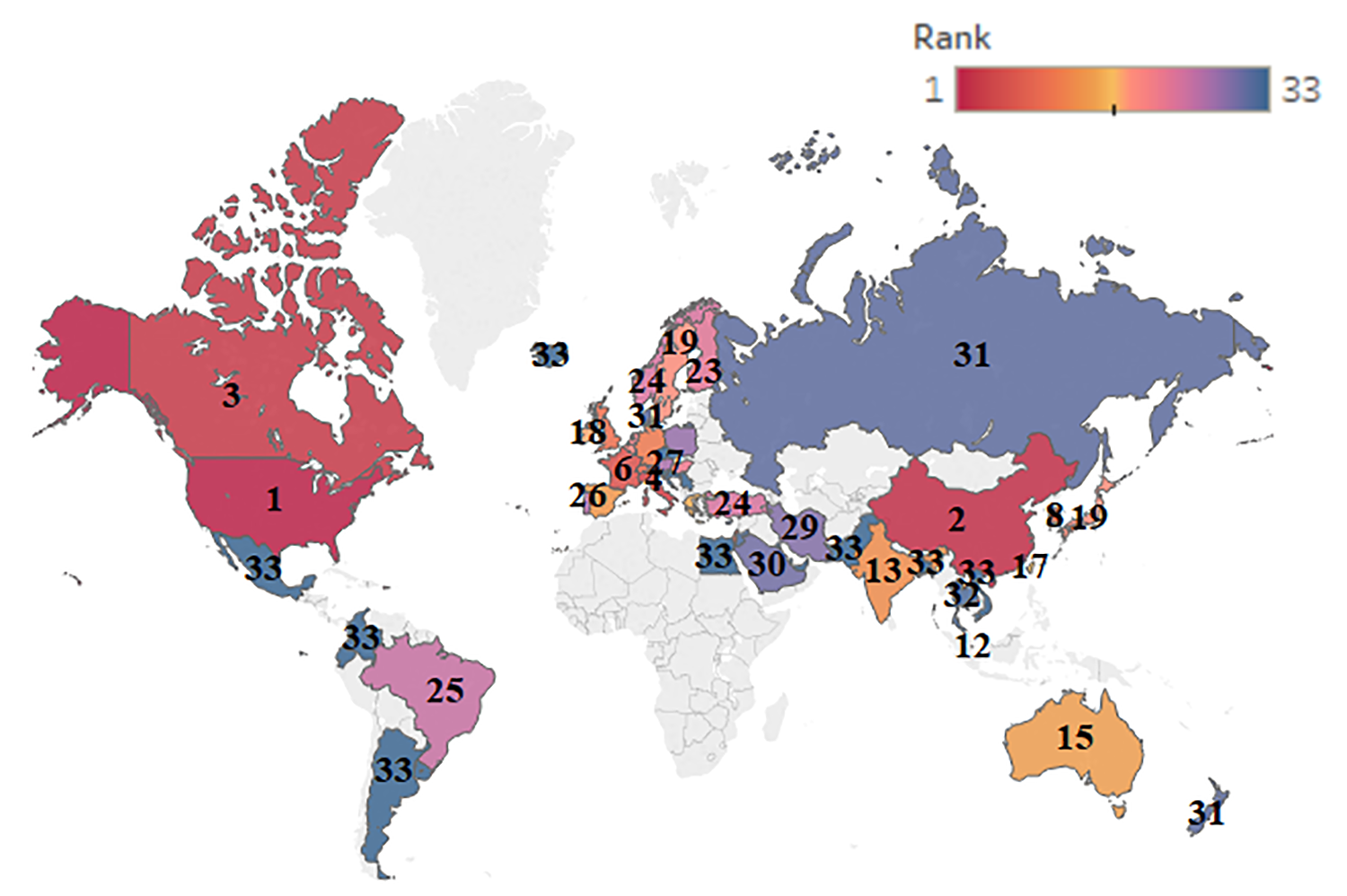}}
	\caption{Rank of different countries in COMST and TON based on publication count. \textit{Although most countries have similar productivity in these two journals, there are notable exceptions where the publication trends are quite dissimilar.}}
	\label{fig:country_top_rank}
\end{figure}

\par The differences in a country's publications in the two journals can partly be attributed to different publication cultures arising from different incentives for faculty promotion/assessment. Some countries in North America and parts of Europe (e.g., USA and UK) give more weight to top-tier conferences (like Sigcomm, NSDI, Infocomm, etc.) in their assessment criteria while many others in parts of Asia and southern Europe (e.g., Pakistan, Malaysia, France, Spain, Italy) emphasize journal publications. In many cases, extended versions of conference papers in the networking domain are published in journals such as TON. COMST, due to its focus on tutorial/survey papers, is more specialized and therefore not relevant for conference paper extensions. Figure \ref{fig:country_top_rank} shows the rank of different countries in COMST and TON based on published articles using a global heat map. Rank of some countries has significantly changed in both journals. Israel was on Rank 7 out of 33 in TON as compared to 27 out of 28 in COMST. Similarly, Pakistan was on Rank 16 out of 28 in COMST as compared to 33 out of 33 in TON. Many countries have not published a single paper in TON but published many papers in COMST. These countries include Ghana, South Africa, Iceland and many more.
\begin{figure}[!h]
	\centering
	\subfloat[COMST]{\includegraphics[width=0.5\textwidth]{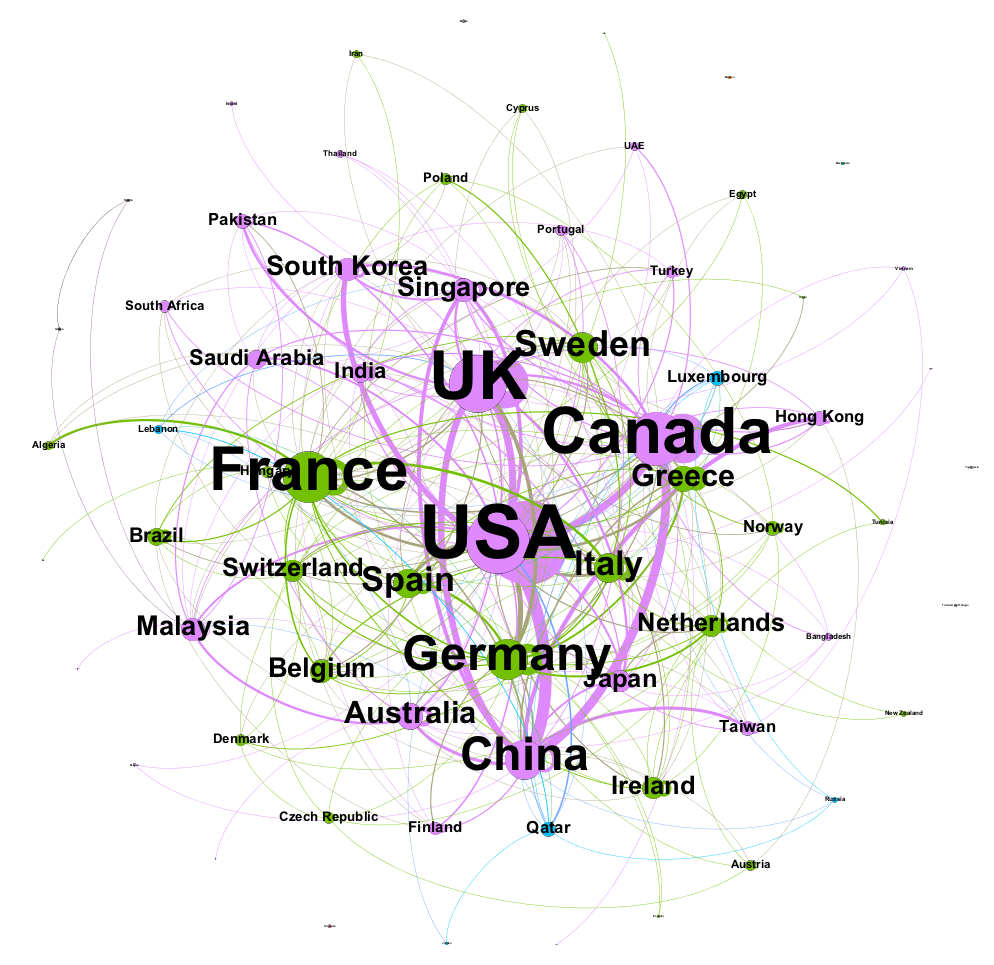}}
	\subfloat[TON]{\includegraphics[width=0.5\textwidth]{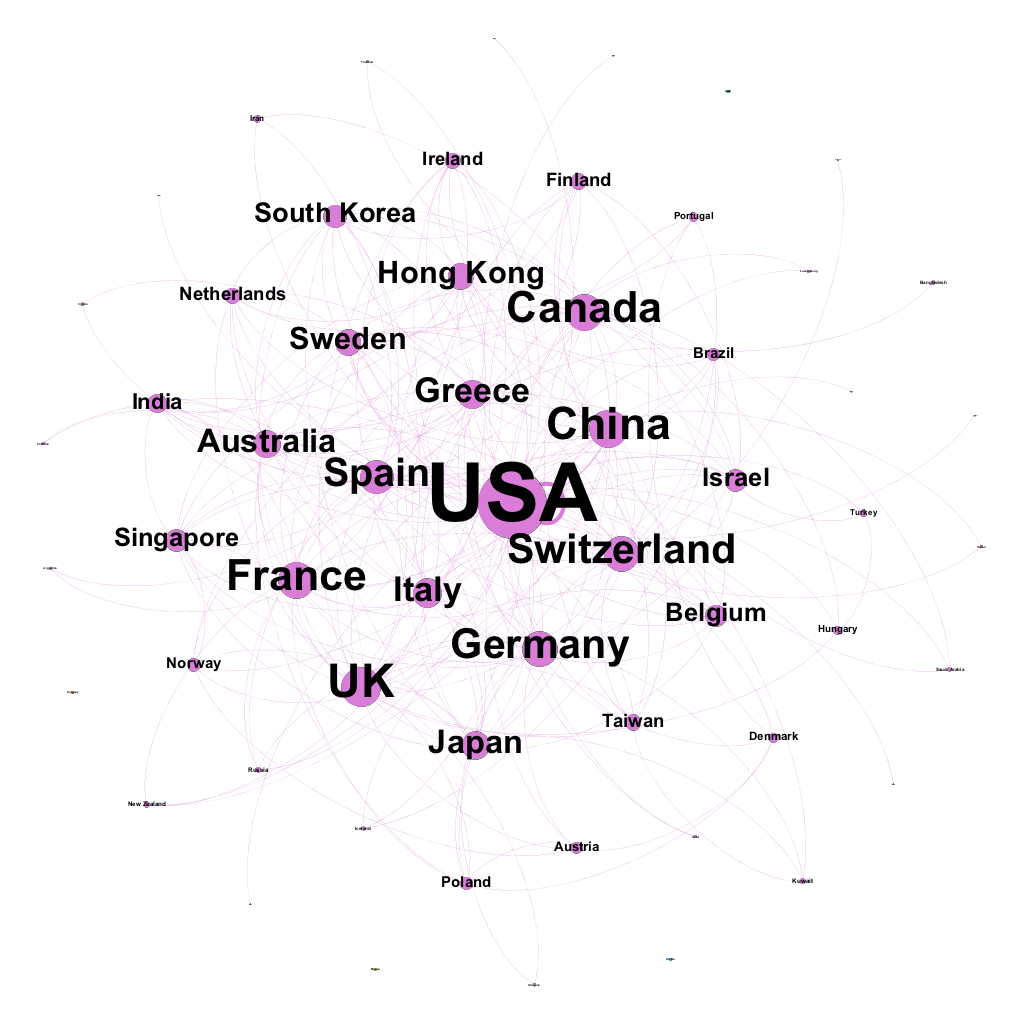}}
	\caption{Co-authorship network among top countries in (a) COMST; (b) TON. Node size indicates the number of links with other nodes in the co-authorship network and the node color represents cluster membership.}
	\label{fig:country_top_gephi}
\end{figure}
We next inspect the collaborations that took place between these countries. Figure \ref{fig:country_top_gephi} shows the co-authorship network of top countries in COMST and TON. In COMST, the top three countries have significant co-authorship activities among themselves, thus they are clustered in a single group. The same pattern is followed by the fourth and fifth most influential countries, which are clustered in one group. In TON, all the major contributing countries are clustered in a single node due to the great publication contribution of the United States. The United States contributed 1,667 of the 2,439 articles in TON. With the advancement of information and communication technologies, researchers from various countries now have new ways to work with each other. Top countries enjoy the share of publication from their authors, and in addition a contribution from authors from collaborating countries.

We next proceed to inspect the productivity rates among countries, specifically in terms of publication and citation count. By using these features, we propose a simple mathematical model for determining the rank of a country in a venue. We kept the highest measurement in each feature as a reference point for the calculation of the ranking score. Normalized Rank Score (NRS) for each country can be calculated by using equation \ref{NRS_equation} where \textit{P} is publication count, \textit{C} is citation count, \textit{hi} is h-index of a country, \textit{$P_{top}$} is maximum publication count, \textit{$C_{top}$} is maximum citation count and \textit{$hi_{top}$} is maximum h-index obtained by a country in a venue.

\begin{equation}%[!h]
NRS = \frac{1}{3}*\Big(\frac{P}{P_{top}}+\frac{C}{C_{top}}+\frac{hi}{hi_{top}}\Big)
\label{NRS_equation}
\end{equation}

\begin{figure}[!h]
	\centering
	\subfloat[COMST]{\includegraphics[width=0.5\textwidth]{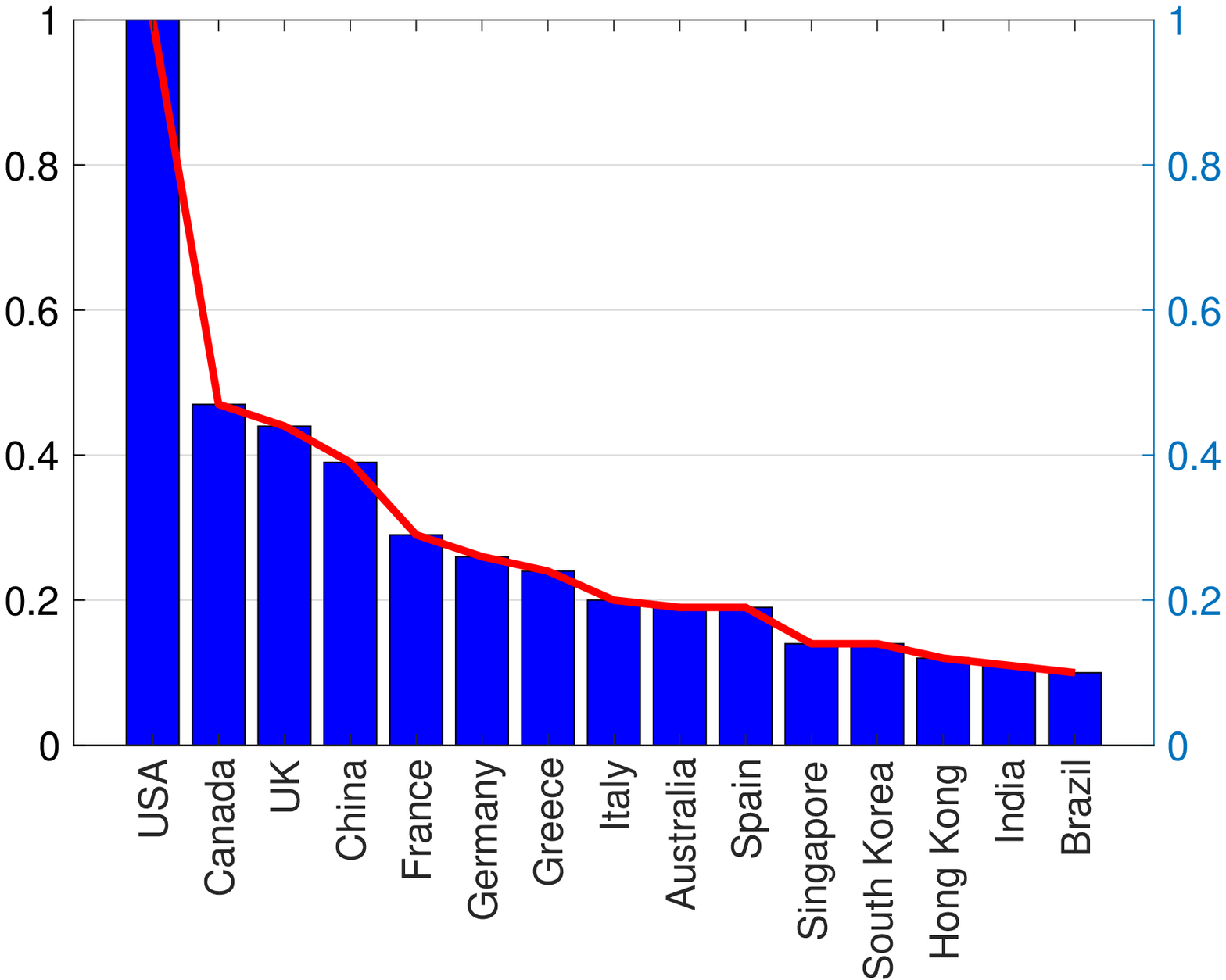}}
	\subfloat[TON]{\includegraphics[width=0.5\textwidth]{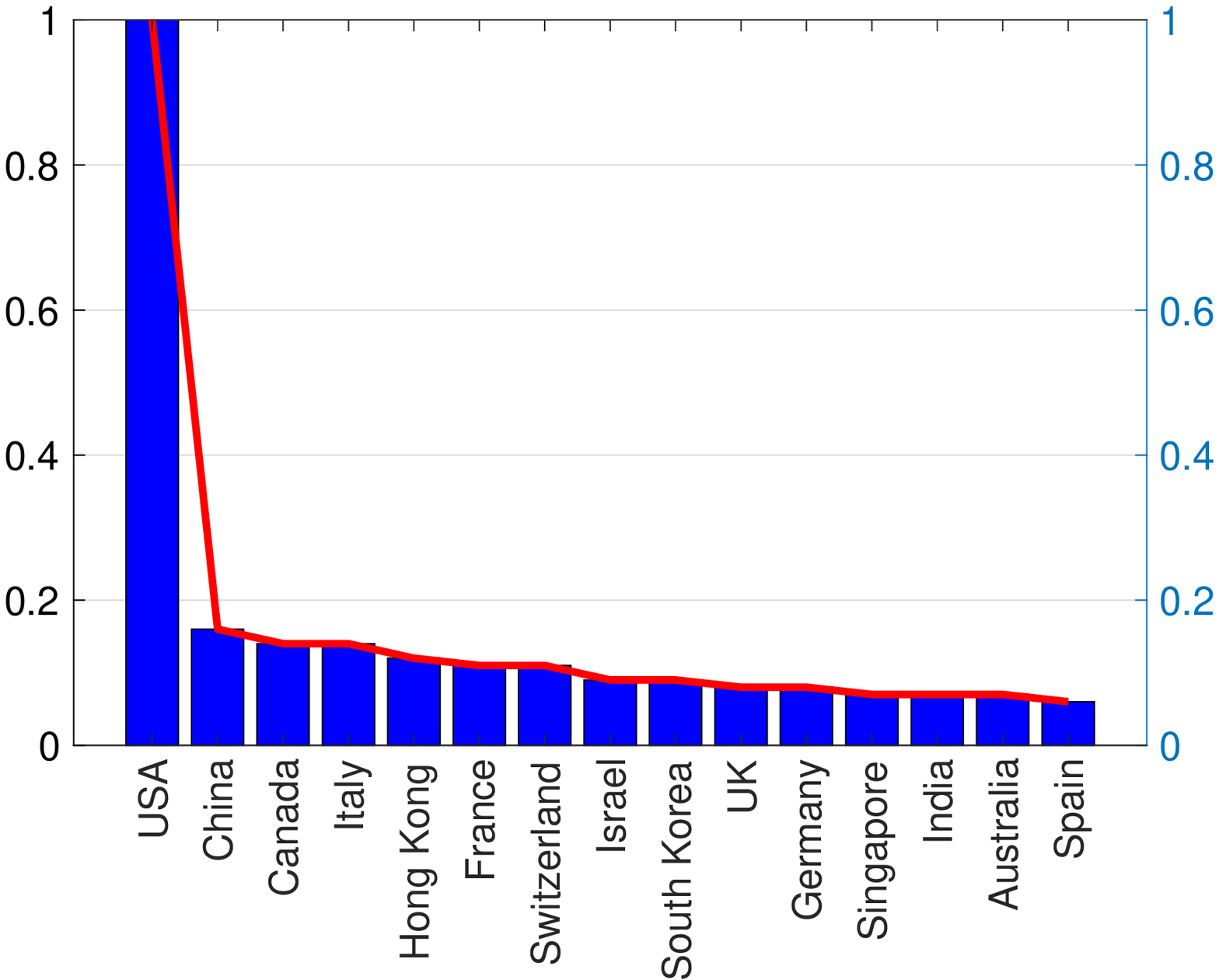}}
	\caption{Rank of countries in COMST and TON based on their publication count, citation count, and h-index. \textit{Country with the highest score in each journal is used as a reference for calculation. USA emerged as the top country in both journals, with the gap being more prominent in TON.}}
	\label{fig:ranks_t}
\end{figure}

\begin{table*}[!h]
	\centering
	\scriptsize
	\caption{Productivity of the top countries in COMST and TON. \textit{By and large the h-index and the citations are highly correlated with the number of publications with some notable exceptions (e.g., Spain has the highest average citations per article in COMST while China despite having many articles in TON has the lowest average citation among the listed countries}).}
	\label{citation_impact_table}
	\begin{adjustbox}{width=1\textwidth}
		\begin{tabular}{m{1.2cm}m{1.4cm}m{1cm}m{1cm}m{0.6cm}m{1.4cm}m{1.4cm}m{1.2cm}m{1.2cm}m{0.7cm}} 
			\hline
			\multicolumn{5}{l}{~ ~ ~ ~ ~ ~ ~ ~ ~ ~ ~ ~ ~ ~ ~ {\centering}\textbf{Rank in COMST}~ ~ ~ ~ ~ ~ ~} & \multicolumn{5}{|l}{ ~ ~ ~ ~ ~ ~ ~ ~ ~ ~ ~ ~ ~ ~ ~ ~ \textbf{Rank in TON}~ ~ ~ ~ ~ ~}  \\ 
			\hline
			Country     & Publications & Total Citations & Avg. Citation & h-index        & Country     & Publications & Total Citations & Avg. Citation & h-index            \\ 
			\hline
			USA         & 254          & 17073           & 67.22            & 72             & USA         & 1666         & 309842          & 185.98           & 120                \\ 
			
			Canada      & 118          & 6216            & 52.68            & 41             & China       & 287          & 11749           & 40.94            & 34                 \\ 
			
			UK          & 116          & 5125            & 44.18            & 40             & Canada      & 153          & 17026           & 111.28           & 34                 \\ 
			
			China       & 102          & 4756            & 46.63            & 35             & Italy       & 133          & 18622           & 140.02           & 34                 \\ 
			
			France      & 66           & 3465            & 52.5             & 29             & Hong Kong   & 131          & 9636            & 73.56            & 31                 \\ 
			
			Germany     & 64           & 2493            & 38.95            & 27             & France      & 108          & 10590           & 98.06            & 26                 \\ 
			
			Greece      & 45           & 2381            & 52.91            & 29             & Israel      & 94           & 8398            & 89.34            & 24                 \\ 
			
			Italy       & 43           & 2677            & 62.26            & 20             & South Korea & 81           & 8597            & 106.14           & 24                 \\ 
			
			Australia   & 37           & 2588            & 69.95            & 20             & Switzerland & 69           & 11819           & 171.29           & 29                 \\ 
			
			Spain       & 35           & 2633            & 75.23            & 20             & UK          & 68           & 10401           & 152.96           & 19                 \\ 
			
			Singapore   & 31           & 991             & 31.97            & 17             & Germany     & 67           & 6067            & 90.55            & 23                 \\ 
			
			South Korea & 31           & 1205            & 38.87            & 16             & Singapore   & 65           & 6120            & 94.15            & 19                 \\ 
			
			India       & 23           & 1190            & 51.74            & 13             & India       & 60           & 6229            & 103.82           & 20                 \\ 
			
			Brazil      & 22           & 1216            & 55.27            & 10             & Spain       & 57           & 2341            & 41.07            & 15                 \\ 
			
			Hong Kong   & 21           & 836             & 39.81            & 16             & Australia   & 55           & 4750            & 86.36            & 20                 \\
			\hline
		\end{tabular}
	\end{adjustbox}
\end{table*}

We calculated ranking scores of top countries in COMST and TON using equation \ref{NRS_equation}. Figure \ref{fig:ranks_t} shows the ranking of different countries in COMST and TON where it is seen that the USA has the maximum ranking score in both the venues. Both the venues are dominated by more or less the same countries with some exceptions---e.g., Israel is among the top-ranked countries publishing in TON but it is not a prominent contributor to COMST. This indicates that different countries can (for various socioeconomic reasons) have incentives to target particular journals. Table \ref{citation_impact_table} shows the impact of the top countries in COMST and TON. For both publication venues, the United States is the highest-ranked contributor with the average citation count per document being higher in TON than in COMST.
\subsection{Author Collaborations} 

\vspace{2mm}
\subsubsection{General Co-Authorship Trends}
Author collaborations is a key ingredient for research productivity \citep{iglivc2017whom, powell2018these}. We next explore the changing trends in co-authorship in COMST and TON over the period 2000 to 2017. We explore how the distribution of collaborating authors changes over time; what kinds of authoring entities (foreign or local authors) have changed in collaborations over time; and whether influential authors tend to collaborate on publications. Note that we use the terms collaboration and co-authorship interchangeably, as it is impossible to identify the exact form of collaboration that took place during the preparation of an article.

%whether authors from the top institutes of the world publish more articles in these publication venues compared to others; who are the most influential authors in COMST and TON

\begin{figure}[!h]
	\centering
	\subfloat[COMST]{\includegraphics[width=0.5\textwidth]{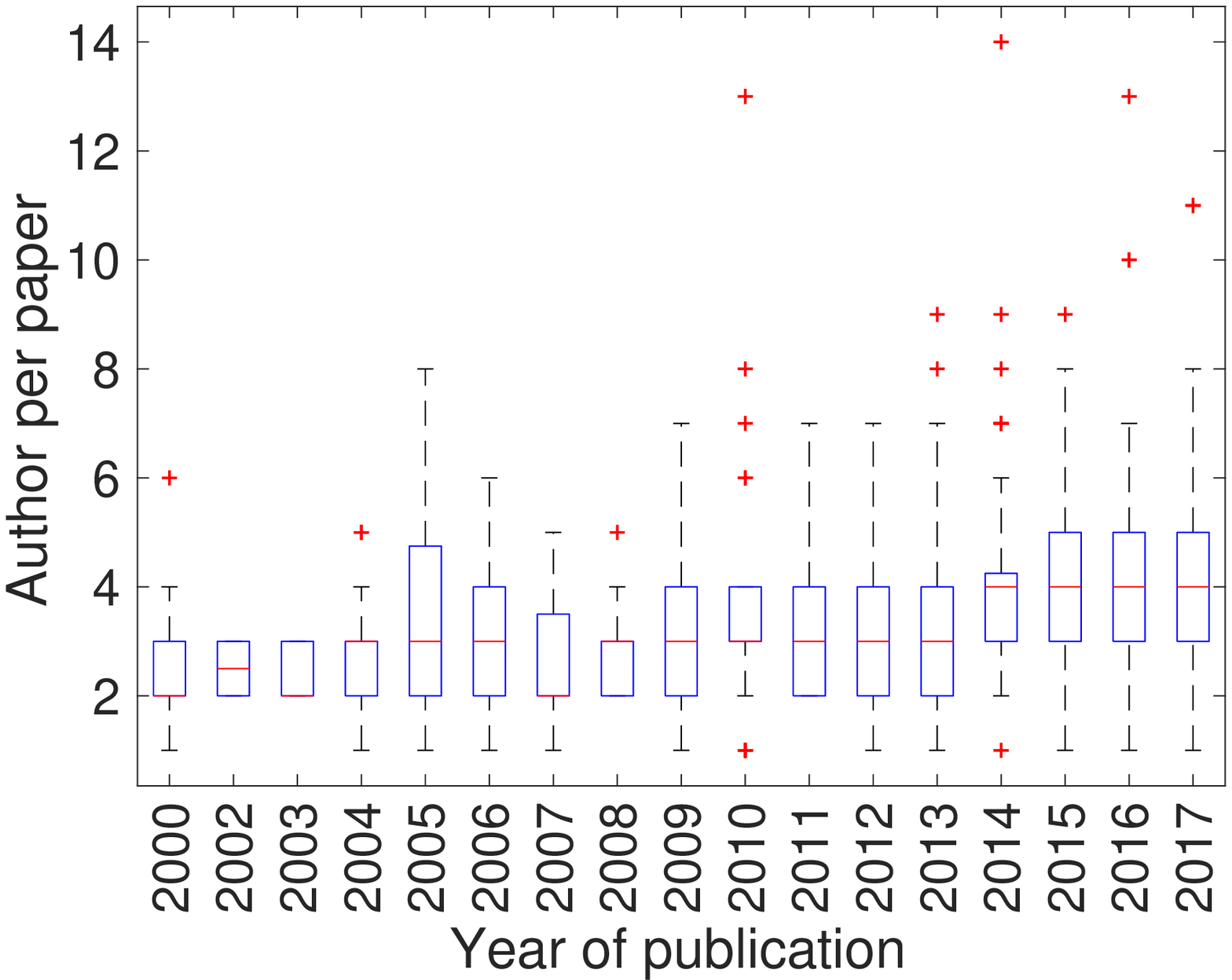}}
	\subfloat[TON]{\includegraphics[width=0.5\textwidth]{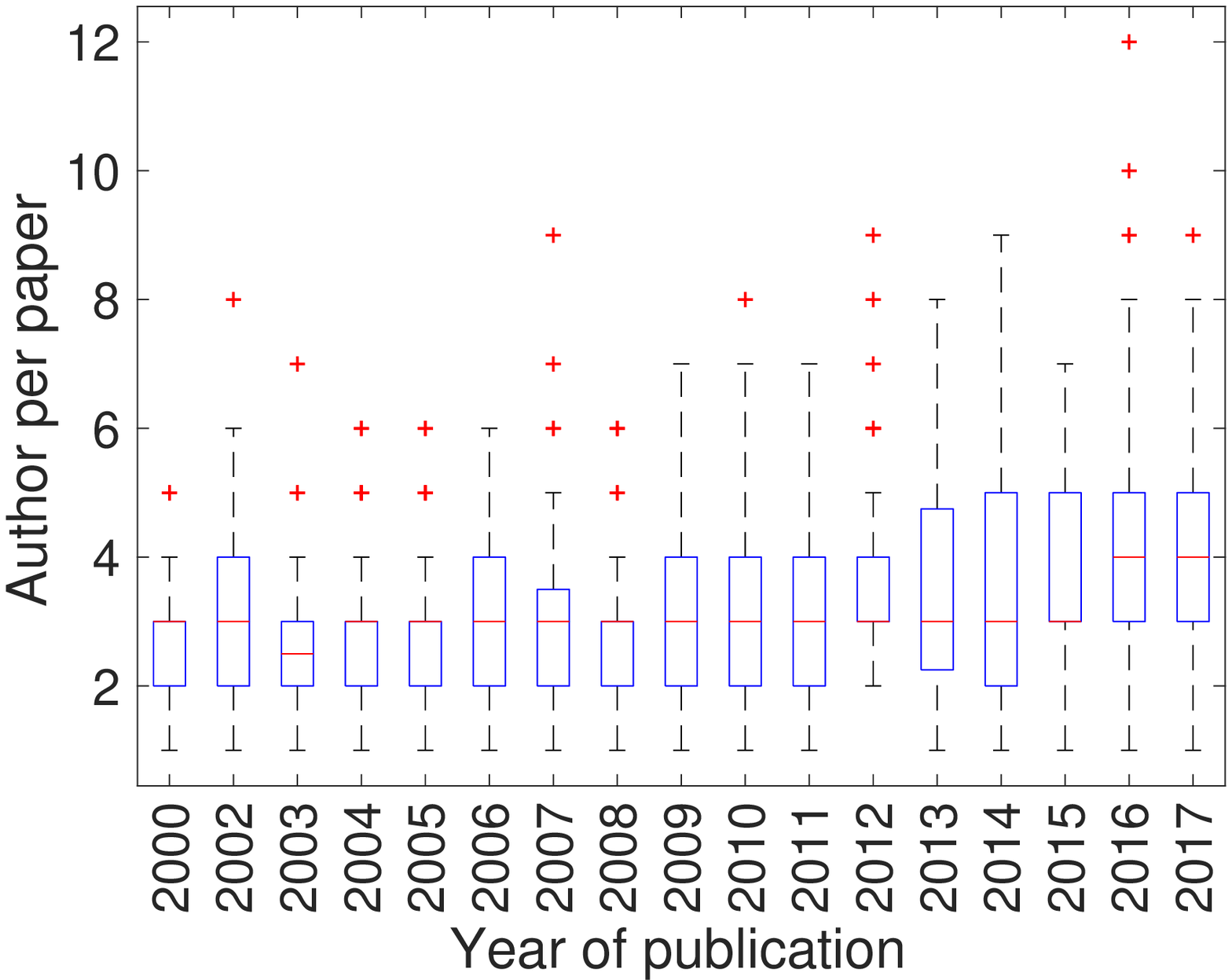}}
	\caption{Distribution of the number of authors per article in COMST and TON throughout 2000--2017. \textit{Tendency for co-authorship is increasing over time in both COMST and TON due to enhancing collaboration between institutes and authors.}}
	\label{fig:authors_year}
\end{figure}

Figure \ref{fig:authors_year} shows the distribution of the number of authors per article in COMST per year. It is clear that the tendency for co-authorship is increasing; in 2000 the median number of authors is 2 for COMST and 3 for TON, compared to 4 and 4 in 2017. Perhaps most noteworthy is the spread of authorship numbers across articles, with a standard deviation of 0.87 in 2000 vs.\ 1.69 in 2017 for TON (similar trends of COMST). The outliers in authorship pattern are clear with 11\% of authorship lists exceeding 8 in 2017 (compared to 2\% prior to 2006). The tendency for co-authorship is increasing over time in both COMST and TON due to enhancing collaboration between institutes and authors. This increasing trends may be a result of several elements which include expanding the number of members in different graphical unit e.g. European Union, cross-country funding, and the arrival of increasing degrees of remote (skype/email) collaboration. 

\vspace{2mm}
\subsubsection{Institutional and Country Based Collaborations}
This subsection presents the varying trends of collaborations among the institutes and countries in COMST and TON over the period from 2000 to 2017. We will address several important questions relating to the collaboration patterns of institutes and countries; how the distribution of collaborating institutes and countries changes over time; the most influential institutes and nations in COMST and TON; and whether influential institutes and nations tend to work as collaborators. To observe collaborative relations among the top researchers in COMST and TON, we generate undirected graphs of co-authors and identify clusters using modularity class partitioning. We used undirected graphs to remove duplicate links among publishing entities. 

\begin{figure}[!h]
	\centering
	\subfloat[COMST]{\includegraphics[width=0.5\textwidth]{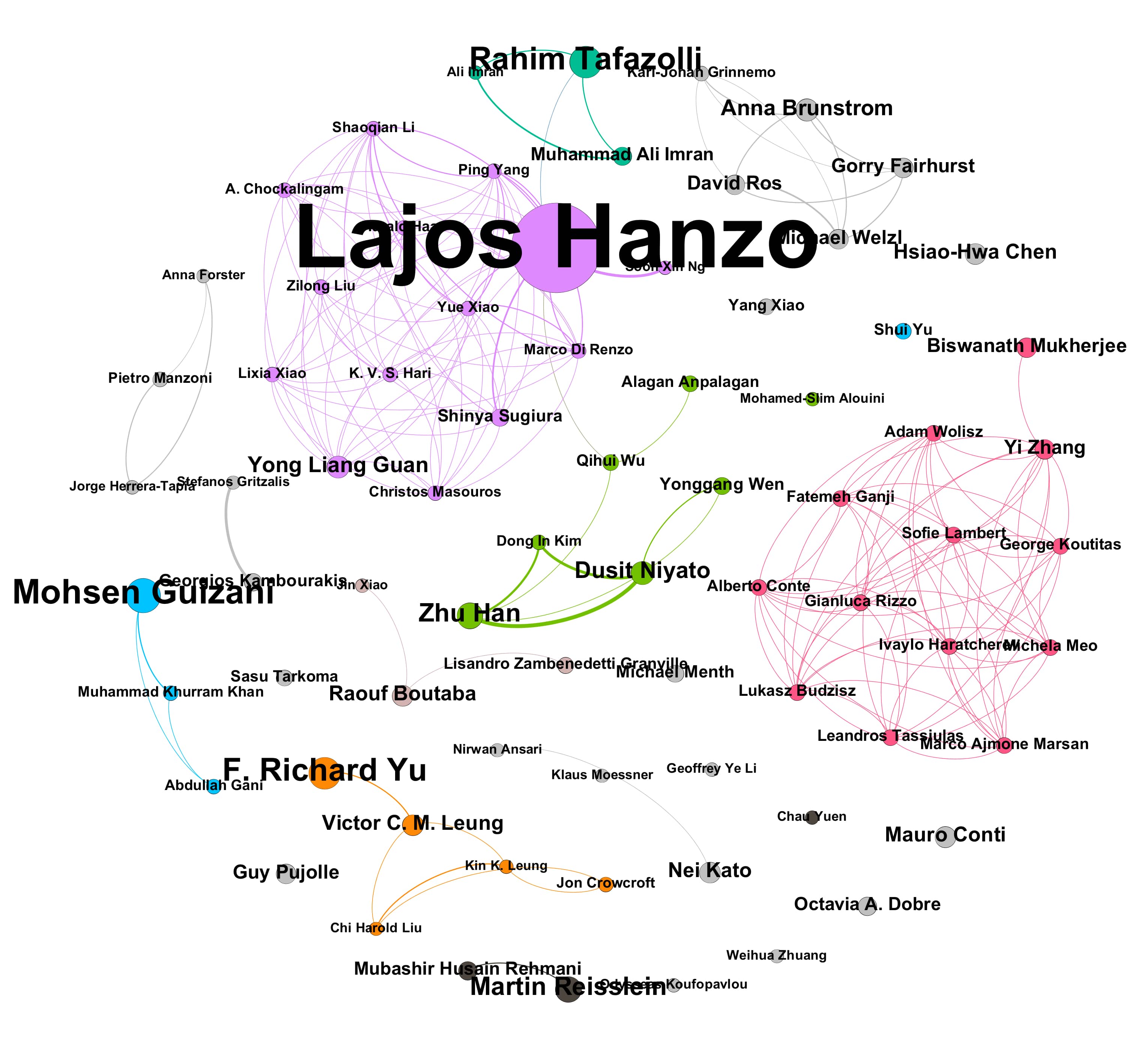}}
	\subfloat[TON]{\includegraphics[width=0.5\textwidth]{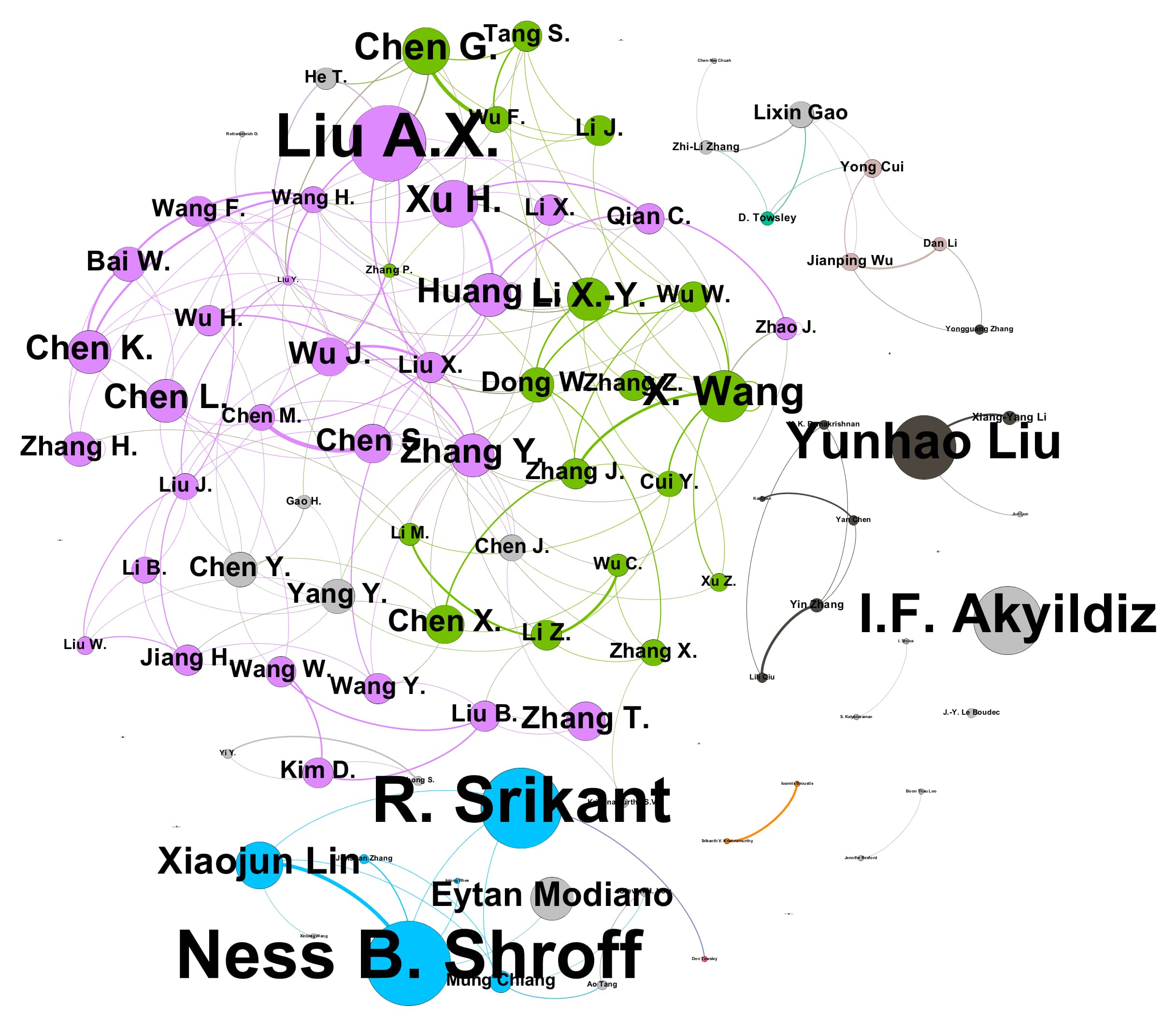}}
	\caption{Co-authorship network among top authors in the field of computer networking (a) COMST; (b) TON. Only those who have authored at least 8 articles are kept in clusters.}
	\label{fig:authors_top_gephi}
\end{figure}

Figure \ref{fig:authors_top_gephi} presents the clusters present in the network. We find 20 different clusters of authors in COMST and 674 clusters in TON. To improve the visualization, we only include authors who have more than eight articles in COMST and TON. After pruning of insignificant clusters, we found 18 clusters of authors in COMST and 15 co-authorship clusters in TON.

\begin{figure}[!h]
	\subfloat[COMST]{\includegraphics[width=0.5\textwidth]{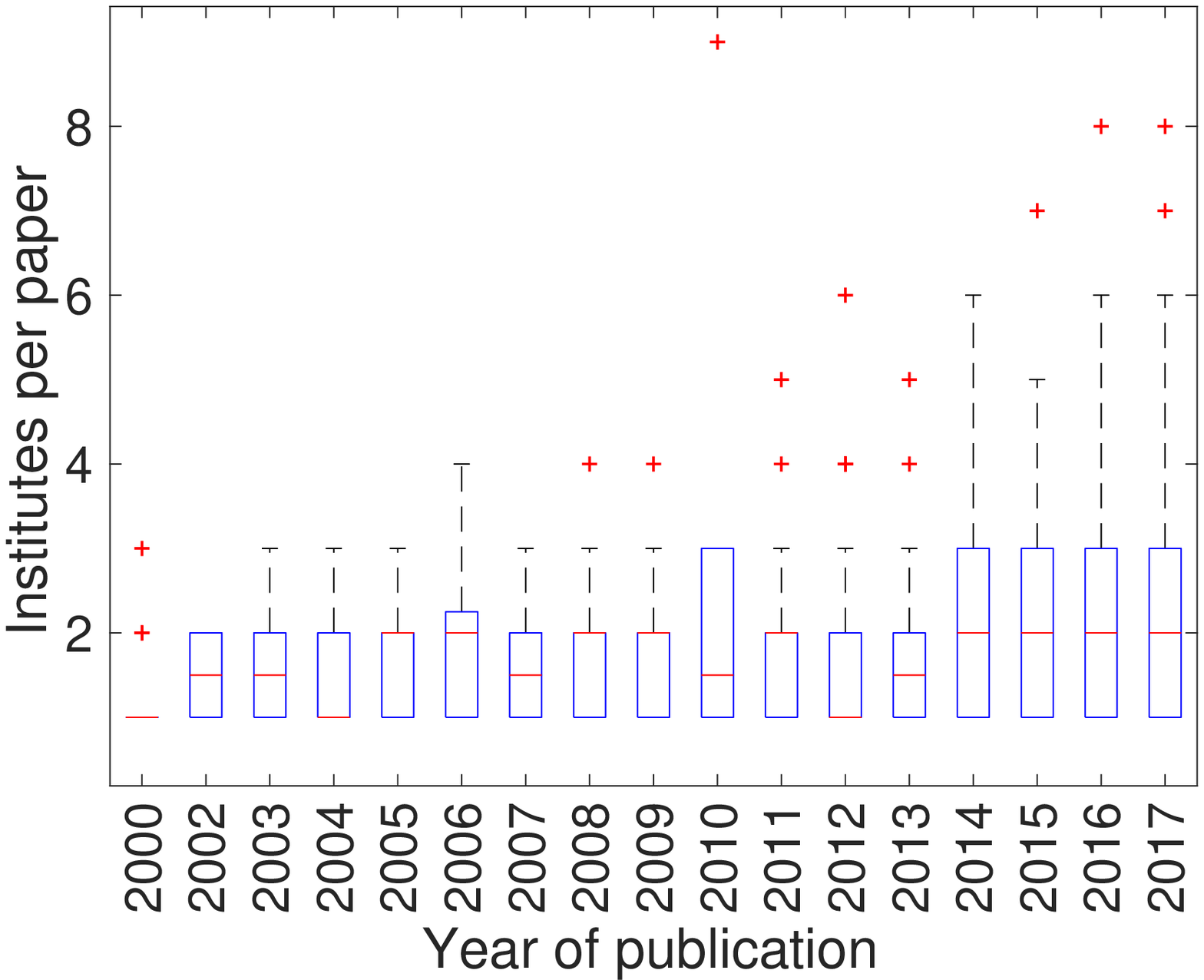}}
	\subfloat[TON]{\includegraphics[width=0.5\textwidth]{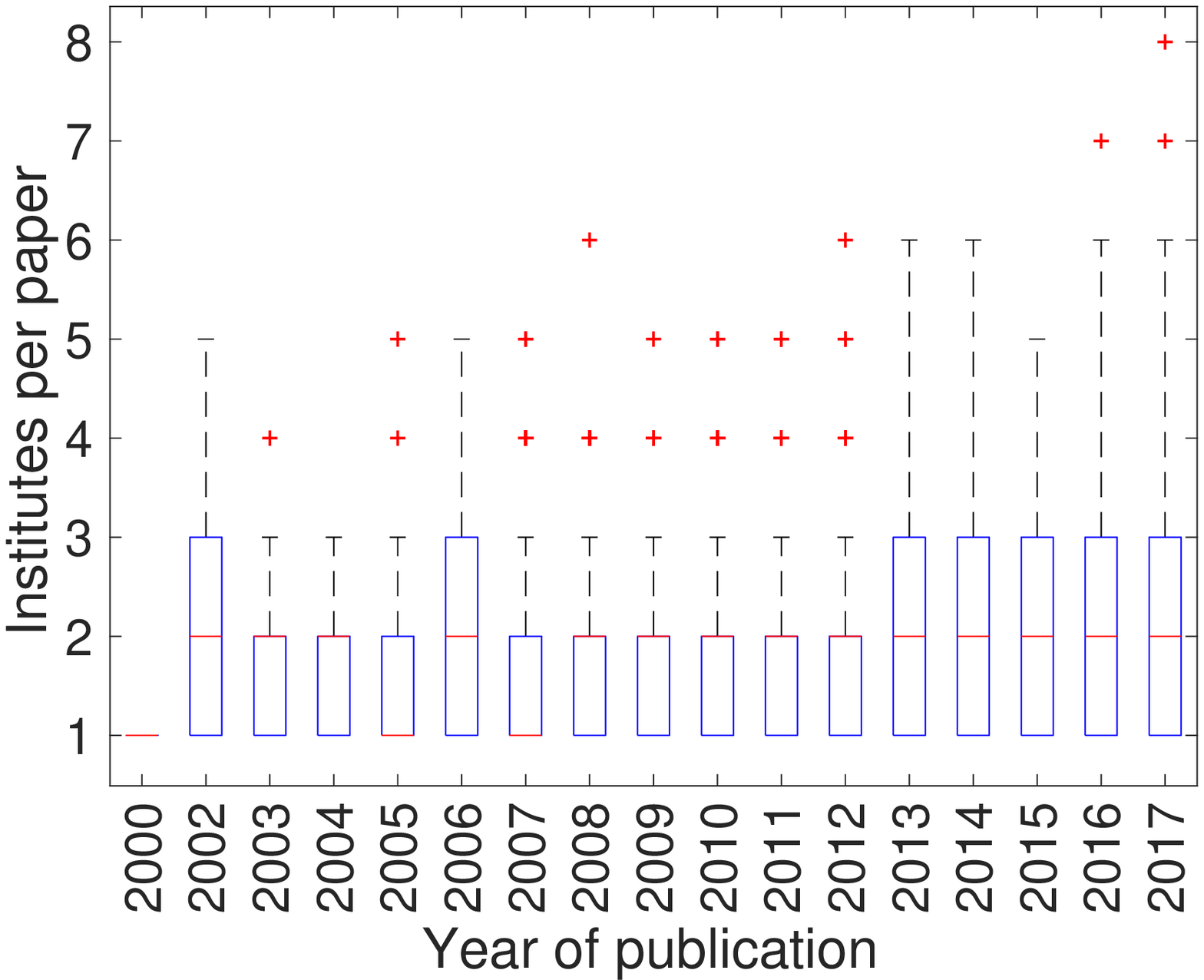}}
	\caption{Distribution of collaborating institutes per article during 2000--2017. \textit{In TON, the number of collaborating institutes increased from earlier years more than in COMST.}}
	\label{fig:insti_t}
\end{figure}

To analyze the behavior of collaborating institutes in COMST and TON, we performed an occurrence-based analysis on the count of collaborating institutes. Figure \ref{fig:insti_t} shows the distribution of the number of collaborating institutes per article in COMST and TON. With the passage of time, more institutes are contributing to COMST and TON articles, showing a trend toward increased collaboration among institutes. In the first 6 years, 37\% of articles have 2 or more contributing institutes in COMST whereas these numbers increased to 49\% in TON in the discussed time period. Overall, 51\% of articles have 2 or more contributing institutes in COMST and 63\% of articles in TON have mentioned multiple contributing institutes during the entire time period. It is clear that the tendency for institutional collaboration is increasing (as in other fields \citep{coccia2016evolution}); the median number of institutes in an article in 2000 is 1 for COMST and 1 for TON, compared to 2 and 2 in 2017. In addition, 13\% of authorship lists exceeding 5 in 2017 (compared to 3\% prior to 2006). In TON, the number of collaborating institutes increased from earlier years more than in COMST. 

%Perhaps most noteworthy is the spread of institutional collaboration numbers across articles, with a standard deviation of 0.00 in 2000 vs.\ 1.35 in 2017 for TON and with a standard deviation of 0.54 in 2000 vs.\ 1.31 in 2017 for COMST. The outliers are clear with 

%This is because TON contains more experimental content in its articles, which requires more empirical results. Researchers from various institutes have a distinct research remit and can produce higher quality results \citep{iglivc2017whom,coccia2016evolution,powell2018these}. 

%Throughout the discussed period, most COMST and TON articles did not have any collaborating foreign (they do not belong to the same country as the lead author) institute and the distribution of the collaborating local institutes was the same as the overall pattern of collaborating institutes. This trend supports general practice in research, whereby authors prefer to work with those from their own institute. In the case of experimental study-based articles, leading authors prefer to work with authors of local institutes so that their experiment can be performed under the same conditions and standardized results can be produced. Iglivc et al. and Smith et al. report similar results in their articles \citep{iglivc2017whom,smith2000collaborative}.

\begin{figure}[!h]
	\begin{center}
		\includegraphics[width=0.7\textwidth]{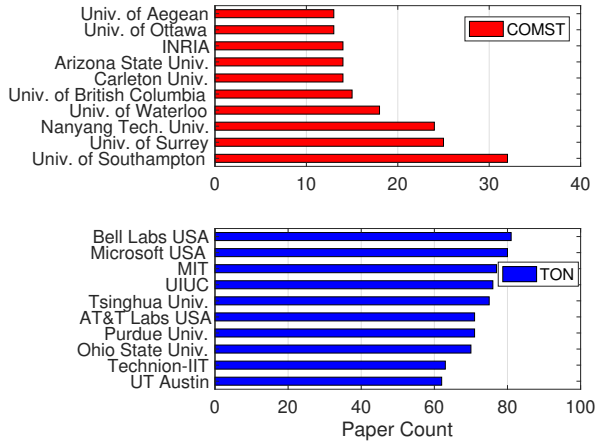}
	\end{center}
	\caption{Most-published institutes during 2000--2017, according to their article count. \textit{In COMST, academic institutes are publishing more whereas in TON, industrial institutes are more significant contributors with the top two contributors being Bell Labs USA and Microsoft USA.}}
	\label{fig:inst_top}
\end{figure}

\begin{figure}[!h]
	\centering
	\subfloat[COMST]{\includegraphics[width=0.5\textwidth]{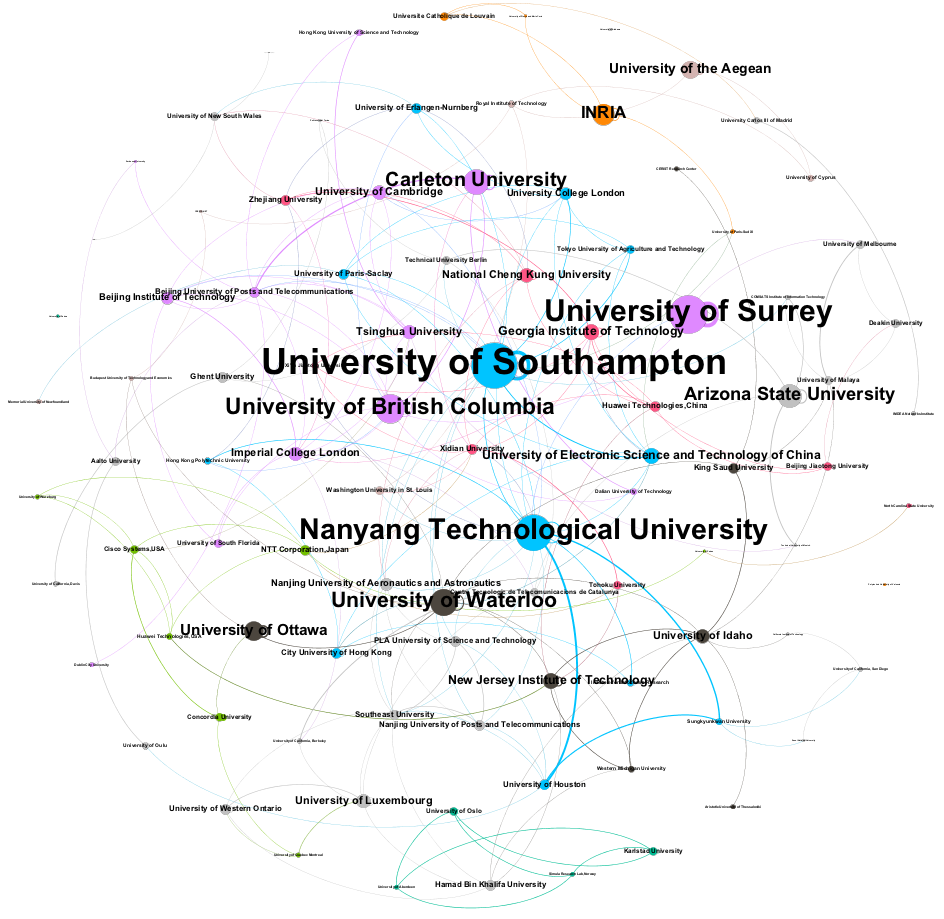}}
	\subfloat[TON]{\includegraphics[width=0.5\textwidth]{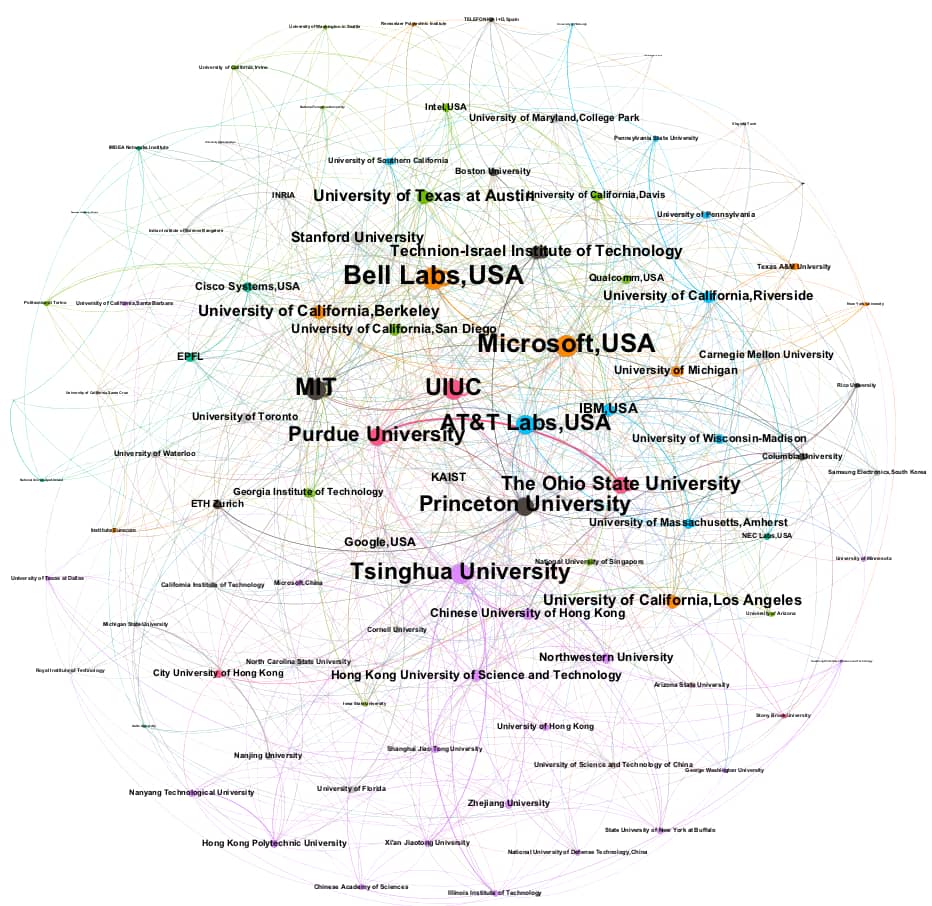}}
	\caption{Co-authorship network among top institutes in (a) COMST and (b) TON (with 10 minimum published articles). \textit{Distinct patterns can be observed in COMST and TON: academic institutes are prominent in COMST whereas clusters involving industrial centers (e.g., Bell Labs, Microsoft, and AT\&T Labs) are prominent in TON.}}
	\label{fig:inst_top_gephi}
\end{figure}

Published research is a crucial factor in determining the quality of education and research at any institute. Figure \ref{fig:inst_top} shows a similar result for the top institutes. We performed a clustering analysis using modularity class algorithm over COMST and TON articles. Figure \ref{fig:inst_top_gephi} shows a similar result for both the COMST and TON datasets. In both, the top publishing institutes are clustered into three groups according to their publishing behavior. In the TON data, Bell Labs and Microsoft, both in the United States, showed a significant co-authorship pattern. Similarly, the Massachusetts Institute of Technology (MIT) and the University of Illinois at Urbana-Champaign  (UIUC) are clustered together, and Tsinghua University is clustered with Princeton University. 

To remove the weak links, in both journals we set the degree threshold to 10. We found 21 different clusters of authors in COMST and 84 clusters in TON. To improve the visualization, we only include authors who have more than eight articles in COMST and TON. After pruning of insignificant clusters, we found 19 clusters of institutes in COMST and 12 co-authorship clusters of institutes in TON. Sudden decrease of a number of clusters in TON shows that there is a high number of institutes who are either new to TON or are not actively publishing in TON. Social network analysis has shown us the hidden relations between the top authors of COMST and TON, and we conclude that most of the top authors (measured by their publication count in COMST and TON) are clustered together because either they have strong collaboration behavior with each other or common co-author in-between.
\subsection{Analysis based on Structural Elements of Article} 
The structural elements of an article consist of the mathematical and graphical parts and the references cited. The mathematical and graphical elements help authors to convey the results related to an article, to discuss problems more precisely and concisely, and the references help readers to find research relating to the article. This sub-section addresses many important bibliometric questions on the structural elements of a research article. These include the distribution of the references in different genres of articles; the relationship between higher numbers of references and the author count of an article; the relationship between the number of references and the number of mathematical and graphical elements; and what kind of graphical and mathematical elements are found more in survey articles than experimental studies, and vice versa.
\begin{figure}[!h]
	\begin{center}
		\includegraphics[width=0.7\textwidth]{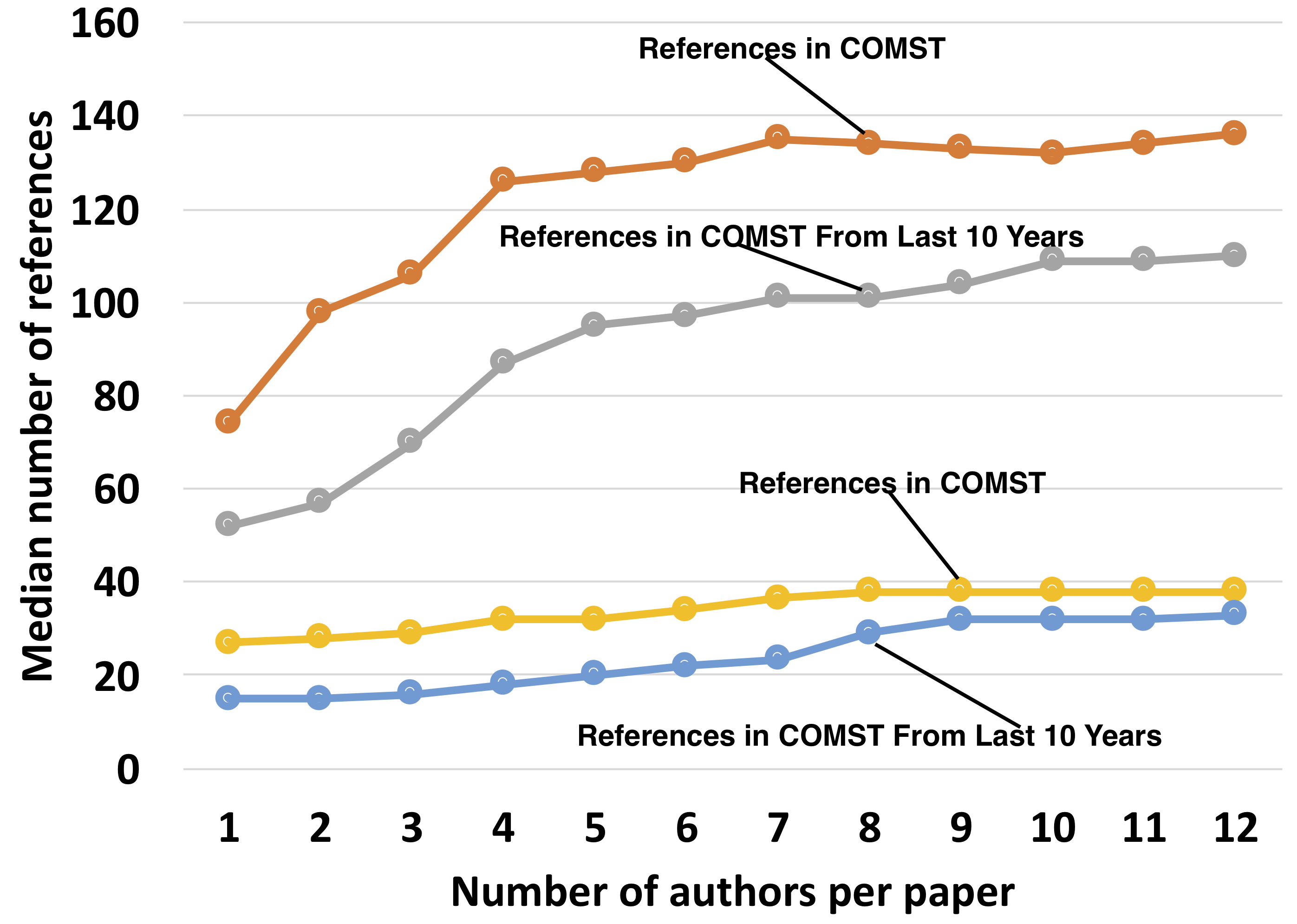}
	\end{center}
	\caption{Median number of references and median number of references used in article published in last ten years. \textit{COMST articles have a high number of references because their very nature requires references to numerous works.}}
	\label{fig:refernces_t}
\end{figure}

Different kinds of articles have varying numbers of references. For instance, survey-based articles have a high number by their nature that requires coverage of a broad area. Figure \ref{fig:refernces_t} shows that articles from a particular number of authors have higher numbers of median references in COMST than in TON. Figure \ref{fig:refernces_t} shows the results for COMST and TON data, where the number of references in COMST and TON goes up with the increasing number of authors. Similar results are reported by Saeed et al., Valenzuela et al. and Zhu et al. in their studies \citep{hassan2017identifying, valenzuela2015identifying, zhu2015measuring}. Figure \ref{fig:refernces_t} also shows that with the increasing number of authors, number of references from the last ten years in a paper also increase in COMST and TON. The data also has some outliers in terms of the number of references and references from the last ten years in a paper. Therefore, we have used median references for analysis because mean is more susceptible to outliers than median \citep{leys2013detecting}.  

%It is also observed that as the number of authors increases, the number of references also increases. This is potentially because more authors can contribute their suggestions or the work crosses multiple sub-disciplines. Most references in an article are usually discussed as related work or in the introduction. Authors usually try to cite recent articles yet to focus more on their own work than that of others. These articles have fewer references than others. \gareth{The above statements need quantification. What fraction of refs are in related work? What fraction of refs are self citations? etc If we can't quantify, we should remove the paragraph}

\begin{figure}[!h]
	\begin{center}
		\includegraphics[width=0.7\textwidth]{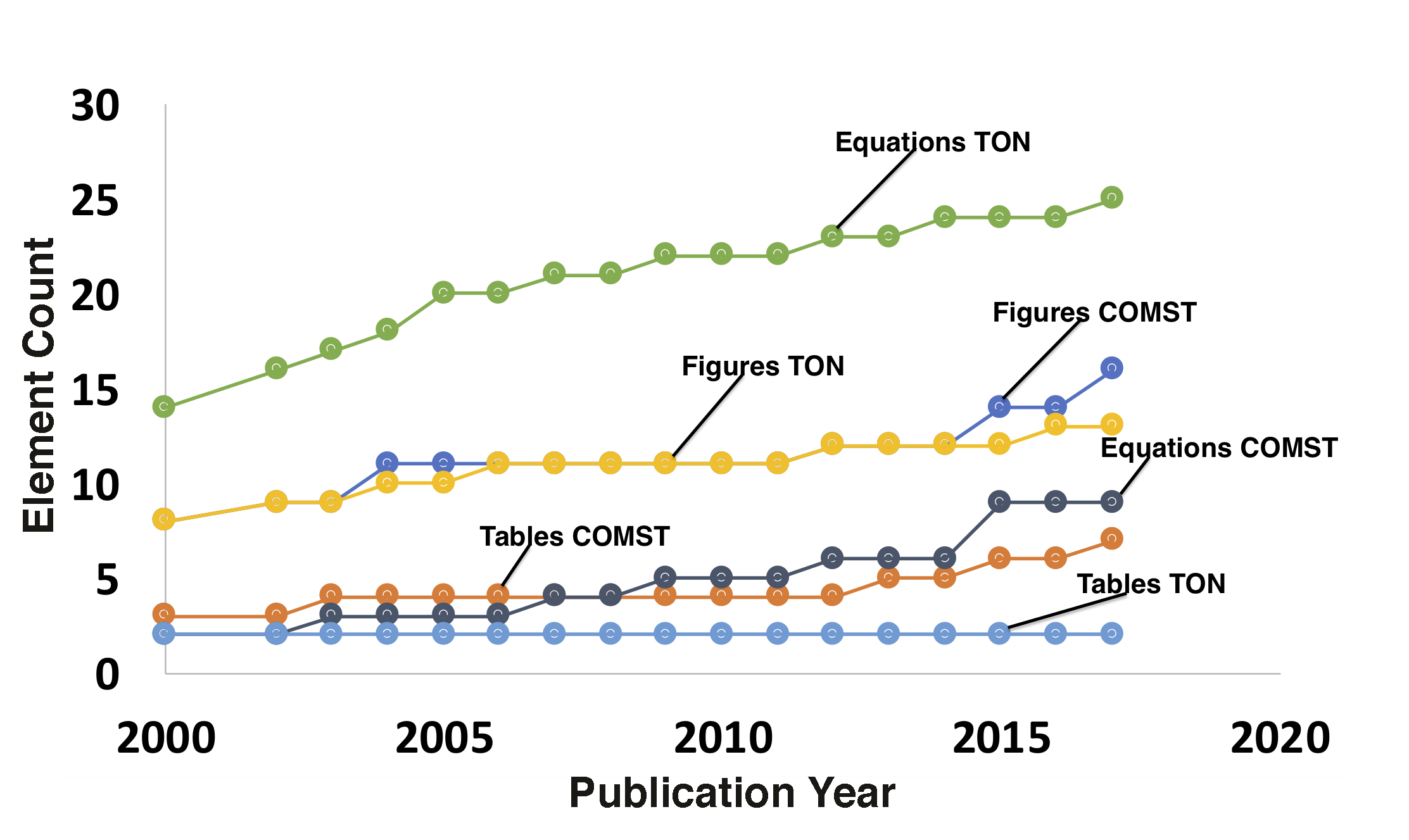}
	\end{center}
	\caption{Average number of mathematical and graphical elements during 2000--2017. \textit{Note that TON tends to have more equations whereas COMST tends to have more tables.}}
	\label{fig:math_t}
\end{figure}

Different types of research articles have different types of structural elements. For example, a survey-based article might have a higher number of graphical elements than mathematical equations, because tutorials can explain topics best using figures and tables. Figure \ref{fig:math_t} presents a breakdown of the average numbers of artifacts per year. In both journals, tables are the least frequently used. COMST has a high number of figures each year, and TON has a high number of equations. This is not surprising, considering the contrasting nature of these two journals.

We also note that the number of references in an article increases with the number of 
authors. Over time this trend is increasing, with the numbers of authors per article growing for both COMST and TON. Moreover, the number of references is higher in COMST articles than in TON articles. This is to be expected, as COMST focus on review and survey articles. Similar trends are send with graphical elements, where COMST exceeds TON. In contrast, TON has more mathematical elements which, again, is to be expected as TON tends to contain experiment-based publications.

\section{Content Based Analysis and Findings}
\label{sec:content}
\par This section contains two types of analysis of COMST and TON: (A) keyword-based analysis, based on index keywords; and (B) readability-based analysis. We address questions such as, what are the popular topics of computer networking research during each year? what topics are discussed by top authors in COMST and TON? and which types of articles are easiest to read?

\begin{figure}[H]
	\begin{center}
		\includegraphics[width=0.7\textwidth]{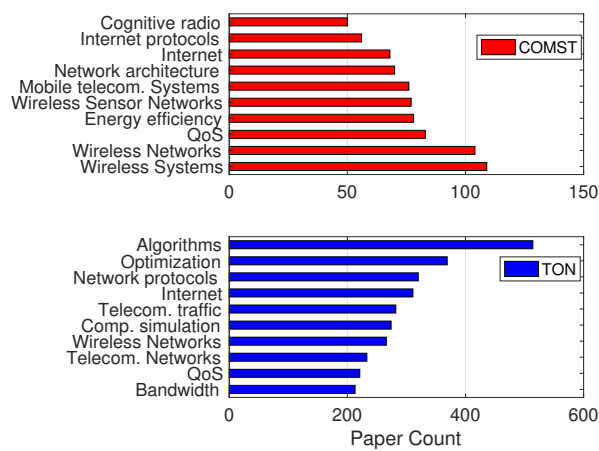}
	\end{center}
	\caption{Most popular topics in COMST and TON and their article count during 2000--2017, in terms of article count (cf. Figure \ref{fig:cited_top_keywords}, in which keywords of the most-cited articles are listed.)}
	\label{fig:keywords_top}
\end{figure}
\subsection{Keyword-based analysis of articles} 
\begin{table}[!h]
	\centering
	\tiny
	\caption{Popular topics extracted from COMST and TON on the basis of indexed keywords. \textit{Topics are largely stable but temporal shifts in trends can be identified (e.g., spike of interest in ``complex networks'' in TON over the last 3 years)}.}
	\label{hot_topics_years}
	\begin{adjustbox}{width=1\textwidth}
		\begin{tabular}{m{.4cm}m{3.6cm}m{3.6cm}}
			\hline
			Year & COMST    & TON                                               \\\cline{1-3}
			2000 & Computer networks, Bandwidth, Telephony                                          & Telecom. traffic, Congestion control (communication) , Algorithms          \\
			2002 & IP networks, Bandwidth, WLAN                                                     & Algorithms, Telecom. traffic, Network protocols                            \\
			2003 & Bandwidth, Web and Internet services, Scalability                                & Algorithms, Telecom. traffic, Bandwidth                                    \\
			2004 & Telecom. traffic, IP networks, Optical fiber networks
			& Computer simulation, Mathematical models, Algorithms\\
			2005 & Mobile ad hoc networks, Cellular network, Complex networks                       & Mathematical models, Computer simulation, Algorithms                       \\
			2006 & Mobile ad hoc networks, Algorithms, Internet                                     & Algorithms, Congestion control (communication), Computer simulation        \\
			2007 & Telecom networks, Mobile ad hoc networks, Service infrastructure 
			& Computer simulation, Telecom. traffic, Optimization                                                                                   \\
			2008 & Network security, Internet, Optimization                                         & Network protocols, MANs, Sensor networks                                   \\
			2009 & Wireless telecommunication systems, Mobile telecommunication systems, Security   & Wireless telecommunication systems, Internet, Optimization                 \\
			2010 & Optimization, Sensor networks, Scheduling                                        & Optimization, Topology, Throughput                                         \\
			2011 & Telecommunication networks, Sensors, Network architecture                        & Optimization, Computer simulation, Approximation algorithms                \\
			
			2012 & Wireless telecommunication systems, Quality of service, Resource allocation      & Optimization, Algorithms, Wireless networks                                \\
			2013 & Energy efficiency, Wireless telecommunication systems, Algorithms                & Algorithms, Optimization, Scheduling                                      \\
			2014 & Wireless telecommunication systems, Complex networks, LTE                        & Wireless networks, Optimization, Electric network topology                 \\
			2015 & Mobile telecommunication systems, Network architecture, Energy efficiency        & Algorithms, Complex networks, Scheduling                                   \\
			2016 & Energy Efficiency, Mobile telecommunication systems, Software-defined networking & Optimization, Complex networks, Software engineering                       \\
			
			2017 & Wireless sensor networks, Bandwidth, Computer architecture                       & Optimization, Polynomial approximation, Complex networks                   \\
			\hline
		\end{tabular}
	\end{adjustbox}
\end{table}
Investigating the popular topics is considered to be one of the best ways of studying the paradigm shifts in any research field. It is helpful in describing the research trends of a field. In this sub-section, we use COMST and TON data to analyze the popular topics in the field of computer networking. We have described the top 10 popular topics discussed in survey-based and experimental studies-based articles in computer networks. This approach provides a holistic overview of research trends in computer networking since it covers both original and survey-based articles.
Figure \ref{fig:keywords_top} represents the most popular topics in computer networking, according to the COMST and TON dataset. COMST contains survey articles and, from 2000 to 2017, it published surveys relating to wireless and mobile communication systems, QoS, and Internet. By contrast, during this period most of the articles published in TON discuss algorithmic and optimization problems relating to computer networking.
Table \ref{hot_topics_years} shows the change over time of popular topics in the field of computer networking, using the COMST and TON datasets. Popular topics mentioned in Table \ref{hot_topics_years} give the approximate overall research trends in the field of computer networking. While there is a lot of stability in the keywords (`wireless networks' is common in COMST and `optimization' and `algorithms' is common in TON, we see over time new topics emerging such as `complex networks' in the last three years of TON publications).
\begin{table}[H]
	\centering
	\tiny
	\caption{Using LDA-based topic modeling to determine 10 most popular topics in COMST and TON. \textit{We see different (more coherent) results using LDA-based topic modeling compared to the keywords-based results in Table \ref{hot_topics_years}.}}
	\label{keyword_LDA}
	\begin{adjustbox}{width=1\textwidth}
		\begin{tabular}{|m{4cm}|m{4cm}|}
			\hline        
			COMST& TON  \\\hline
			[attack, detect, social, privacy, threat, anonymous, vulnerable, trust, category, protect] & [algorithm, problem, optimization, schedule, achieve, policy, solution, distribution, wireless, propose]  \\\hline
			[spectrum, optics, radio, cognition, band, sensor, model, cellular, fiber, availability]&[queue, congest, fair, buffer, stabilize, class, loss, converge, arrive, parameter] \\\hline
			[mobile, scheme, multimedia, content, access, satellite, delivery, solution, device, difference] &[detect, attack, estimate, accuracy, identify, filter, memory, trace, aggregate, acute]   \\\hline
			[smart, data, grid, energy, power, center, secure, manage, consumption, trust]& [switch, energy, power, consumption, spectrum, synchronize, architecture, device, cell, input]  \\\hline
			[protocol, wireless, design, sensor, node, control, propose, route, optimize, algorithm]&[approximate, compute, bound, graph, case, path, general, topology, maximum, scheme]   \\\hline
			[protocol, route, node, sensor, application, propose, mobile, design, wireless, research] & [node, sensor, energy, data, wireless, distribute, protocol, attack, transmission, power]  \\\hline
			[video, sensor, multicast, data, local, content, wireless, application, multimedia, stream] &[schedule, delay, throughput, packet, queue, rate, policy, bound, buffer, scheme]   \\\hline
			[network, protocol, route, application, node, survey, control, propose, design, sensor]& [algorithm, problem, optimize, schedule, policy, delay, perform, rate, achieve, bound]  \\\hline
			[compute, application, model, system, technique, cloud, local, resource, environment, method] & [node, wireless, mobile, channel, transmiss, energy, protocol, propose, use, power]  \\\hline
			[system, technique, communication, channel, wireless, transmission, perform, design, code, signal]& [control, allocate, network, user, resource, provide, service, fair, bandwidth, algorithm]  \\\hline
		\end{tabular}
	\end{adjustbox}
\end{table}
One limitation of the analysis above is that it is based on stipulated keywords, which may exclude pertinent topics. Hence, we use Latent Dirichlet Allocation (LDA) to identify important themes within the article's body. LDA takes raw text, the number of topics and a dictionary of words as the input, and outputs the most significant topics with words from the raw data \citep{blei2003latent}. We kept the number of latent output topics to 10 and iterated our algorithms 400 times on our dataset in order to achieve converged results. Table \ref{keyword_LDA} shows the results of LDA on the COMST and TON datasets. It can be seen that the results are different from those of the results for keywords in Table \ref{hot_topics_years} and refer to different topics such as smart grid, sensor networks, cognitive radios for COMST and optimization algorithms, congestion control solutions, approximation algorithms for TON.
\subsection{Keyword co-occurrence analysis}
\begin{figure}[!h]
	\centering
	\subfloat[COMST]{\includegraphics[width=0.5\textwidth]{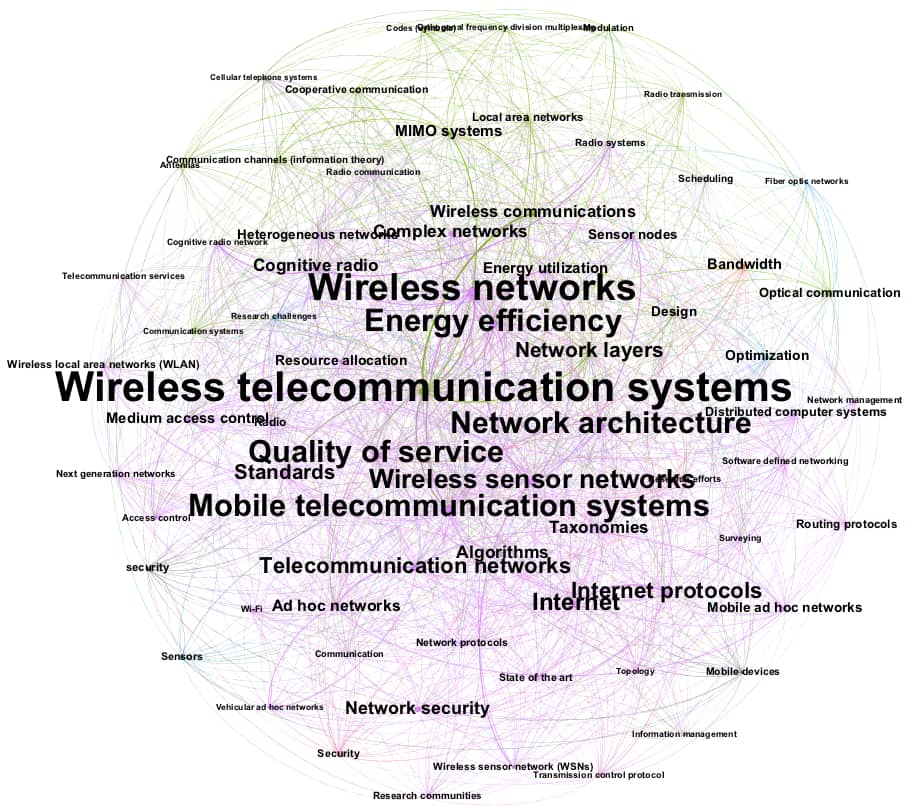}}
	\subfloat[TON]{\includegraphics[width=0.5\textwidth]{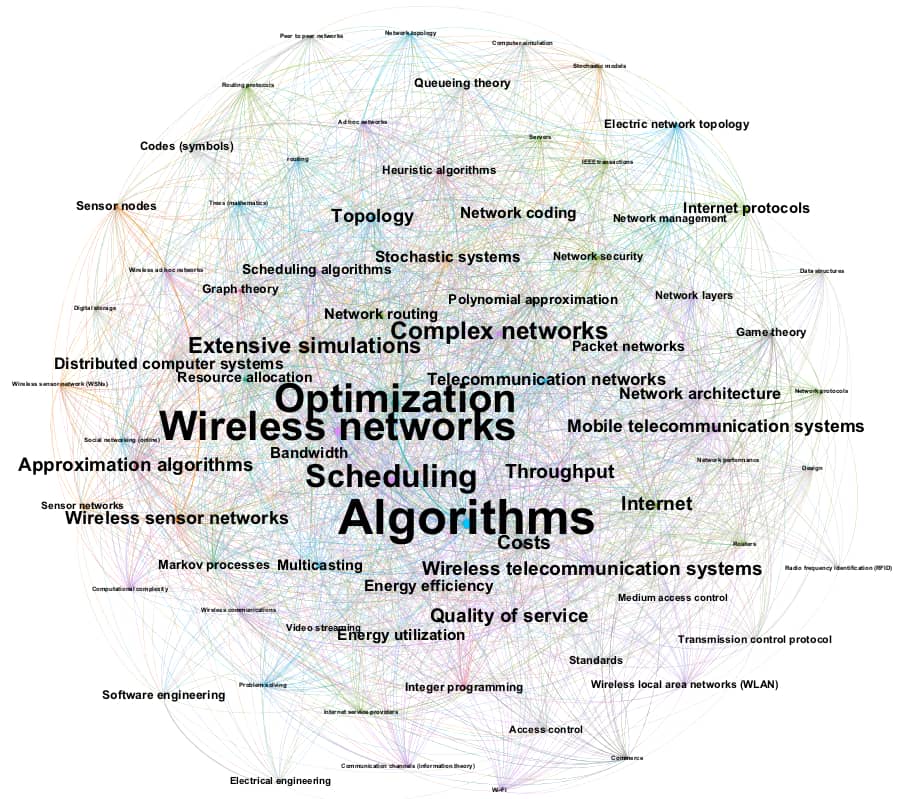}}
	\caption{Keyword co-occurrence network in which the node size indicates the number of links with other nodes and node color represents cluster membership. \textit{It can be noted that COMST (TON) keywords are typically biased towards problems and network types (solutions and techniques)}.}
	\label{fig:keyword_top_gephi}
\end{figure}

Keyword co-occurrence analysis helps researchers to find a publication venue's most common topics. These analyses also help researchers to find topics and domains that are strongly related to each other. Figure \ref{fig:keyword_top_gephi} is the term co-occurrence map for COMST and TON.

There is limited overlap in the keywords used in the top-cited articles in TON and COMST. The keywords in COMST are biased towards problems and those in TON towards techniques/solutions.

Terms in a larger font size have a higher co-occurrence than other keywords in the graphs. In COMST, frequently co-occurring terms are "Wireless Telecommunication Systems", "Wireless Networks", "Quality of Service", "Energy Efficiency", "Mobile Telecommunication Systems", and so on. In TON, the most frequently co-occurring terms are "Optimization", "Algorithms", "Wireless Networks", "Scheduling", and so on. Top keywords (measured on publication count) in both the venues are clustered in the same groups and have stronger links with each other than with unpopular keywords. This trend shows that in both venues, there are only some top keywords (measured on publication count) which are discussed in most of the articles. The results also show that in most of the articles in COMST and TON, top keywords co-occur with each other.

%Another well-known method to extract latent topics from the raw text is topic modeling. One of the best-known algorithms for topic modeling is Latent Dirichlet Allocation (LDA) (discussed earlier in Section \ref{sec:methodology}). Topic modeling done using LDA showed similar results to keyword-based analysis in both COMST and TON.

We also observe several other trends that are noteworthy. For example, in COMST, authors mostly discuss network configurations (e.g. WSN) and problems (e.g. scheduling, energy efficiency), whereas in TON it is the techniques (such as optimization, algorithms) that are emphasized. We note that the Keyword co-occurrence-based analysis also helps researchers to establish the topics and domains that are strongly related to each other. Our findings are that the most popular keyword terms in COMST and TON relate to \textit{problems} (quality of service, energy efficiency etc.) and \textit{techniques} (optimization, algorithms etc.), respectively.

\section{Citation Based Analysis and Findings}
\label{sec:citation}
Citations are used to investigate the contributions of an author, organization, country or publication venue. Citation analysis is an effective tool to rank the productivity of various research bodies. In this section, we address some important bibliometric questions using citation data from COMST and TON articles, such as who are the most-cited authors in COMST and TON; whether they have the same h-index as the most-published authors in COMST; whether increasing the number of authors affects the number of citations of an article; the most-cited keywords in COMST and TON; and whether a larger number of mathematical and graphical elements in an article increases its citation count.

\subsection{Citation Based Analysis of Different Research Entities}
In computer networking, some authors play more significant roles in advancements of the field than others. It is worth observing the impact and usability of their research. 
\begin{figure}[!h]
	\begin{center}
		\includegraphics[width=0.7\textwidth]{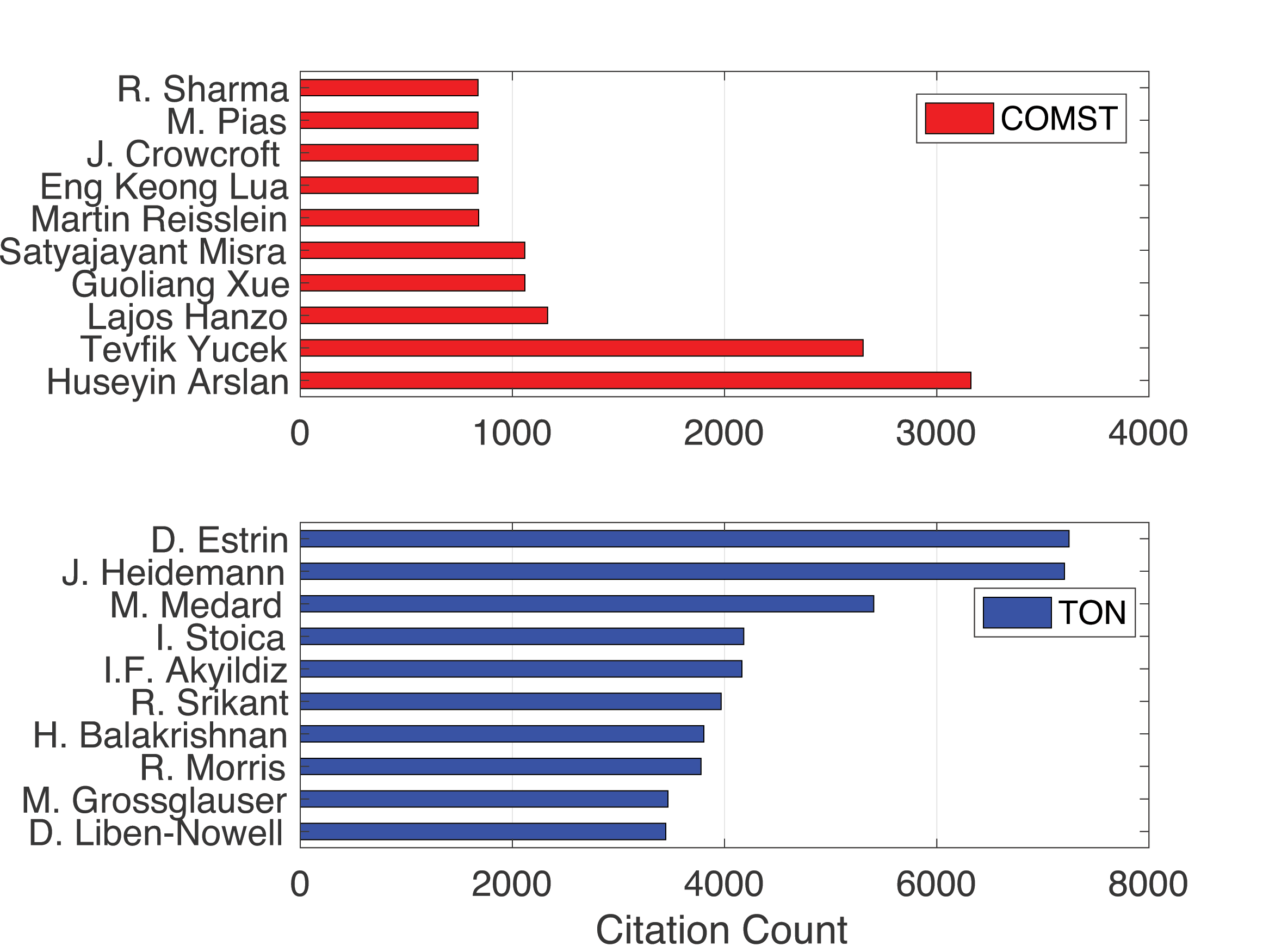}
	\end{center}
	\caption{Most-cited authors. \textit{We see that the most-cited TON articles tend to have more citations even though COMST on average are cited more; cf. Table I, which shows that COMST (TON) on average has 67 (37) citations.}}
	\label{fig:cited_top}
\end{figure}
Figure \ref{fig:cited_top} shows the most-cited authors in COMST and TON from 2000 to 2017. From Figure \ref{fig:cited_top} and Figure \ref{fig:authors_top}, it can be observed that the top most-published authors and the top most-cited authors in COMST and TON are entirely different. Citations do not entirely represent the significance of the research undertaken by a researcher. There are many parameters to analyze its significance, but the h-index is the most widely used, and it is a better measure of an author's significance in a field than a simple citation count. 

Figure \ref{fig:cited_top_hindex} shows the authors in COMST and TON with the highest h-index, and how the top ten highest publication counts are from the top ten authors with the highest h-index in COMST and TON. The data confirms that the top authors (measured by publication count) are the ones who have significant research contributions in terms of publication count as well as citation count.

\begin{figure}[!h]
	\begin{center}
		\includegraphics[width=0.7    \textwidth]{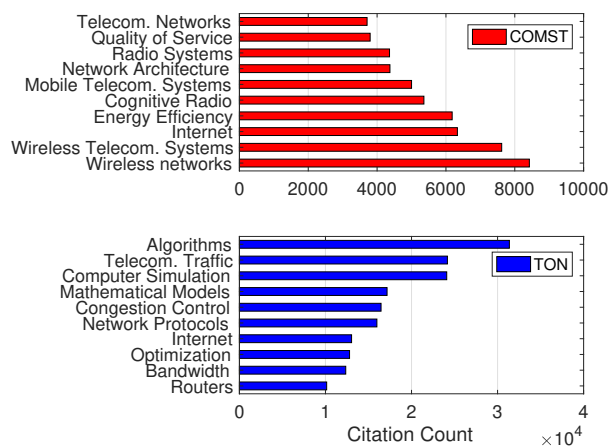}
	\end{center}
	\caption{Most-cited keywords. \textit{It is noticeable that there is negligible overlap in the keywords used in TON and COMST.}}
	\label{fig:cited_top_keywords}
\end{figure}

\begin{figure}[H]
	\centering
	\subfloat[COMST]{\includegraphics[width=0.4\textwidth]{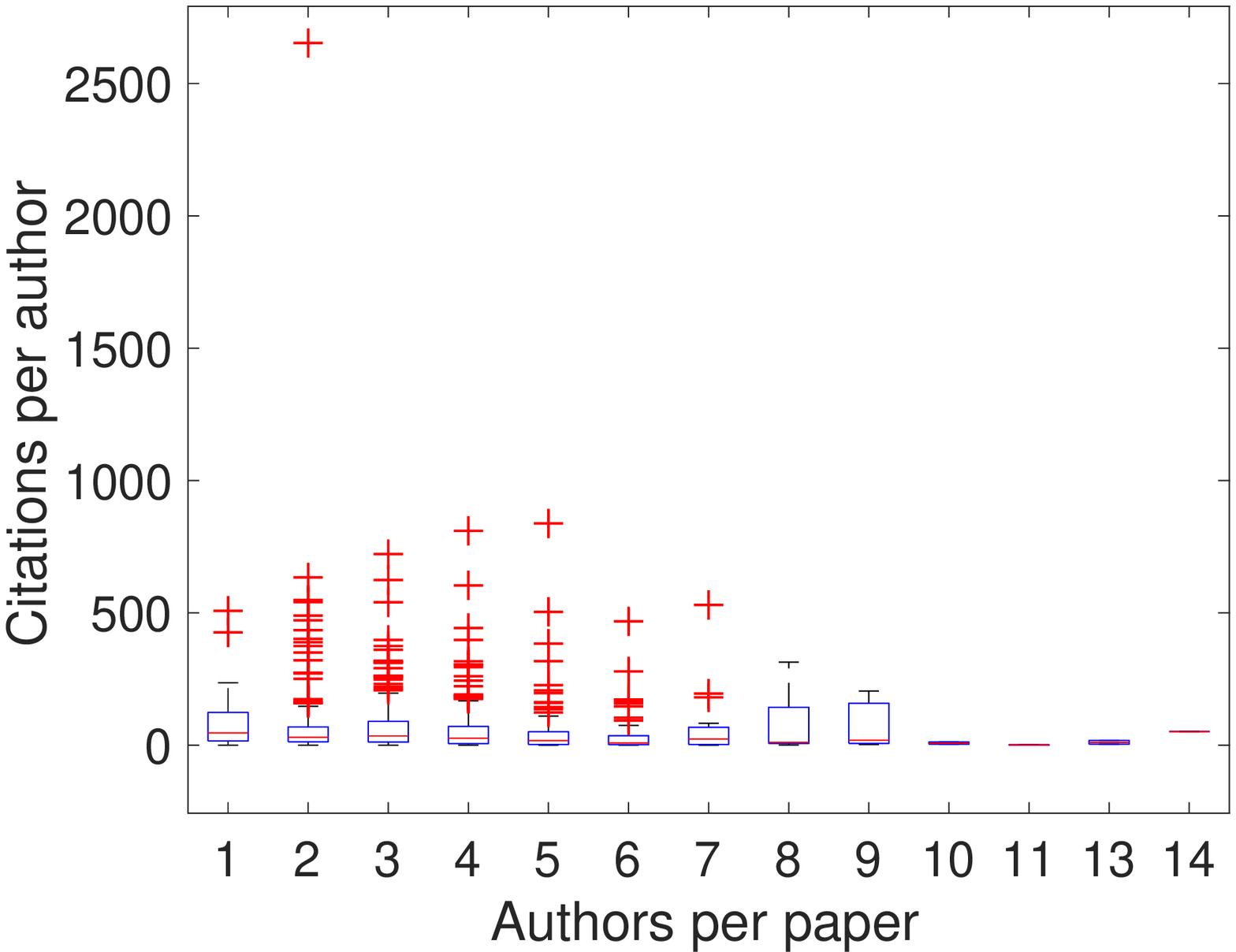}}
	\subfloat[TON]{\includegraphics[width=0.4\textwidth]{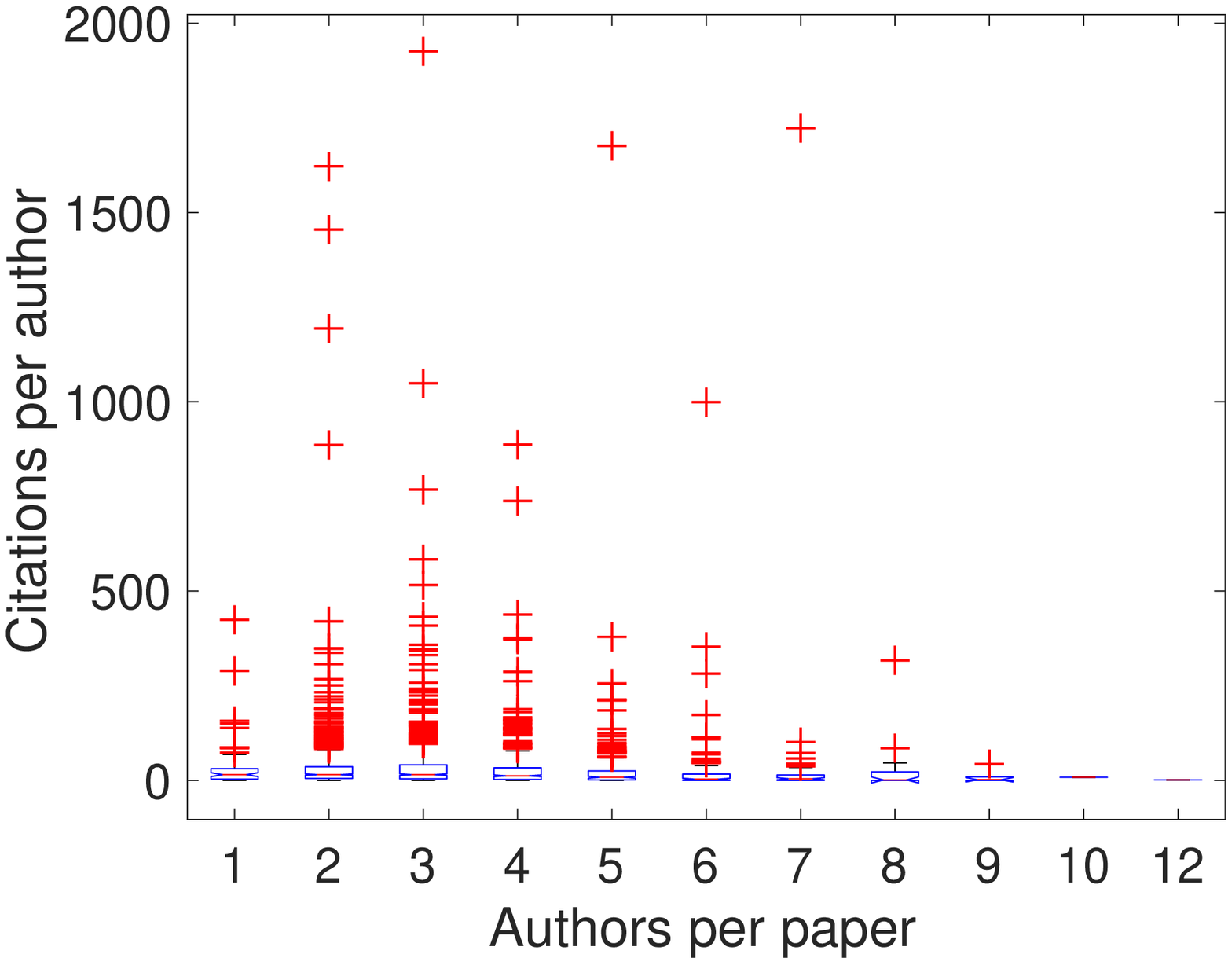}}
	\caption{Number of citations per article in COMST and TON with respect to the number of authors. \textit{The most-cited articles typically have a moderate number of authors.}}
	\label{fig:cite_t}
\end{figure}

Figure \ref{fig:cited_top_keywords} shows the impact of the top countries in COMST and TON. For both publication venues, the United States is the most prominent contributor. Figure \ref{fig:cite_t} presents the citation counts for each journal based on how many authors are on the article. We see that TON articles tend to have higher citation counts than survey-based articles when we consider the top-cited articles but on average COMST articles are cited more (see Table I, in which it is shown than COMST have on average 67 citations compared to 37 for TON). The higher citations of COMST articles on average likely stems from their citations in many topic-specific articles as a general resource. 

\subsection{Impact of Different Attributes On Article's Citation Count}
\par Different parameters of an article have a different impact on its citation count. The feature ranking of parameters can be performed by various methods such as PCA, SVD, and Random Forest. To measure the impact of these parameters on the citation count in our dataset, we used the Extremely Randomized Trees classifier, which is a variant of Random Forest. It computes the importance of a feature using Gini or average decay in impurity, which gives the impact of a feature on the label of a dataset. A higher value from the ExtraTree Classifier for a feature indicates greater importance for that feature with respect to the dependent variable (class label) \citep{geurts2006extremely}. Table \ref{cited_feature_rank} shows the impact of each feature on a dependent variable (class label). Results from Table \ref{cited_feature_rank} show that citations of the papers are more dependent on structural elements of paper as compared to the author based elements of the paper.
%\waleed{PCA arrange the feature based on their significance but does not return the significance value of each feature. On the other hand, Extremely Randomized Trees classifier compute the value of significance of a feature on the class label.}

% \subsection{Discussion on Citation Based Analysis}

% Citations are used to investigate the contributions of an author and are an effective tool to rank the productivity of various research bodies. 

% \par In COMST and TON, the most-cited authors are those with the highest h-index, which shows that the prominent researchers are those who are actively undertaking research in the field. We also found that COMST publishes experimental-based articles with a higher total and the average number of citations than survey-based articles. 
\begin{table}[H]
	\centering
	\footnotesize
	\caption{Impact of different features, based on their citation, on scale 0 to 1 for both COMST and TON}
	\label{cited_feature_rank}
	\begin{adjustbox}{width=1\textwidth}
		\begin{tabular}{cc}
			\hline
			\textbf{Feature Name} & \textbf{Impact (Gini Impurity Index)}\\
			\hline
			Number of Figures                                               & 0.07  \\ 
			
			Coleman-Liau Readability Test                                                    & 0.07  \\ 
			
			SMOG Readability Test                                                           & 0.07  \\ 
			
			No. of words in article title                                                     & 0.07  \\ 
			
			Special Sections on Pitfalls                                    & 0.07  \\ 
			
			Number of local authors (with reference to the first author's country)& 0.07  \\ 
			
			Number of foreign authors (with reference to the first author's country)                                 & 0.07  \\ 
			
			No. of Equations                                                & 0.06  \\ 
			
			Flesch-Kincaid Ease Readability Test                                            & 0.06  \\ 
			
			Number of references (i.e., articles cited in the article)                         & 0.06  \\ 
			
			Number of local authors                                         & 0.06  \\ 
			
			Number of Tables                                                & 0.05  \\ 
			
			Number of authors                                               & 0.05  \\ 
			No. of Institutions                                             & 0.04  \\ 
			
			Number of Institution from same countries as lead author        & 0.04  \\ 
			
			Number of Participating Countries                               & 0.03  \\ 
			
			Flesch-Kincaid Grade                                            & 0.02  \\ 
			
			Number of references (i.e., articles cited in the article) from last 10 years articles                   & 0.02  \\ 
			\hline
		\end{tabular}
	\end{adjustbox}
\end{table}

\section{Comparison Between Top Journals and Conferences in Computer Networking}
\label{sec:confcomp}

The previous sections have explored computer networking research soley through the lens of journals. Although important, computer networking stands out as a discipline that also values conference publications. Thus, we next proceed to compare the previously observed trends within journal publishing against that seen for conferences. For this, we select two top conference in computer networking: ACM SIGCOMM\footnote{http://www.sigcomm.org/} and IEEE INFOCOM\footnote{http://www.ieee-infocom.org/}.

In this section, we analyze SIGCOMM and INFOCOM based on the different key parameter such as author productivity, content-based analysis, and citations and compare them with COMST and TON. 
\subsection{Research Productivity of Authors}
\begin{figure}[!h]
	\centering
	\subfloat[Journals]{\includegraphics[width=0.45\textwidth]{Authors_t1.png}}
	\subfloat[Conferences]{\includegraphics[width=0.45\textwidth]{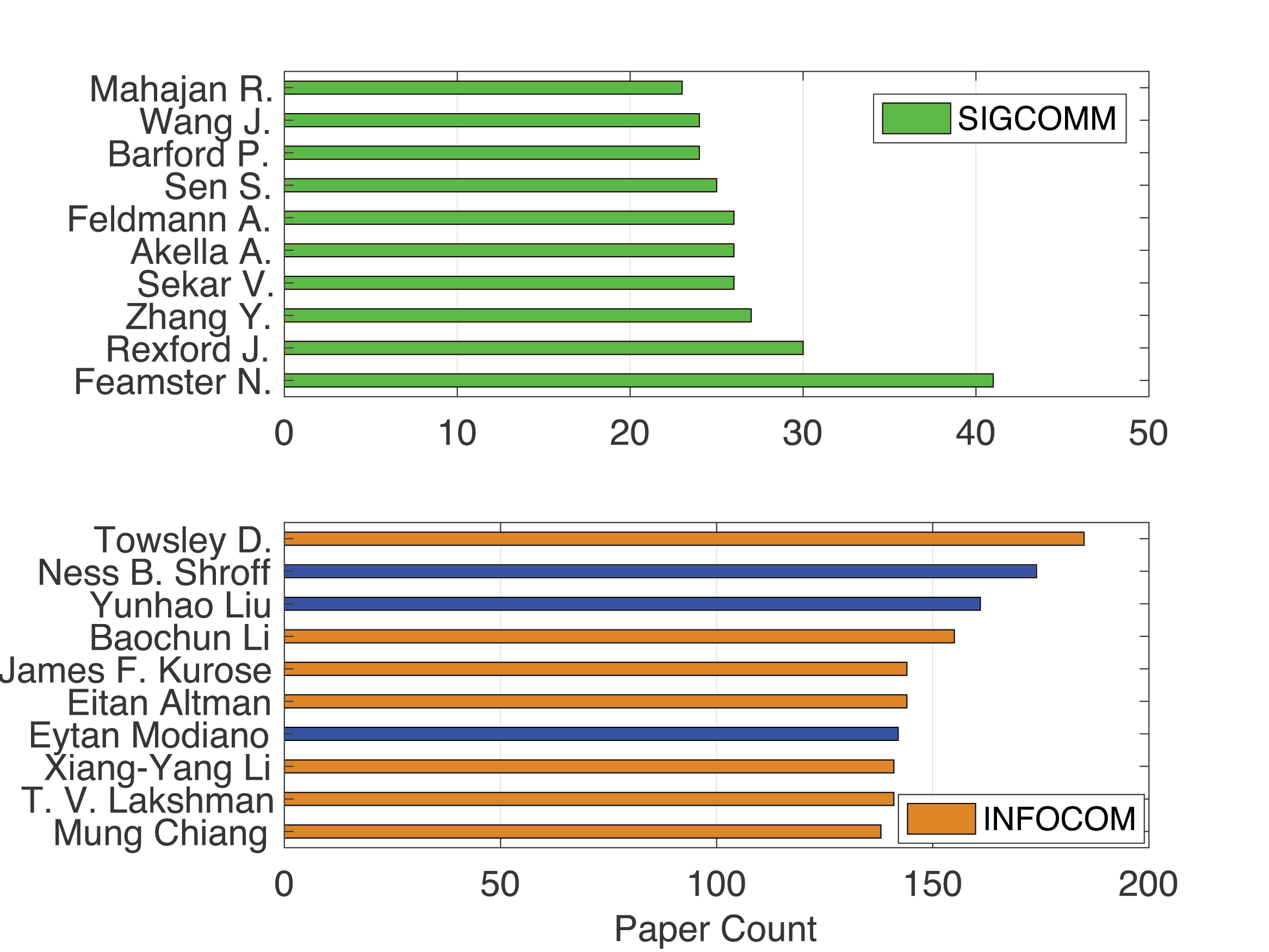}}
	\caption{Most-published authors during 2000--2017, according to article count. (a) Journals; (b) Conferences. Same color bars represent the overlapping authors among different venues. \textit{Interestingly, there is an overlap between the top authors of TON and INFOCOM which shows the prominent authors in TON and INFOCOM.}}
	\label{fig:top_authors_c}
\end{figure}

As publication count is one of the simplest metric to analyze the research productivity of authors, we analyze the top authors in all of the four top venues based on their publication count. Figure \ref{fig:top_authors_c} shows the top published authors in all COMST, TON, SIGCOMM, and INFOCOM. The analysis shows that Ness B. Shroff of The Ohio State University, Yunhao Liu of Tsinghua University and Eytan Modiano of Massachusetts Institute of Technology are the overlapping most-published authors in TON and INFOCOM and emerged as most prolific common authors in these two venues. Furthermore, there is no overlap between COMST and the other three venues.

\begin{figure}[!h]
	\centering
	\includegraphics[width=0.5\textwidth]{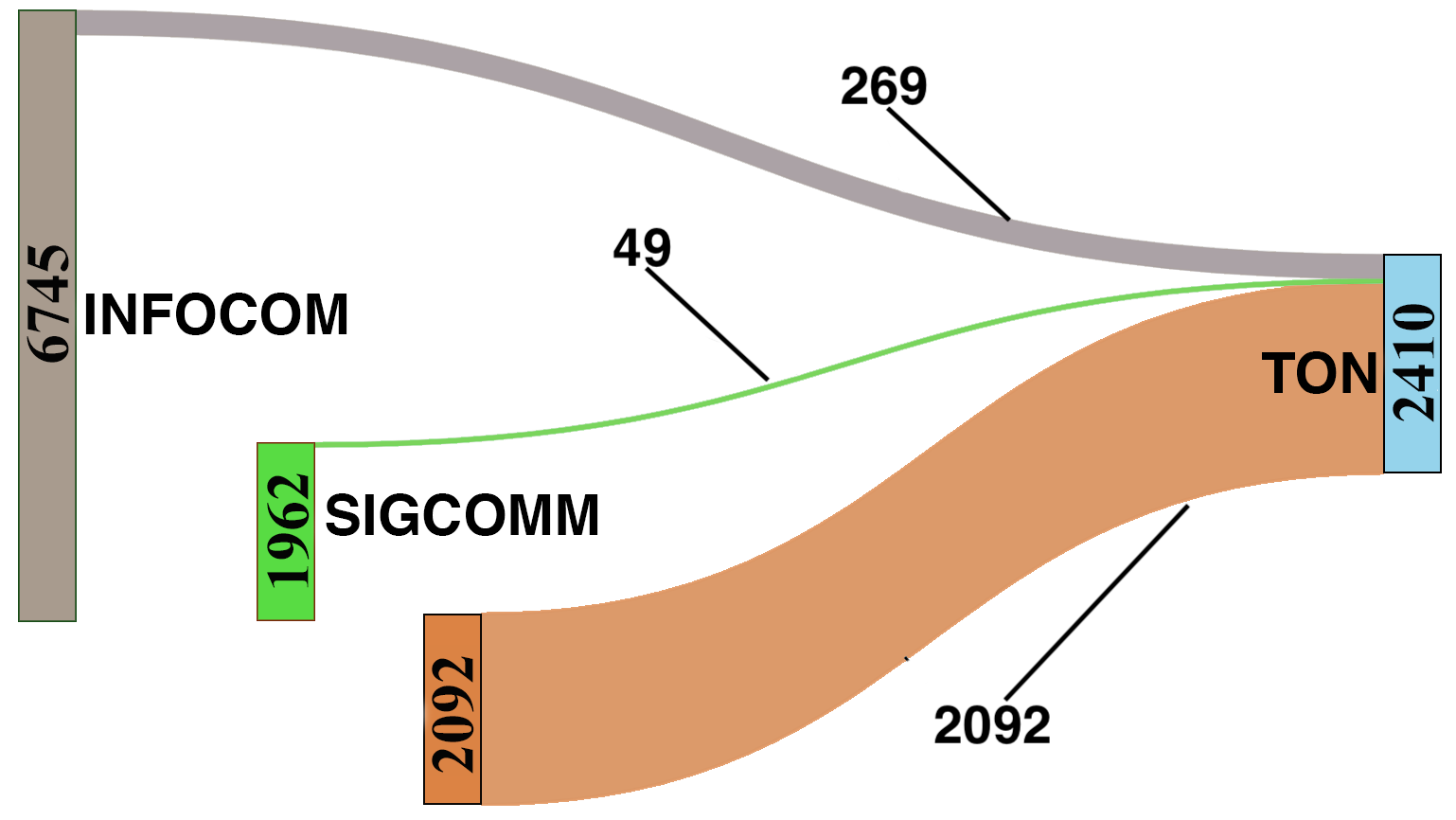}
	\caption{The flow of publications from top conferences to top journals in all of the four venues during 2000--2017. \textit{Interestingly, more extended version of articles from INFOCOM than SIGCOMM are published in TON.}}
	\label{fig:top_flow_authors_c}
\end{figure}

\par TON is one of the most reputed journals in computer networking and many authors extend their work, published in different conferences, to publish in TON. Figure \ref{fig:top_flow_authors_c} shows the number of articles published in TON whose prequel work is published in either INFOCOM and SIGCOMM. We found out that 269 out of 2410 (~10\%)  articles of TON have their prequel work published in INFOCOM. Similarly, 69 out of 2410 articles of TON are the sequel of the work published in SIGCOMM. There is no overlap between SIGCOMM and INFOCOM. Similarly, COMST has no intersection with any of the other venues.

\par
\begin{figure}[H]
	\centering
	\subfloat[COMST]{\includegraphics[width=0.45\textwidth]{COMST_authors_years1.eps}}
	\subfloat[TON]{\includegraphics[width=0.45\textwidth]{authors_years_TON1.eps}}\\
	\subfloat[SIGCOMM]{\includegraphics[width=0.45\textwidth]{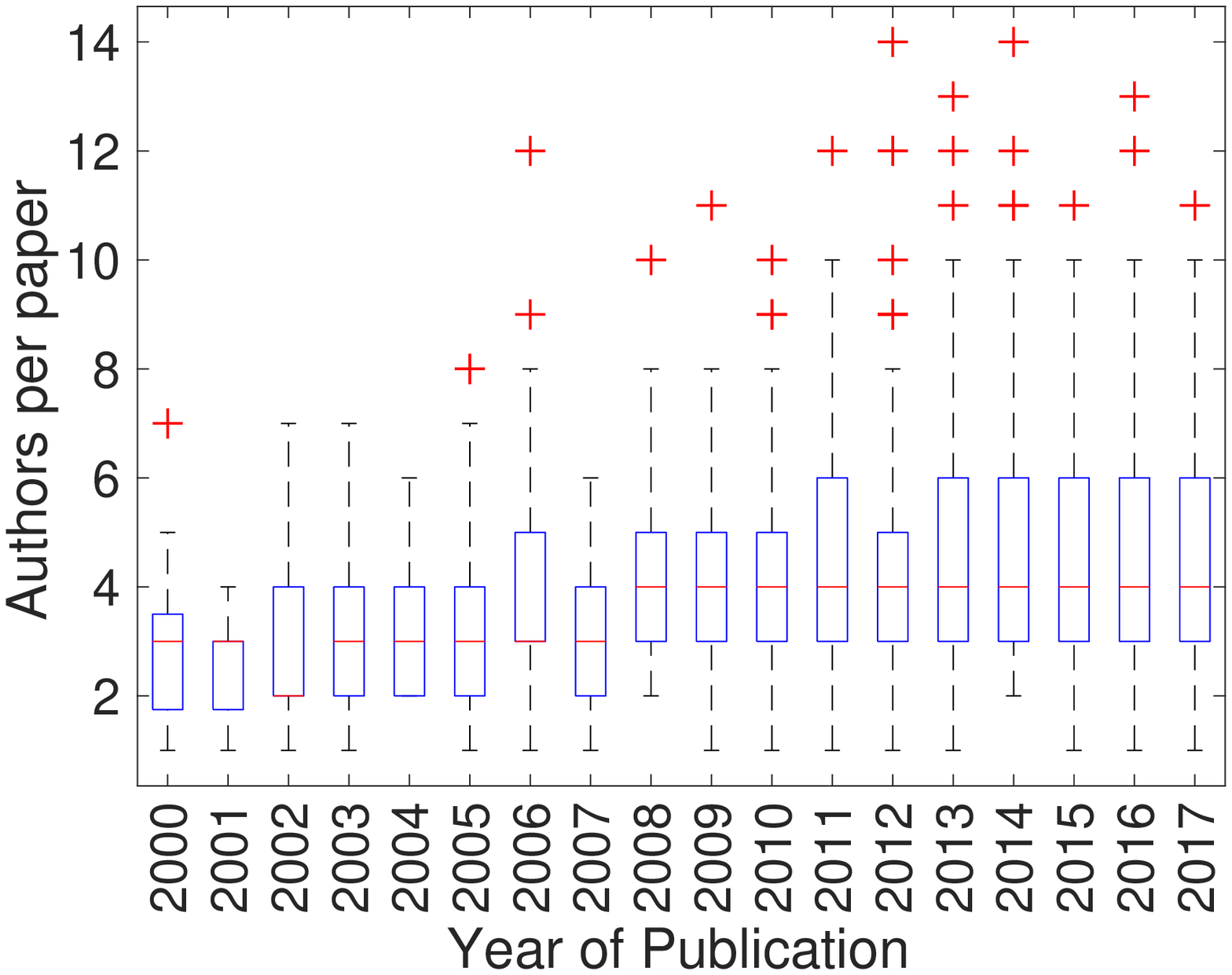}}
	\subfloat[INFOCOM]{\includegraphics[width=0.45\textwidth]{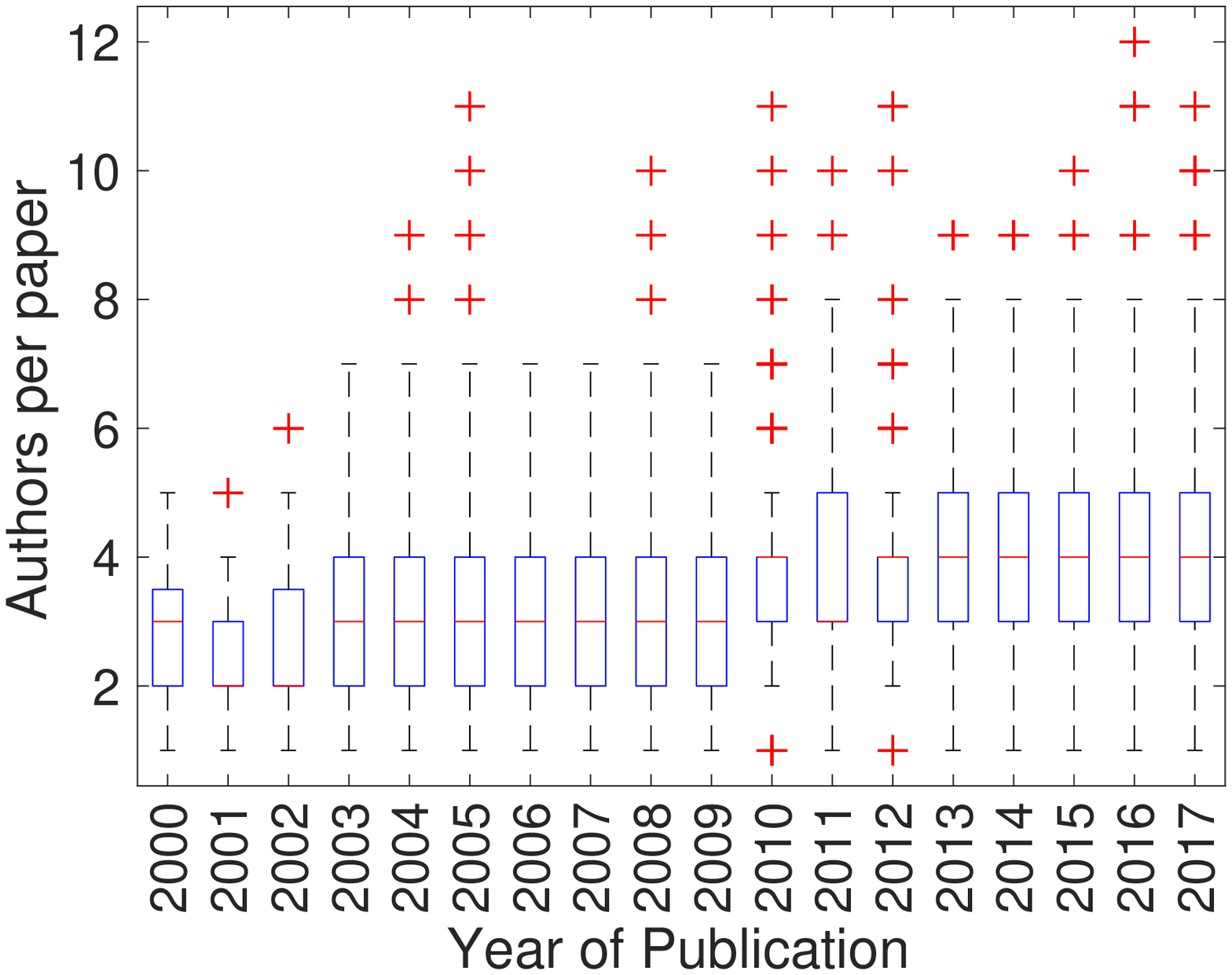}}
	\caption{Distribution of the number of authors per article throughout 2000--2017. \textit{Tendency for co-authorship is increasing over time in both all of the four venues (COMST, TON, SIGCOMM, and INFOCOM) due to enhancing collaboration between institutes and authors.}}
	\label{fig:top_authors_years_c}
\end{figure}

 We have explored the changing trends in co-authorship in SIGCOMM and INFOCOM over the period 2000 to 2017 and compared them with discussed journals. We explore how the distribution of collaborating authors changes over time. 
Figure \ref{fig:top_authors_years_c} shows the distribution of the number of authors per article in COMST per year.

It is clear that the tendency for co-authorship is increasing; in 2000 the median number of authors is 2 for COMST and 3 for TON, compared to 4 and 4 in 2017. Perhaps most noteworthy is the spread of authorship numbers across articles, with a standard deviation of 0.87 in 2000 vs.\ 1.69 in 2017 for TON (similar trends of COMST). Similarly, in SIGCOMM and INFOCOM, the tendency for co-authorship is increasing by the passage of time; in 2000 the median number of authors is 3 for SIGCOMM and 3 for INFOCOM, compared to 4 and 4 in 2017. Again, one of the most worth observing trends is the spread of authorship across time duration with a standard deviation of 1.94 in 2000 vs.\ 2.52 in 2017 for SIGCOMM (standard deviation of 0.95 in 2000 vs.\ 1.65 in 2017 for INFOCOM). One more surprising fact is the comparison between the spread of authorship of journals and conferences. Top conferences in computer networking show the higher spread of authorship across the years as compared to journals.
\par
\begin{figure}[H]
	\centering
	\includegraphics[width=0.7\textwidth]{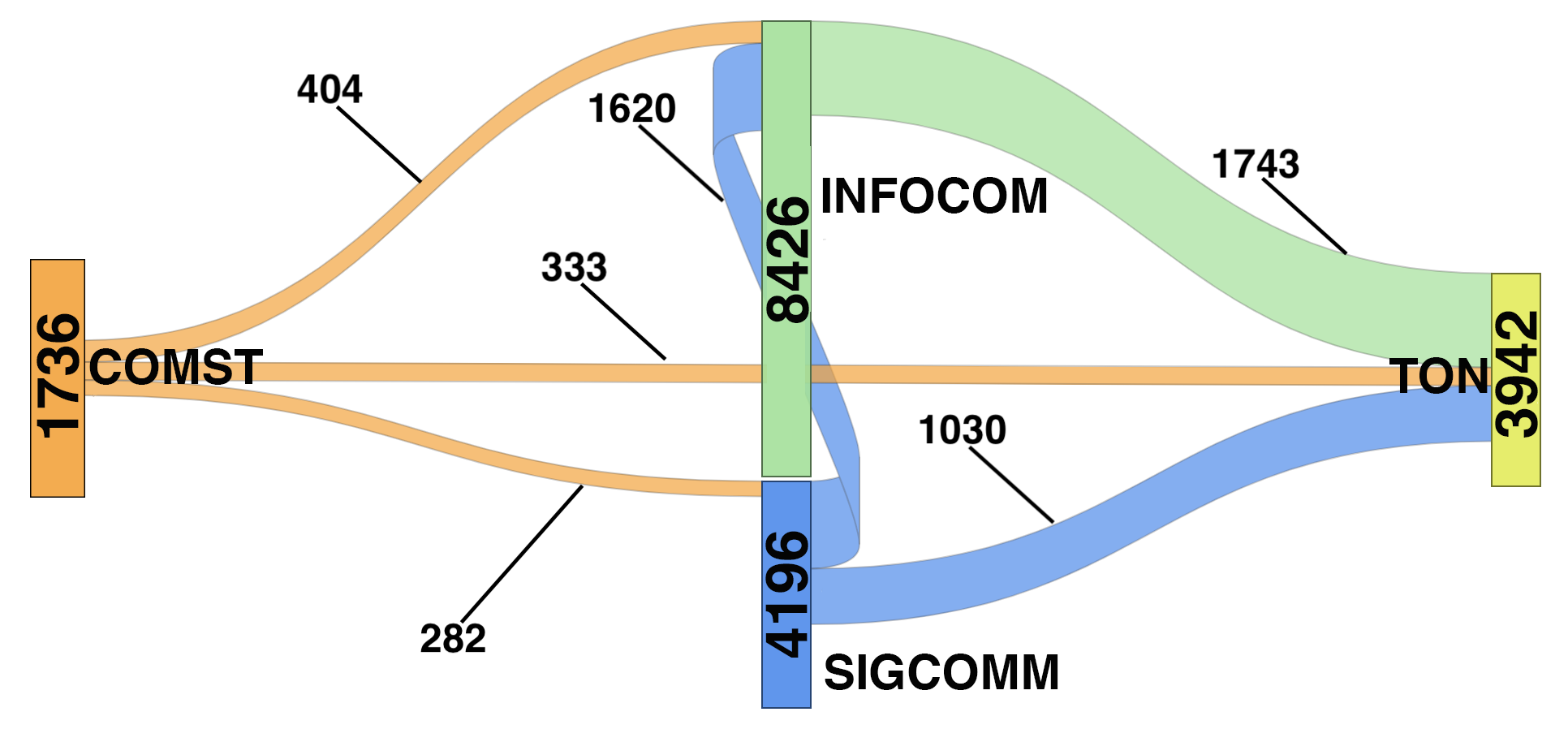}
	\caption{The flow of authors in all of the four venues during 2000--2017. Flows of authors, shown in the figure, are undirected. \textit{Interestingly, a large number of authors are publishing in all genres of venues.}}
	\label{fig:top_flow_com_authors_c}
\end{figure}

 Each venue in every domain has a handful of common authors and this trend is also similar in computer science. Figure \ref{fig:top_flow_com_authors_c} shows the number common authors among all of the venues during 2000-2017. From results present in this figure, we can observe that SIGCOMM and TON have the highest percentage of common authors among all of the venues.

\subsection{Country Based Productivity Analysis}

\begin{figure}[!h]
	\centering
	\subfloat[SIGCOMM]{\includegraphics[width=0.5\textwidth]{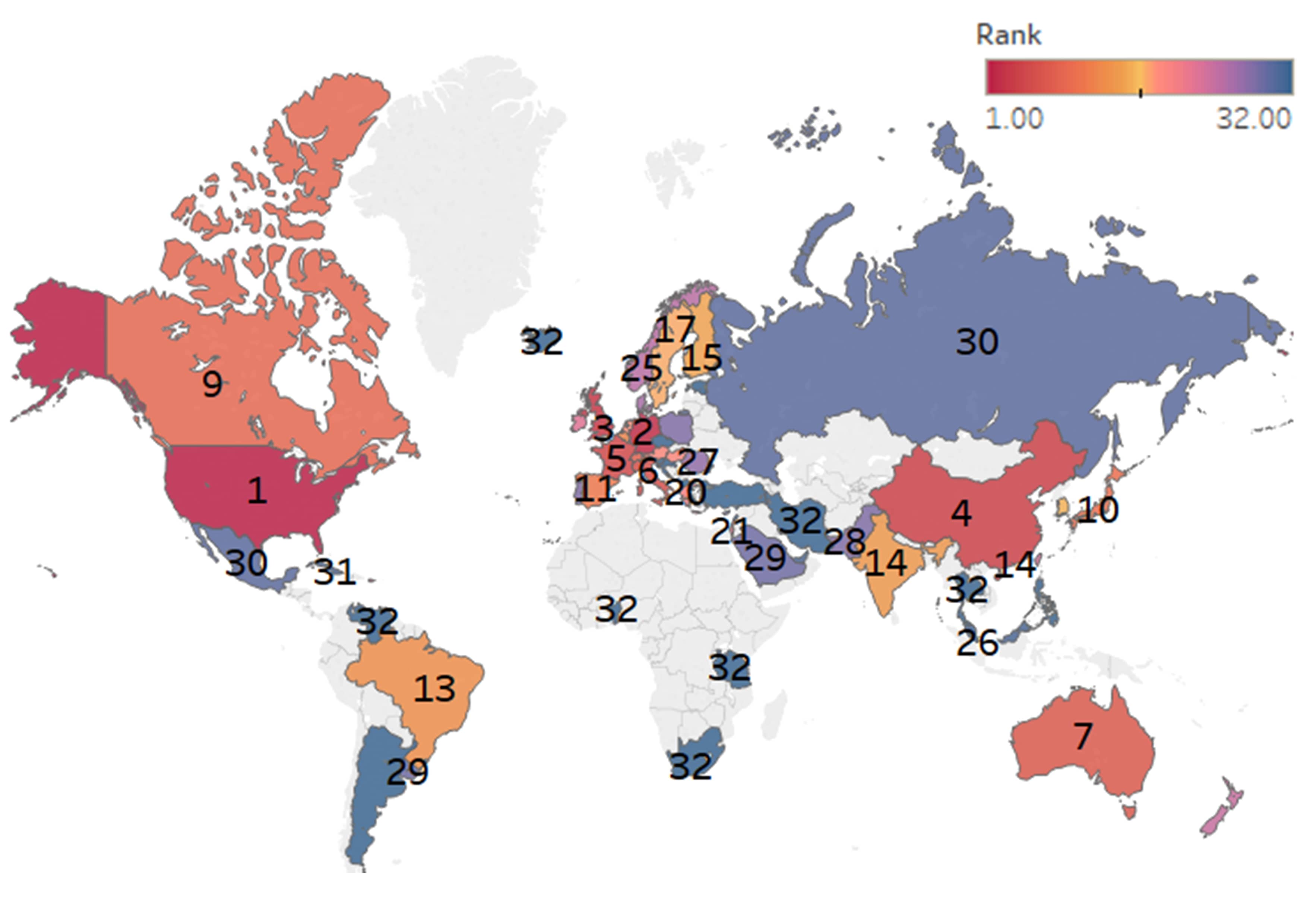}}
	\subfloat[INFOCOM]{\includegraphics[width=0.5\textwidth]{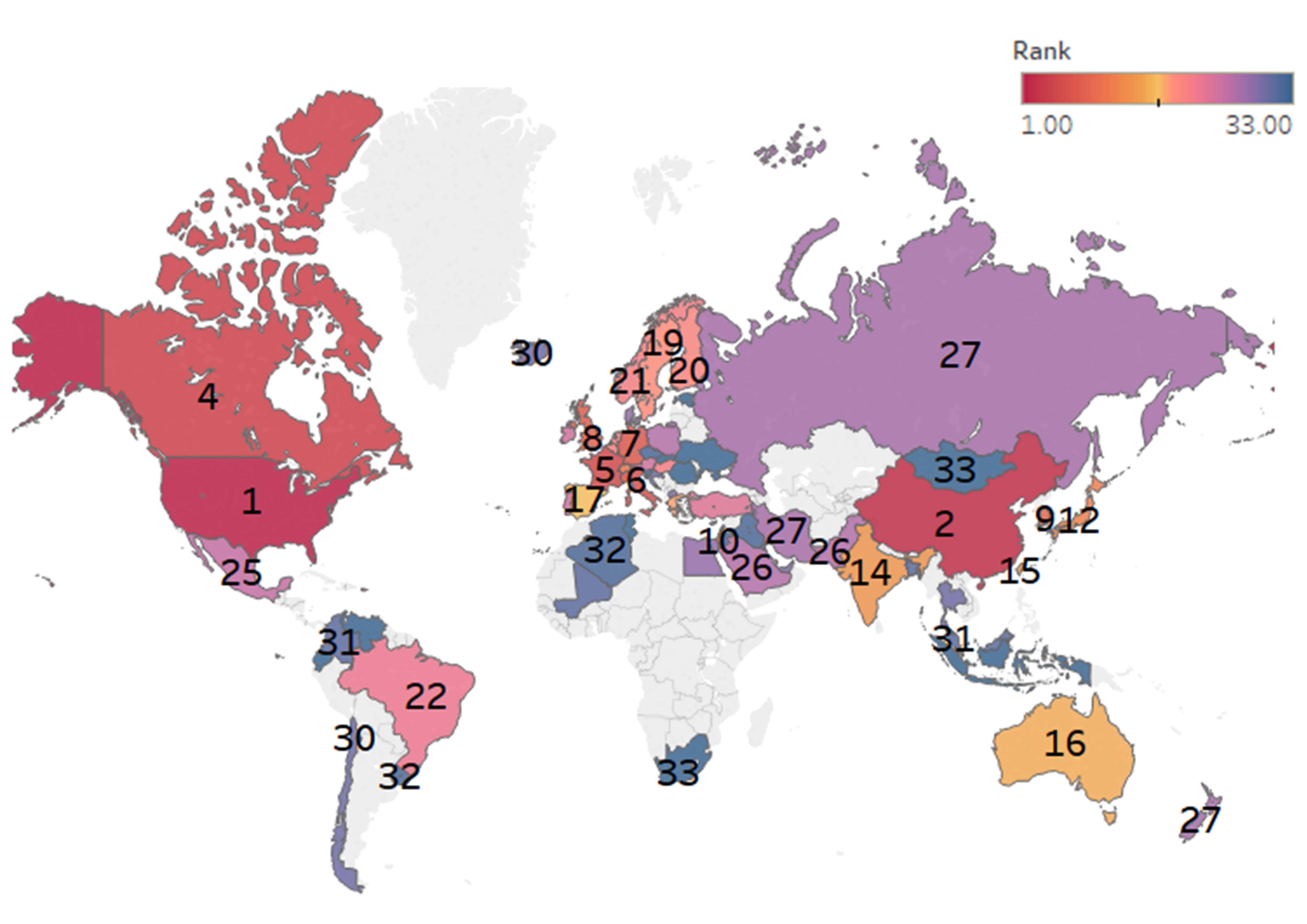}}
	\caption{Rank of different countries in SIGCOMM and INFOCOM based on publication count. \textit{Although most countries have similar productivity in these two journals, there are notable exceptions where the publication trends are quite dissimilar.}}
	\label{fig:country_top_rank_c}
\end{figure}

In a research domain, some countries play a pivotal role in driving the ongoing advancements in that field. 
Figure \ref{fig:country_top_rank_c} shows the rank of a contributing country in SIGCOMM and INFOCOM using a global heat map. Similar to COMST and TON, the United States is in the highest position in SIGCOMM and INFOCOM in terms of publication count. Other top countries include Canada, China, France, and the United Kingdom in SIGCOMM. In INFOCOM, top countries remained the same but Hong Kong replaced the United Kingdom in the list of top countries. There is also a noticeable change in rank of China in SIGCOMM and INFOCOM. In INFOCOM, China is second ranked, but loses its position in SIGCOMM and moves to the forth rank. Similar trends are observed in COMST and TON as well, which are shown in Figure \ref{fig:country_top_rank}.

\begin{figure}[!h]
	\centering
	\subfloat[SIGCOMM]{\includegraphics[width=0.5\textwidth]{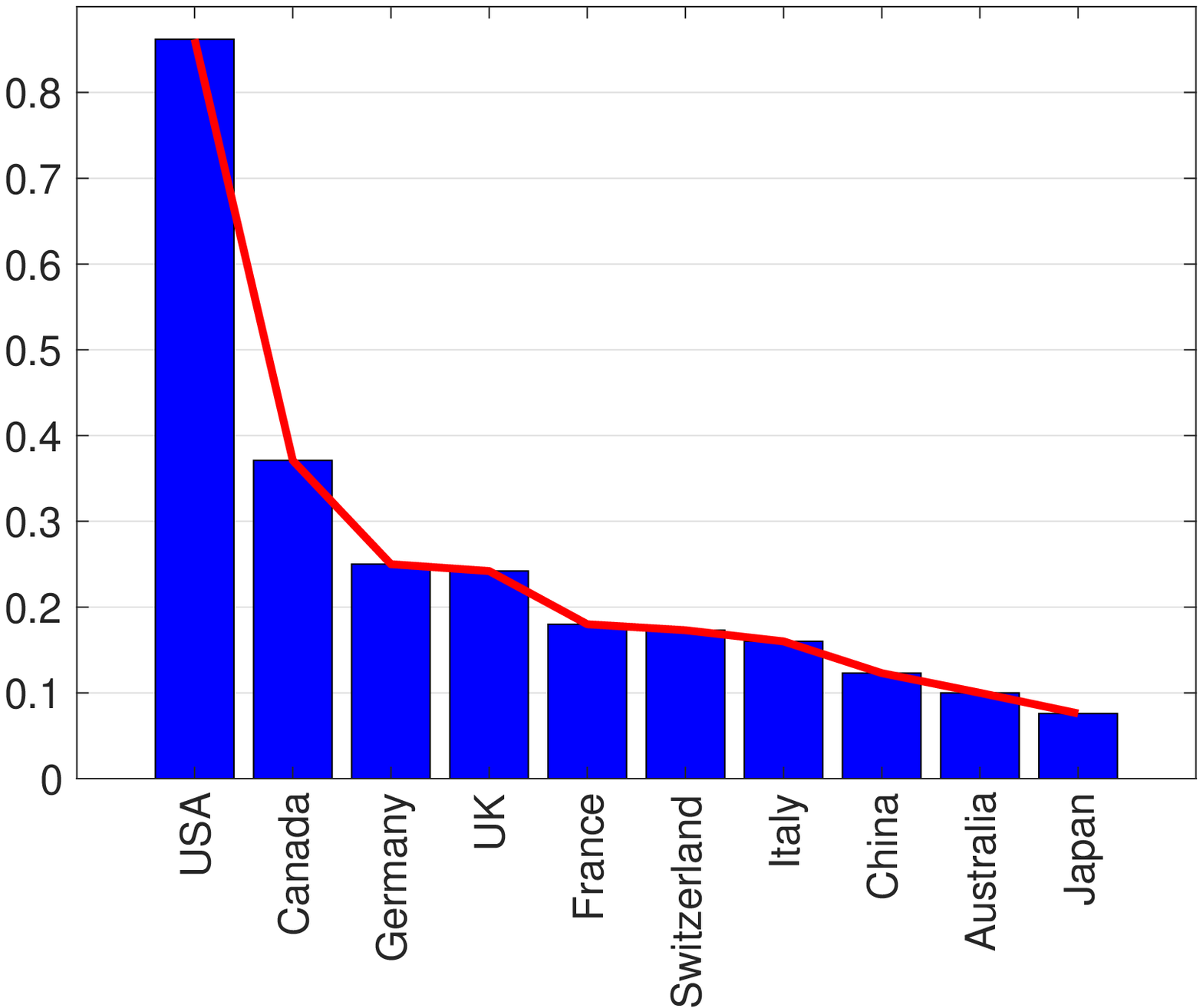}}
	\subfloat[INFOCOM]{\includegraphics[width=0.5\textwidth]{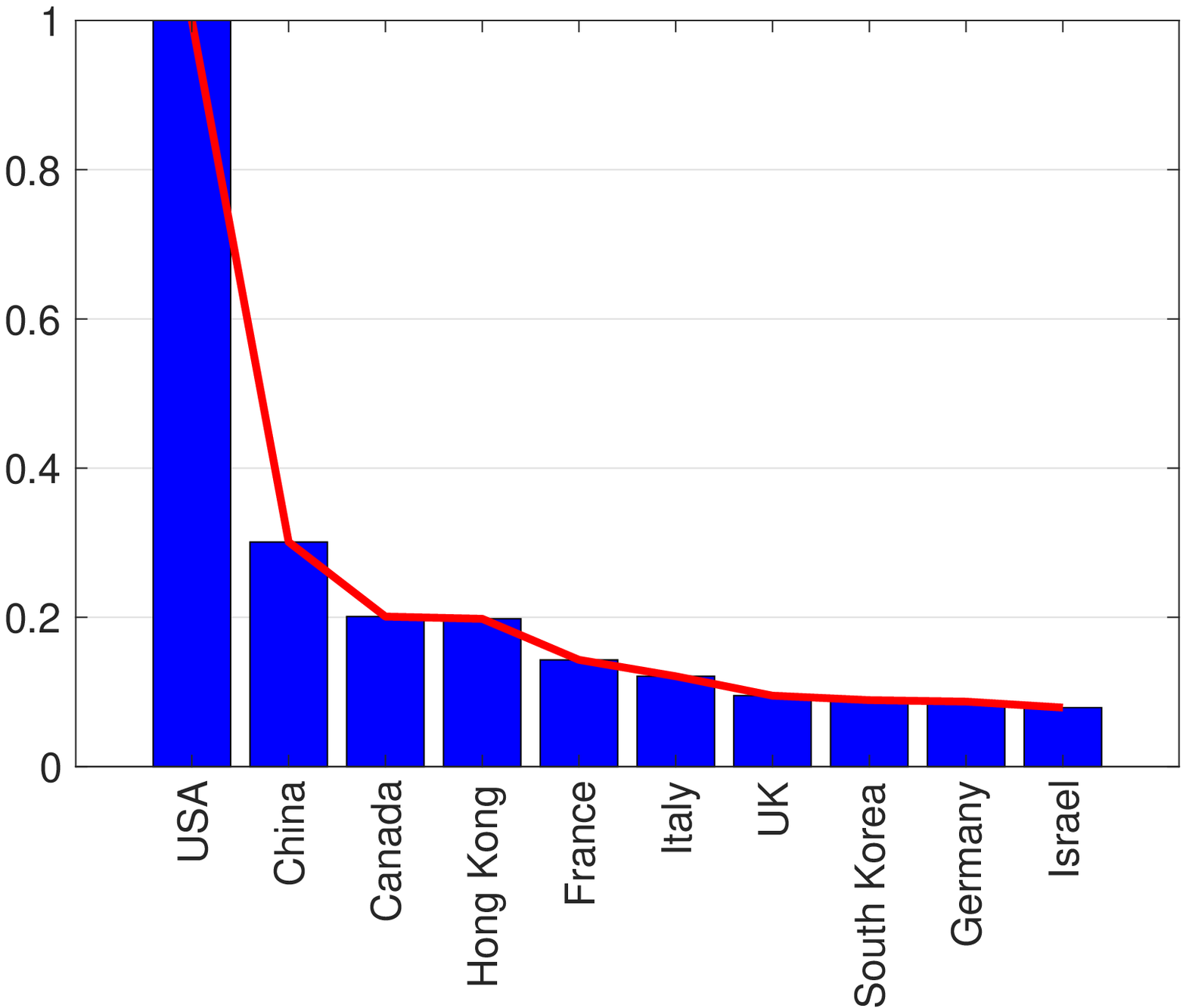}}
	\caption{Rank of countries in SIGCOMM and INFOCOM based on their publication count, citation count, and h-index. \textit{Country with the highest score in each venue is used as reference for calculation. USA emerged as the top country in all of three venues, with the gap being most prominent in INFOCOM.}}
	\label{fig:ranks_t_c}
\end{figure}
We calculated ranking scores of top countries in SIGCOMM and INFOCOM using equation \ref{NRS_equation}. Figure \ref{fig:ranks_t_c} shows the ranking of different countries in SIGCOMM and INFOCOM where it can be seen that the USA has the maximum ranking score in both the venues. Both the venues are dominated by more or less the same countries with some exceptions---e.g., Hong Kong is among the top-ranked countries publishing in INFOCOM but it is not a prominent contributor to SIGCOMM. Similar results are observed for COMST and TON in Figure \ref{fig:ranks_t}.
\subsection{Citation Based Analysis of Authors}
\par
\begin{figure}[!h]
	\centering
	\subfloat[Journals]{\includegraphics[width=0.5\textwidth]{author_cited1.png}}
	\subfloat[Conferences]{\includegraphics[width=0.5\textwidth]{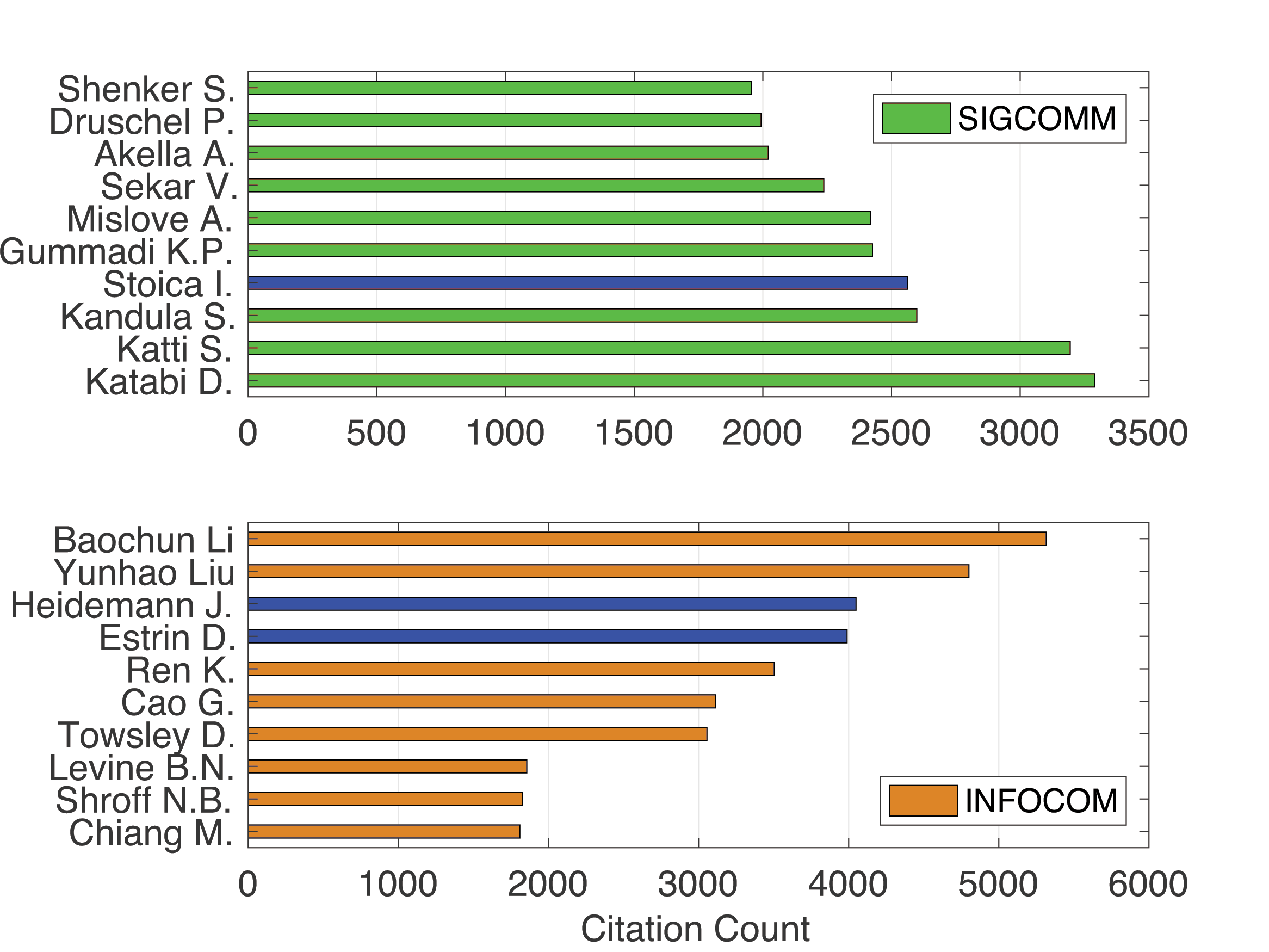}}
	\caption{most-cited authors during 2000--2017, according to citation count. (a) Journals; (b) Conferences. Same color bars represent the overlapping authors among different venues. \textit{Interestingly, there is an overlap between the top cited authors of TON and INFOCOM which shows the common authors with most highly usable research in both venues.}}
	\label{fig:top_cite_authors_c}
\end{figure}

 Citations of an author is a good parameter to analyze the impact and usability of research done by that researcher. It is worth doing the analysis of top-cited authors in all of these four venues. Figure \ref{fig:top_cite_authors_c} shows the most-cited authors in all four venues. Interestingly, there is an overlap between the top cited authors of TON and INFOCOM which shows the common authors with most highly usable research in both venues. It is also worth noting that the most-published authors in all venues are not the ones with highly usable and cited research except a few exceptions.
\begin{figure}[!h]
	\centering
	\subfloat[Journals]{\includegraphics[width=0.5\textwidth]{h-index1.png}}
	\subfloat[Conferences]{\includegraphics[width=0.5\textwidth]{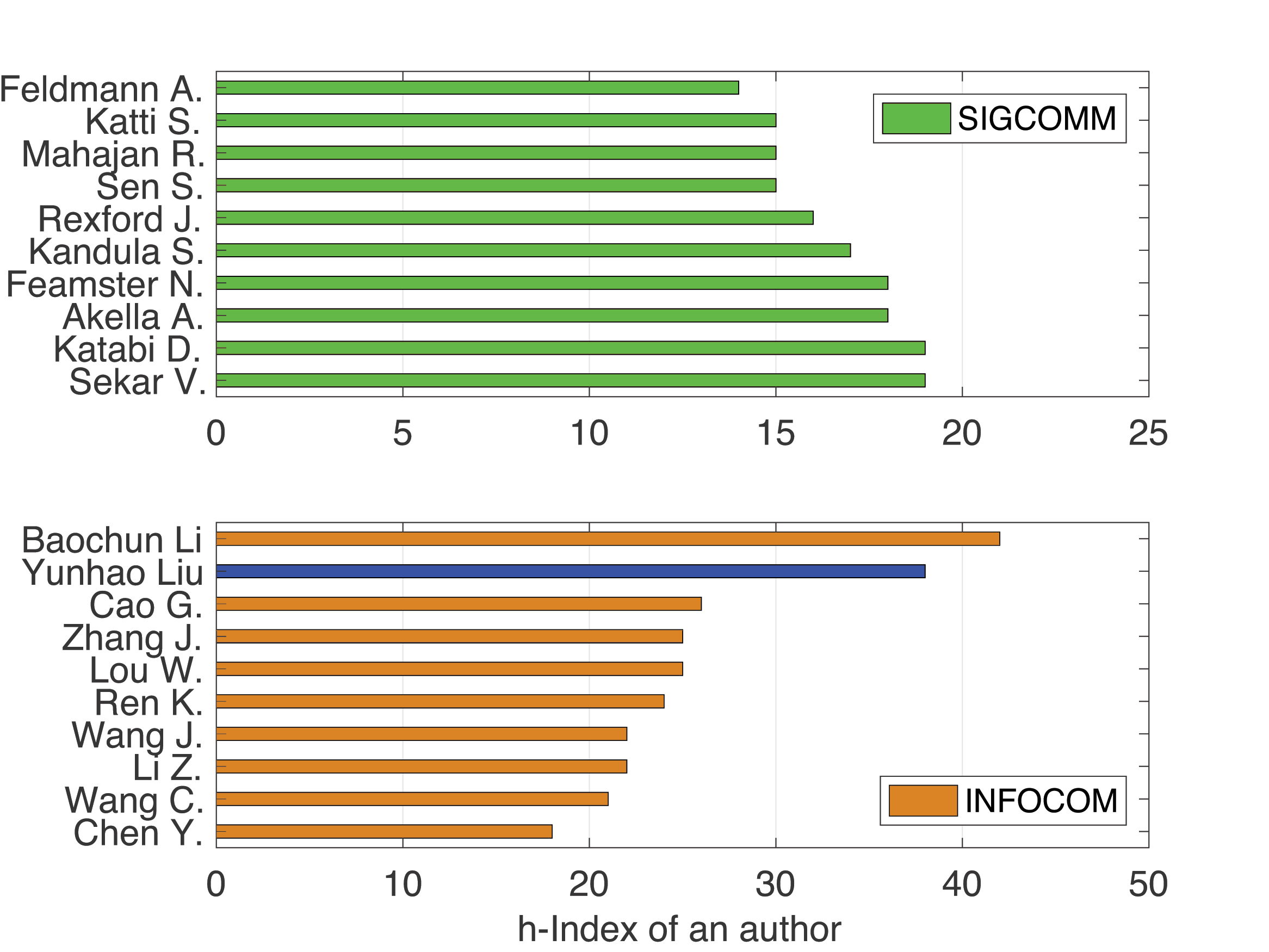}}
	\caption{Ten authors with the highest h-index during 2000--2017, (a) Journals; (b) Conferences. Same color bars represent the overlapping authors among different venues. \textit{Interestingly in journals (COMST and TON), the top 10 most-published list and the top 10 authors with the highest h-index are almost identical but in conferences (SIGCOMM and INFOCOM), the trend is not true as the most-published authors and the authors with the highest h-index are not same.}}
	\label{fig:top_hindex_authors_c}
\end{figure}

\par h-Index is one of the good bibliographic metrics to analyze the research activeness through usable research of an author. Figure \ref{fig:top_hindex_authors_c} shows the ten authors with the highest h-index. Data from all of these four venues shows some interesting results. Surprisingly, in journals (COMST and TON), the top 10 most-published list and the top 10 authors with the highest h-index are almost identical but in conferences (SIGCOMM and INFOCOM), this trend is not true as the most-published authors and authors with the highest h-index are not same. For top conferences, this data shows that the authors with top publication count are not the ones with a balanced contribution of publication count and citation count.  
\begin{figure}[!h]
	\centering
	\subfloat[COMST]{\includegraphics[width=0.5\textwidth]{box_plot_citation_comst.eps}}
	\subfloat[TON]{\includegraphics[width=0.5\textwidth]{box_plot_citation_ton.eps}}\\
	\subfloat[SIGCOMM]{\includegraphics[width=0.5\textwidth]{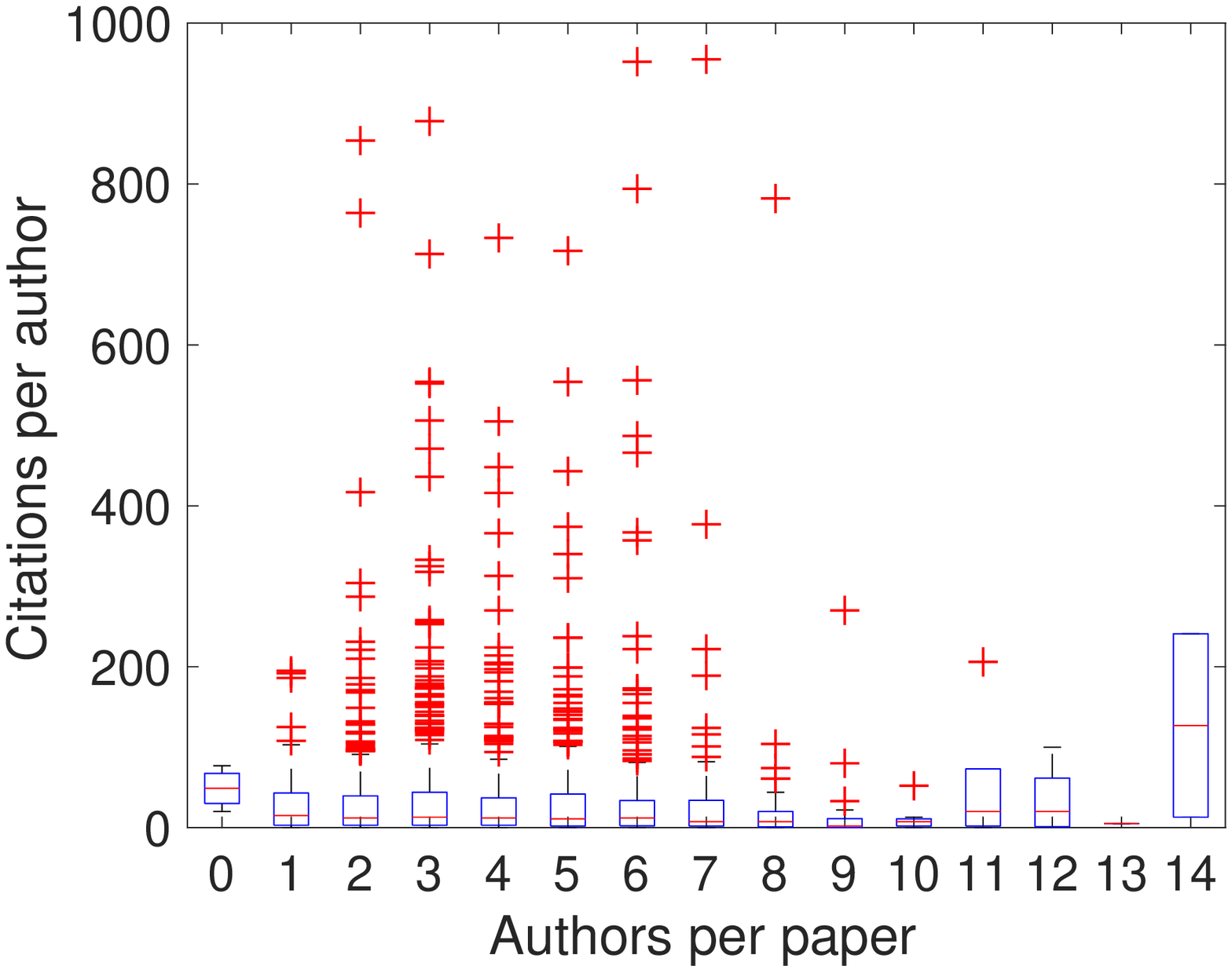}}
	\subfloat[INFOCOM]{\includegraphics[width=0.5\textwidth]{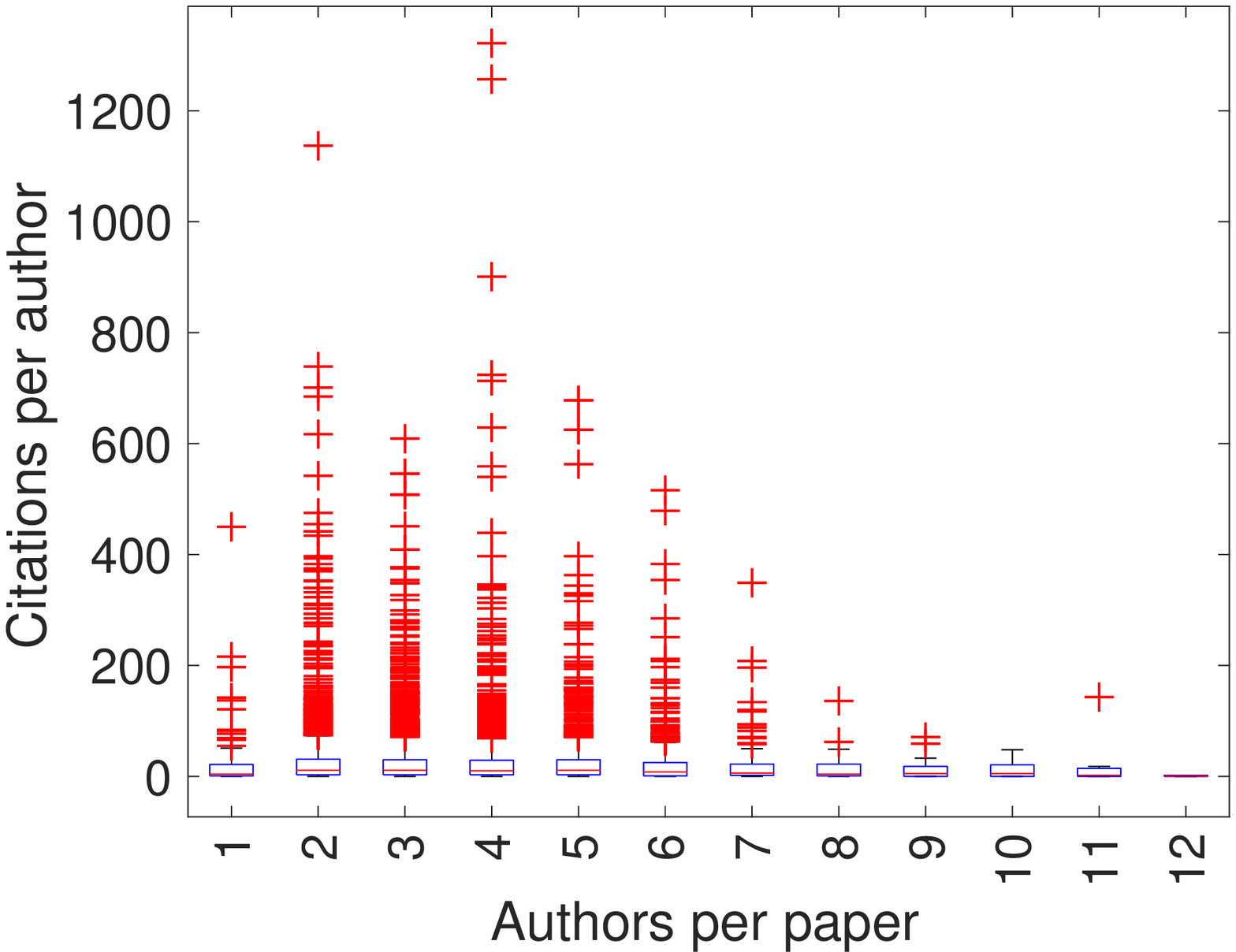}}
	\caption{The number of citations per article in all of the four venues with respect to the number of authors. \textit{The most-cited articles typically have a moderate number of authors.}}
	\label{fig:top_authors_years_cite_c}
\end{figure}

\par Figure \ref{fig:top_authors_years_cite_c} presents the citation counts for each journal and conference based on how many authors are on the article. We see that TON articles tend to have higher citation counts than survey-based articles when we consider the top-cited articles but on average COMST articles are cited more (see Table I, in which it is shown that COMST has on average 67 citations compared to 37 for TON). The higher citations of COMST articles on average likely stems from their citations in many topic-specific articles as a general resource. Similarly, INFOCOM articles tend to have higher citation count across the time duration.
\newpage

\subsection{Keyword Based Analysis}
\par
\begin{figure}[!h]
	\centering
	\subfloat{\includegraphics[width=.8\textwidth]{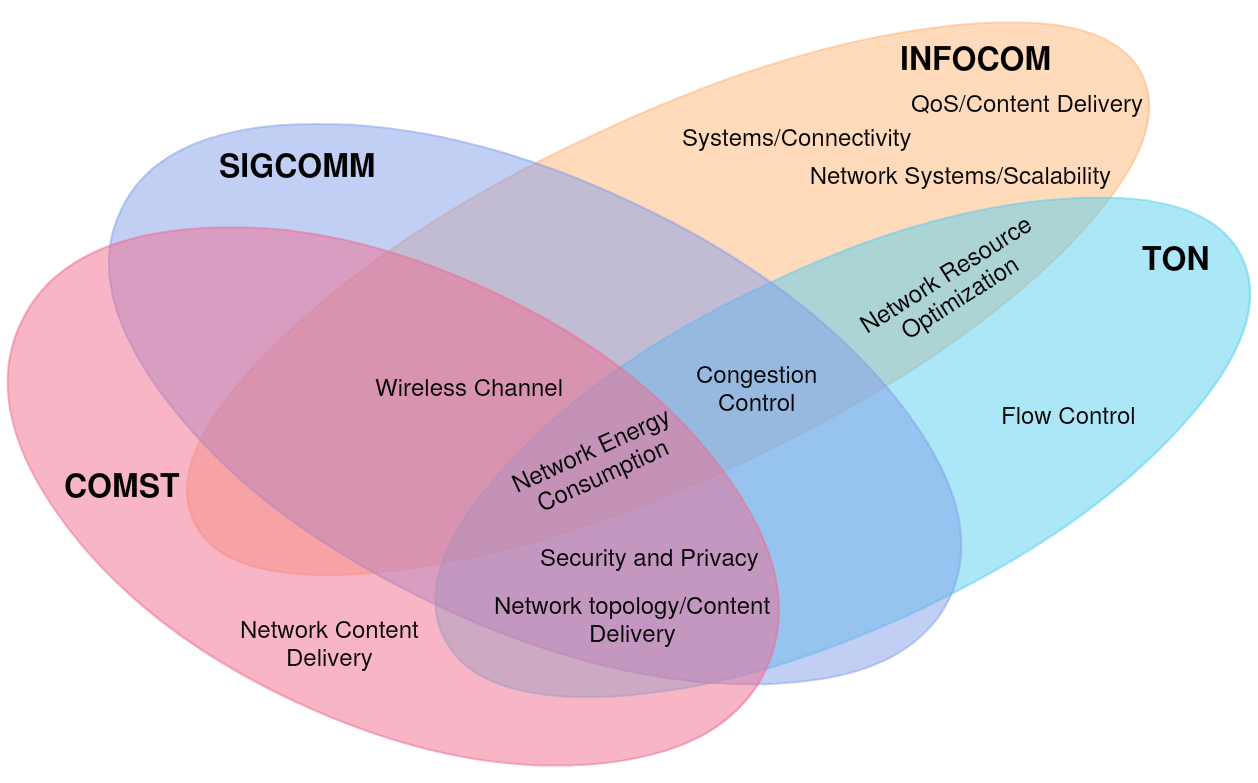}}
	\caption{Distribution of top categories discussed in all of the four venues.\textit{These categories are derived from latent topics extracted from all of the four venues.}}
	\label{fig:keyword_top_tm_c}
\end{figure}

 Investigating popular topics is considered to be one of the best ways of studying the paradigm shifts in any research field. It is helpful in describing the research trends of a field. In this section, we investigate such paradigm shift in journals and conferences and analyze the overlapping between those two genres.
To perform keyword-based analysis, we use Latent Dirichlet Allocation (LDA). LDA takes raw text, the number of topics and a dictionary of words as the input, and outputs the most significant topics with words from the raw data. We kept the number of latent output topics to 10 and iterated our algorithms 400 times on our dataset in order to achieve converged results. Furthermore, we categorized the top latent topics extracted from all datasets into 11 main categories. Top topics in all of these four venues are discussed mainly from these categories. Figure \ref{fig:keyword_top_tm_c} shows the overlap between these categories in all of the four venues. Table \ref{keyword_LDA_c} shows the results of LDA on the COMST, TON, SIGCOMM and INFOCOM datasets.

\newpage
\begin{table}[H]
	\centering
	\caption{Using LDA-based topic modeling to determine 10 most popular topics in all four venues. \textit{We see coherent results using LDA-based topic modeling as it reveals the topics hidden in actual text.}}
	\label{keyword_LDA_c}
	\begin{adjustbox}{width=1\textwidth}
		\begin{tabular}{|l|l|}
			\hline
			\textbf{Category}                     & \textbf{Latent Topic}                                                                                                                                                                                                                                                                                                                                                                                                                                                                                                                        \\ \hline
			System/Connectivity                 & \begin{tabular}[c]{@{}l@{}} {[sensor, deploy system, sense, coverage, tag, propose, local, detect, use]} (INFOCOM) \\{[route, path, network, traffic, link, forward, use, protocol, propose, failure]} (INFOCOM)                                                                                                                                                                                                                                                                                                                                                    \end{tabular} \\ \hline
			Security and Privacy                     & \begin{tabular}[c]{@{}l@{}} {[attack, detect, social, privacy, threat, anonymous, vulnerable, trust, category, protect]} (COMST) \\{[detect, attack, estimate, accuracy, identify, filter, memory, trace, aggregate, acute]} (TON)\\{[attack, traffic, network, detect, flow, anomaly, defense, data, sample, system]} (SIGCOMM)                                                                                                                                                                                                                                      \end{tabular} \\ \hline
			Network resource optimization                 & \begin{tabular}[c]{@{}l@{}} {[algorithm, problem, optimization, schedule, achieve, policy, solution, distribution, wireless, propose]} (TON) \\{[algorithm, problem, optimize, schedule, policy, delay, perform, rate, achieve, bound]} (TON)\\{[control, allocate, network, user, resource, provide, service, fair, bandwidth, algorithm]} (TON)\\{[schedule, delay, policy, algorithm, time, optimal, perform, system, bound, queue]} (INFOCOM)                                                                                                                       \end{tabular} \\ \hline
			Wireless Channel                & \begin{tabular}[c]{@{}l@{}} {[spectrum, optics, radio, cognition, band, sensor, model, cellular, fiber, availability]} (COMST)\\{[system, technique, communication, channel, wireless, transmission, perform, design, code, signal]} (COMST)\\{[wireless, channel, transmission, network, protocol, code, scheme, throughput, receive, rate]} (INFOCOM) \\{[spectrum, user, game, channel, sensor, cooperate, secondary, primary, radio, cognitive]} (INFOCOM) \\{[wireless, use, communication, channel, device, throughput, radio, receiver, design, signal]} (SIGCOMM) \end{tabular} \\ \hline
			Congestion Control           & \begin{tabular}[c]{@{}l@{}} {[queue, congestion, fair, buffer, stabilize, class, loss, converge, arrive, parameter]} (TON)\\{[flow, packet, rate, network, traffic, control, congestion, loss, fair, switch]} (INFOCOM) \\{[flow, traffic, control, congestion, network, packet, provide, perform, application, user]} (SIGCOMM)                                                                                                                                                                                                                                      \end{tabular} \\ \hline
			Network Content Delivery  & \begin{tabular}[c]{@{}l@{}} {[mobile, scheme, multimedia, content, access, satellite, delivery, solution, device, difference]} (COMST)                                                                                                                                                                                                                                                                                                                                                                                                                            \end{tabular} \\ \hline
			System Energy Consumption          & \begin{tabular}[c]{@{}l@{}} {[smart, data, grid, energy, power, center, secure, manage, consumption, trust]} (COMST)\\{[switch, energy, power, consumption, spectrum, synchronize, architecture, device, cell, input]} (TON)\\{[energy, device, power, communication, mobile, consumption, propose, system, paper, smart]} (INFOCOM)\\ {[protocol, wireless, design, sensor, node, control, propose, route, optimize algorithm]} (COMST)                                                                                                                                \end{tabular} \\ \hline
			Network topology/Content Delivery             & \begin{tabular}[c]{@{}l@{}} {[node, sensor, energy, data, wireless, distributed, protocol, attack, transmission, power]} (TON) \\{[video, sensor, multicast, data, local, content, wireless, application, multimedia, stream]} (COMST) \\{[network, protocol, route, application, node, survey, control, propose, design, sensor]} (COMST) \\{[route, path, network, topology, use, node, protocol, show, packet, router]} (SIGCOMM)                                                                                                                                    \end{tabular} \\ \hline
			Flow Control                 & \begin{tabular}[c]{@{}l@{}} {[schedule, delay, throughput, packet, queue, rate, policy, bound, buffer, scheme]} (TON)                                                                                                                                                                                                                                                                                                                                                                                                                                             \end{tabular} \\ \hline
			Network Systems/Scalability                 & \begin{tabular}[c]{@{}l@{}} {[content, cache, data, storage, file, request, distributed, scalability, system, server]} (INFOCOM)                                                                                                                                                                                                                                                                                                                                                                                                                                  \end{tabular} \\ \hline
			QoS/Content Delivery                          & \begin{tabular}[c]{@{}l@{}} {[user, video, network, stream, service, peer, system, social, qualities, provide]} (INFOCOM)                                                              \end{tabular} \\ \hline
		\end{tabular}
	\end{adjustbox}
\end{table}

\section{Future Directions}
\label{sec:future_directions}
Our study provides a methodology and framework for performing a comprehensive bibliometric analysis on computer networking research and the public release of a comprehensive dataset. Future research of this study can be extended in several directions, some of which we highlight below:
\begin{itemize}
    \item This work can be followed up with a more comprehensive analysis on a larger set of related journals and conferences in the field of computer networking;
    \item Future researchers can also explore using  data from, and integrating with, popular conference management systems (EDAS, HotCRP, EasyChair, etc.)
    \item This study can be extended by work that finds correlation of publications in computer networking literature with the priorities defined by major global research funding agencies;
    \item A comparison of computer networking with other fields (e.g. machine learning, artificial intelligence, network science) can be performed and differences in publication trends (such as citations, h-index) can be identified.
\end{itemize}
\section{Conclusions}
\label{sec:conclusion}

In this paper, we have performed an in-depth bibliometric study of the publication trends in computer networking literature using article content and metadata of four important computer networking periodicals---IEEE Communications Surveys and Tutorials (COMST), IEEE/ACM Transactions on Networking (TON), ACM Special Interest Group on Data Communications (SIGCOMM), and IEEE International Conference on Computer Communications (INFOCOM)---gathered over the time period 2000--2017.  Our work extends the state of the art in bibliometric analysis of computer networking literature by presented comprehensive analyses that shed light on the publication patterns in these journals including which kinds of articles are published where; how are journal and conference publications different in this area; and which different authors, institutes, and countries have been successful in these venues (and how).  Although we cannot make strong claims about causality or the parameters responsible for the acceptance/rejection of an article since we did not have access to missing data (rejected articles), we believe that our analyses provide an insightful look into the publication culture in the networking community and can help develop a more nuanced understanding of this research field especially in the light of the limited existing bibliometric work that focused on the computer networking community. In this regard, we have also publicly shared our dataset that includes content, metadata, and citation-related information related to the articles published from 2000 to 2017 in COMST, TON, SIGCOMM, and INFOCOM as our contribution to the research community. \footnote{https://github.com/waleediqbal411/Scientometrics-paper-data2019}

 %We also compared the results from COMST and TON with SIGCOMM and INFOCOM (another dataset gathered for duration 2000--2017) and shed light over the common and contrast results among these top journals and top conferences. We found out that in COMST and TON, most published authors are the ones with the highest h-index but this trend is not true in case of SIGCOMM and INFOCOM. SIGCOMM and INFOCOM also have a higher spread of authorship as compared to TON and COMST throughout the studied time period. 

%\input{methodology.tex}

\bibliographystyle{spbasic_updated}      % basic style, author-year citations
\bibliography{sample_library}   % name your BibTeX data base

\begin{thebibliography}{35}
\providecommand{\natexlab}[1]{#1}
\providecommand{\url}[1]{{#1}}
\providecommand{\urlprefix}{URL }
\expandafter\ifx\csname urlstyle\endcsname\relax
  \providecommand{\doi}[1]{DOI~\discretionary{}{}{}#1}\else
  \providecommand{\doi}{DOI~\discretionary{}{}{}\begingroup
  \urlstyle{rm}\Url}\fi
\providecommand{\eprint}[2][]{\url{#2}}

\bibitem[{Bartneck and Hu(2009)}]{bartneck2009scientometric}
Bartneck C, Hu J (2009) Scientometric analysis of the {CHI} proceedings. In:
  Proceedings of the SIGCHI conference on human factors in computing systems,
  ACM, pp 699--708

\bibitem[{Blei et~al.(2003)Blei, Ng, and Jordan}]{blei2003latent}
Blei DM, Ng AY, Jordan MI (2003) Latent {D}irichlet allocation. Journal of
  machine Learning research 3(Jan):993--1022

\bibitem[{Blondel et~al.(2008)Blondel, Guillaume, Lambiotte, and
  Lefebvre}]{blondel2008fast}
Blondel VD, Guillaume JL, Lambiotte R, Lefebvre E (2008) Fast unfolding of
  communities in large networks. Journal of statistical mechanics: theory and
  experiment 2008(10):P10,008

\bibitem[{Borgatti et~al.(2009)Borgatti, Mehra, Brass, and
  Labianca}]{borgatti2009network}
Borgatti SP, Mehra A, Brass DJ, Labianca G (2009) Network analysis in the
  social sciences. science 323(5916):892--895

\bibitem[{Chiu and Fu(2010)}]{chiu2010publish}
Chiu DM, Fu TZ (2010) Publish or perish in the internet age: a study of
  publication statistics in computer networking research. ACM SIGCOMM Computer
  Communication Review 40(1):34--43

\bibitem[{Choi et~al.(2011)Choi, Yi, and Lee}]{choi2011analysis}
Choi J, Yi S, Lee KC (2011) Analysis of keyword networks in mis research and
  implications for predicting knowledge evolution. Information \& Management
  48(8):371--381

\bibitem[{Coccia and Wang(2016)}]{coccia2016evolution}
Coccia M, Wang L (2016) Evolution and convergence of the patterns of
  international scientific collaboration. Proceedings of the National Academy
  of Sciences 113(8):2057--2061

\bibitem[{Coleman and Liau(1975)}]{coleman1975computer}
Coleman M, Liau TL (1975) A computer readability formula designed for machine
  scoring. Journal of Applied Psychology 60(2):283

\bibitem[{Didegah and Thelwall(2018)}]{didegah2018co}
Didegah F, Thelwall M (2018) Co-saved, co-tweeted, and co-cited networks.
  Journal of the Association for Information Science and Technology

\bibitem[{Fernandes and Monteiro(2017)}]{fernandes2017evolution}
Fernandes JM, Monteiro MP (2017) Evolution in the number of authors of computer
  science publications. Scientometrics 110(2):529--539

\bibitem[{Flittner et~al.(2018)Flittner, Mahfoudi, Saucez, W{\"a}hlisch,
  Iannone, Bajpai, and Afanasyev}]{flittner2018survey}
Flittner M, Mahfoudi MN, Saucez D, W{\"a}hlisch M, Iannone L, Bajpai V,
  Afanasyev A (2018) A survey on artifacts from {CoNEXT, ICN, IMC, and SIGCOMM
  Conference}s in 2017. ACM SIGCOMM Computer Communication Review 48(1):75--80

\bibitem[{Geurts et~al.(2006)Geurts, Ernst, and Wehenkel}]{geurts2006extremely}
Geurts P, Ernst D, Wehenkel L (2006) Extremely randomized trees. Machine
  learning 63(1):3--42

\bibitem[{Hamadicharef(2012)}]{hamadicharef2012scientometric}
Hamadicharef B (2012) Scientometric study of the {IEEE} transactions on
  software engineering 1980-2010. In: Proceedings of the 2011 2nd International
  Congress on Computer Applications and Computational Science, Springer, pp
  101--106

\bibitem[{Hassan et~al.(2017{\natexlab{a}})Hassan, Akram, and
  Haddawy}]{hassan2017identifying}
Hassan SU, Akram A, Haddawy P (2017{\natexlab{a}}) Identifying important
  citations using contextual information from full text. In: Proceedings of the
  17th ACM/IEEE Joint Conference on Digital Libraries, IEEE Press, pp 41--48

\bibitem[{Hassan et~al.(2017{\natexlab{b}})Hassan, Imran, Gillani, Aljohani,
  Bowman, and Didegah}]{hassan2017measuring}
Hassan SU, Imran M, Gillani U, Aljohani NR, Bowman TD, Didegah F
  (2017{\natexlab{b}}) Measuring social media activity of scientific
  literature: an exhaustive comparison of scopus and novel altmetrics big data.
  Scientometrics 113(2):1037--1057

\bibitem[{Heilig and Vo{\ss}(2014)}]{heilig2014scientometric}
Heilig L, Vo{\ss} S (2014) A scientometric analysis of cloud computing
  literature. IEEE Transactions on Cloud Computing 2(3):266--278

\bibitem[{Hirsch(2005)}]{hirsch2005index}
Hirsch JE (2005) An index to quantify an individual's scientific research
  output. Proceedings of the National academy of Sciences of the United States
  of America 102(46):16,569

\bibitem[{Igli{\v{c}} et~al.(2017)Igli{\v{c}}, Doreian, Kronegger, and
  Ferligoj}]{iglivc2017whom}
Igli{\v{c}} H, Doreian P, Kronegger L, Ferligoj A (2017) With whom do
  researchers collaborate and why? Scientometrics 112(1):153--174

\bibitem[{Kincaid et~al.(1975)Kincaid, Fishburne~Jr, Rogers, and
  Chissom}]{kincaid1975derivation}
Kincaid JP, Fishburne~Jr RP, Rogers RL, Chissom BS (1975) Derivation of new
  readability formulas (automated readability index, fog count and {Flesch}
  reading ease formula) for navy enlisted personnel. Naval Technical Training
  Command Millington TN Research Branch, Tech rep

\bibitem[{Leys et~al.(2013)Leys, Ley, Klein, Bernard, and
  Licata}]{leys2013detecting}
Leys C, Ley C, Klein O, Bernard P, Licata L (2013) Detecting outliers: Do not
  use standard deviation around the mean, use absolute deviation around the
  median. Journal of Experimental Social Psychology 49(4):764--766

\bibitem[{McLaughlin(1969)}]{mc1969smog}
McLaughlin GH (1969) {SMOG} grading---a new readability formula. Journal of
  reading 12(8):639--646

\bibitem[{Narin et~al.(1994)Narin, Olivastro, and
  Stevens}]{narin1994bibliometrics}
Narin F, Olivastro D, Stevens KA (1994) Bibliometrics/theory, practice and
  problems. Evaluation review 18(1):65--76

\bibitem[{Nattar(2009)}]{nattar2009indian}
Nattar S (2009) Indian journal of physics: A scientometric analysis.
  International Journal of Library and Information Science 1(4):043--61

\bibitem[{Nobre and Tavares(2017)}]{nobre2017scientific}
Nobre GC, Tavares E (2017) Scientific literature analysis on big data and
  internet of things applications on circular economy: a bibliometric study.
  Scientometrics 111(1):463--492

\bibitem[{Paul and Girju(2009)}]{paul2009topic}
Paul M, Girju R (2009) Topic modeling of research fields: An interdisciplinary
  perspective. In: Proceedings of the International Conference RANLP-2009, pp
  337--342

\bibitem[{Powell(2018)}]{powell2018these}
Powell K (2018) These labs are remarkably diverse--here's why they're winning
  at science. Nature 558(7708):19

\bibitem[{Rajendran et~al.(2011)Rajendran, Jeyshankar, and
  Elango}]{rajendran2011scientometric}
Rajendran P, Jeyshankar R, Elango B (2011) Scientometric analysis of
  contributions to journal of scientific and industrial research. International
  Journal of Digital Library Services 1(2):79--89

\bibitem[{Savi{\'c} et~al.(2017)Savi{\'c}, Ivanovi{\'c}, and
  Surla}]{savic2017analysis}
Savi{\'c} M, Ivanovi{\'c} M, Surla BD (2017) Analysis of intra-institutional
  research collaboration: a case of a {S}erbian faculty of sciences.
  Scientometrics 110(1):195--216

\bibitem[{Serenko et~al.(2009)Serenko, Bontis, and
  Grant}]{serenko2009scientometric}
Serenko A, Bontis N, Grant J (2009) A scientometric analysis of the proceedings
  of the {McMaster} world congress on the management of intellectual capital
  and innovation for the 1996-2008 period. Journal of Intellectual Capital
  10(1):8--21

\bibitem[{Valenzuela et~al.(2015)Valenzuela, Ha, and
  Etzioni}]{valenzuela2015identifying}
Valenzuela M, Ha V, Etzioni O (2015) Identifying meaningful citations. In: AAAI
  Workshop: Scholarly Big Data

\bibitem[{Wagner et~al.(2017)Wagner, Whetsell, and
  Leydesdorff}]{wagner2017growth}
Wagner CS, Whetsell TA, Leydesdorff L (2017) Growth of international
  collaboration in science: revisiting six specialties. Scientometrics
  110(3):1633--1652

\bibitem[{Waheed et~al.(2018)Waheed, Hassan, Aljohani, and
  Wasif}]{waheed2018bibliometric}
Waheed H, Hassan SU, Aljohani NR, Wasif M (2018) A bibliometric perspective of
  learning analytics research landscape. Behaviour \& Information Technology pp
  1--17

\bibitem[{Weatherburn(1949)}]{weatherburn1949first}
Weatherburn CE (1949) A first course mathematical statistics, vol 158. CUP
  Archive

\bibitem[{Yin and Zhi(2017)}]{yin2017dancing}
Yin Z, Zhi Q (2017) Dancing with the academic elite: a promotion or hindrance
  of research production? Scientometrics 110(1):17--41

\bibitem[{Zhu et~al.(2015)Zhu, Turney, Lemire, and Vellino}]{zhu2015measuring}
Zhu X, Turney P, Lemire D, Vellino A (2015) Measuring academic influence: Not
  all citations are equal. Journal of the Association for Information Science
  and Technology 66(2):408--427

\end{thebibliography}
\newpage
\end{document}